\documentclass[hyper,11pt,paper]{article}
\usepackage{jheppub}
\usepackage[usenames,dvipsnames,table]{xcolor}
\usepackage{graphicx,amsmath,amssymb,multirow,array,bm,mathrsfs}
\usepackage{epsf,amsfonts}
\usepackage[numbers,sort&compress]{natbib}
\newcommand{\beq}{\begin{equation}}
\newcommand{\eeq}{\end{equation}}

\newcommand{\ab}[2]{^{#1}_{\phantom{#1}#2}}

\newcommand{\spp}{L}

\newcommand{\prn}[1]{\left ( #1 \right )}
\newcommand{\brk}[1]{\left [ #1 \right ]}

\newcommand{\half }{\frac{1}{2}}

\newcommand{\form}[1]{\bm{#1}}
\newcommand{\diffF}{{\delta_\chi}}

\newcommand{\lieD}{\pounds}

\newcommand{\hodge}{{}^\star}

\newcommand{\fTheta}{\form{\Theta}}

\newcommand{\Mag}{\mathcal{M}}
\newcommand{\fJ}{\form{J}}

\newcommand{\fA}{\form{A}}

\newcommand{\fF}{\form{F}}

\newcommand{\fB}{\form{B}}

\newcommand{\El}{E}
\newcommand{\fE}{\form{\El}}

\newcommand{\fGamma}{\form{\Gamma}}

\newcommand{\fR}{\form{R}}

\newcommand{\BR}{{B_R}}
\newcommand{\fBR}{\form{B}_R}

\newcommand{\ER}{{\El_R}}
\newcommand{\fER}{\form{\El}_R}

\newcommand{\muR}{{\mbox{\large$\mu$}_R}}

\newcommand{\fu}{\form{u}}
\newcommand{\acc}{a}
\newcommand{\fa}{\form{\acc}}
\newcommand{\fomega}{\form{\omega}}

\newcommand{\fP}{{\form{\mathcal{P}}}}

\newcommand{\ICS}{{\form{I}}_{CS}}

\newcommand{\VP}{{\form{V}}_{\fP}}
\newcommand{\WCS}{{\form{W}}_{CS}}

\title{Anomaly inflow and thermal equilibrium}

\author[a,b]{Kristan Jensen,}
\author[c]{R. Loganayagam}
\author[d]{Amos Yarom}

\affiliation[a]{Department of Physics and Astronomy, University of Victoria, Victoria, BC V8W 3P6, Canada}
\affiliation[b]{C. N. Yang Institute for Theoretical Physics, SUNY, Stony Brook, NY 11794-3840, United States}
\affiliation[c]{Junior Fellow, Harvard Society of Fellows, Harvard University, Cambridge, MA 02138.}
\affiliation[d]{Department of Physics, Technion, Haifa 32000, Israel}

\emailAdd{kristanj@insti.physics.sunysb.edu}
\emailAdd{nayagam@gmail.com}
\emailAdd{ayarom@physics.technion.ac.il}

\abstract{Using the anomaly inflow mechanism, we compute the flavor/Lorentz non-invariant contribution to the partition function in a background with a $U(1)$ isometry. This contribution is a local functional of the background fields. By identifying the $U(1)$ isometry with Euclidean time we obtain a contribution of the anomaly to the thermodynamic partition function from which hydrostatic correlators can be efficiently computed. Our result is in line with, and an extension of, previous studies on the role of anomalies in a hydrodynamic setting. Along the way we find simplified expressions for Bardeen-Zumino polynomials and various transgression formulae.}

\preprint{YITP-SB-13-35}

\begin{document}
\maketitle

\section{Introduction} \label{S:intro}

Anomalies are a fascinating and unavoidable feature of quantum field theory. Their presence has been studied in great detail over the last forty-odd years leading to an improved understanding of the behavior of quantum field theories in general (via, e.g., the `t Hooft anomaly matching condition, or the Green-Schwarz mechanism) in addition to observable phenomena as predicted by the standard model (such as the pion decay rate). Quite surprisingly, little is known about the effect of anomalies in thermodynamic states or in configurations which are close to thermodynamic equilibrium.

Indeed, the past several years have seen a small revolution in our understanding of the dynamics induced by anomalies. Much of this success has been in the context of anomaly-induced response, by which we mean the part of the thermodynamic response of symmetry currents and energy-momentum that owes its existence to the presence of anomalies \cite{Bhattacharyya:2007vs,Erdmenger:2008rm,Banerjee:2008th,Torabian:2009qk,Son:2009tf,Kharzeev:2009p,Lublinsky:2009wr,Neiman:2010zi,Bhattacharya:2011tra,Kharzeev:2011ds,Loganayagam:2011mu,Neiman:2011mj,Dubovsky:2011sk,Kimura:2011ef,Lin:2011aa,Gao:2012ix,Banerjee:2012iz,Jensen:2012jy,Jain:2012rh,Valle:2012em,Bhattacharyya:2012xi,Golkar:2012kb,Jensen:2012kj,Chen:2012ca,Manes:2012hf,Loganayagam:2012zg,Bhattacharyya:2013ida,Valle:2013aia,Megias:2013uua}. This effect has been studied in diverse arenas, from relativistic hydrodynamics to the physics of topologically non-trivial edge states (see, for instance, \cite{Qi:arXiv1008.2026}). For instance, anomaly-induced transport leads to the manifestation of the chiral magnetic effect and chiral vortical effect in hydrodynamics \cite{Kharzeev:2010gr}.

At nonzero temperature, a generic field theory has a finite number of gapless degrees of freedom which correspond to the relaxation of conserved quantities. Hydrodynamics is the universal long-wavelength effective theory which describes the evolution of those gapless modes as well as their response to a (slowly varying) background gauge field and metric. We may take the fields of hydrodynamics to be the parameters which describe the equilibrium state: a temperature $T$, a local rest frame characterized by a timelike vector $u^{\mu}$ satisfying $u^2=-1$ which we refer to as the velocity field, and if the theory includes a conserved charge, a local chemical potential $\mu$. These (classical) fields are referred to as the hydrodynamic variables.

In what follows, we will be interested in a certain subset of solutions to the hydrodynamic equations of motion (i.e., energy and charge conservation) which we will refer to as hydrostatic configurations. Roughly speaking, hydrostatic configurations may be thought of as time-independent solutions to the hydrodynamic equations of motion in the presence of a slowly varying time independent background gauge field and metric. We refer the reader to Section~\ref{S:minimalhydrostatics} for a more precise definition of hydrostatic configurations and Section~\ref{S:hydrostatics} for an extensive discussion. The virtue of hydrostatic configurations, in the present context, is that their physical content is captured by a generating functional $W_{QFT}$ which is local in the background sources~\cite{Banerjee:2012iz,Jensen:2012jh}.  In other words, the entire dependence of the stress tensor and charge current on the hydrodynamic fields and sources is captured by $W_{QFT}$. All of the anomaly-induced response studied in the literature can be characterized by its effect on correlation functions in a hydrostatic configuration i.e., by its effect on variations of $W_{QFT}$ with respect to the background gauge field and metric.\footnote{This statement is true with a judicious choice of hydrodynamic frame. There are additional subtleties associated with the commonly used Landau frame as discussed in~\cite{Bhattacharyya:2013ida}. } For instance, one can argue that the zero frequency two point function of the covariant current and stress tensor of a $3+1$ dimensional theory with a $U(1)^3$ anomaly characterized by a coefficient $c_{_A}$ and a mixed gauge-gravitational anomaly characterized by a coefficient $c_m$ are given uniquely by~\cite{Amado:2011zx,Golkar:2012kb,Jensen:2012kj}
\beq
	\label{E:4dCVE}
	\langle J_{cov}^i(\vec{k})T_{cov}^{0j}(-\vec{k})\rangle =  i (8 \pi^2 c_m  T^2 + 3 c_{_A} \mu^2) \epsilon^{ijk}k_k + O(k^2)\,,
\eeq
in the small momentum limit.\footnote{Our conventions are such that for a left Weyl fermion with positive unit charge $c_{_A} = 1/(24 \pi^2)$ and $c_m = 1/(192 \pi^2)$.} This statement can be rephrased more elegantly in terms of the helicity~\cite{Loganayagam:2012zg}. Given the momentum operator $\vec{P}$ and  the angular momentum operator $\vec{L}$,  the thermal helicity  $\langle \vec{L}\cdot\vec{P} \rangle$ per unit volume   of a 4d quantum field theory is given by  $-6T(8 \pi^2 c_m  T^2 \mu +  c_A \mu^3)$. We emphasize that the relation~\eqref{E:4dCVE} is valid only for normal fluids. The thermal helicity for fluids which have gapless degrees of freedom beyond the hydrodynamic modes (such as superfluids or zero-temperature Fermi liquids) takes a different form \cite{Lin:2011aa,Bhattacharya:2011tra,Bhattacharyya:2012xi}. We refer the reader to Section~\ref{S:conclusion} and~\cite{Jensen:2013vta} for further discussion. 

In the presence of anomalies the variation of $W_{QFT}$ with respect to sources is not gauge invariant. Therefore, it is useful to decompose $W_{QFT}$ into a gauge-invariant term and a non-gauge-invariant term which we will refer to as an anomalous term,\footnote{Strictly speaking $W_{anom}$ should be understood as a representative of an equivalence class. Any two representatives of this equivalence class differ by the addition of a covariant term.} 
\begin{equation}
	W_{QFT} = W_{gauge-invariant}+W_{anom}\,.
\end{equation}
Since $W_{anom}$ is responsible for the non-gauge-invariance of $W_{QFT}$ it depends explicitly on the anomalies of the theory. Somewhat surprisingly, there are also contributions to $W_{gauge-invariant}$ which are uniquely fixed by the anomalies. For instance, in the $3+1$ dimensional example described in Equation~\eqref{E:4dCVE} the $T^2$ term is of the latter type while the $\mu^2$ term is of the former type~\cite{Jensen:2012kj}. To make the distinction between the two types of contributions of anomalies to the generating function more explicit we split $W_{gauge-invariant}$ into a contribution which is completely fixed by the anomalous content of the theory which we refer to as $W_{transcendental}$ and a non-anomalous contribution, $W_{gauge-invariant} = W_{transcendental}+W_{non-anomalous}$. All known contributions of the anomaly to hydrostatic configurations, to date, are completely determined by $W_{anom}$ and $W_{transcendental}$.  For theories in $1+1$ and $3+1$ dimensions both contributions to the generating function have been computed \cite{Son:2009tf,Neiman:2010zi,Landsteiner:2011cp,Manes:2012hf,Jain:2012rh,Banerjee:2012iz,Jensen:2012kj}. In this work we will obtain an explicit expression for $W_{anom}$ in arbitrary dimensions, and formulate it in such a way so that the evaluation of $W_{transcendental}$ can be simplified. We leave a study of $W_{transcendental}$ to a future publication.

To be more precise, in this work we will construct $W_{anom}$ for global anomalies by which we mean conservation laws which become anomalous only in the presence of external background fields. Thus, we exclude from our analysis anomalies associated with gauge symmetries such as the Axial-Vector-Vector anomaly of the standard model. For anomalies associated with gauge symmetries expressions of the form \eqref{E:4dCVE} generally receive quantum corrections~\cite{Jensen:2013vta,Golkar:2012kb,Hou:2012xg}. 

In a general state it is impossible to write down a local Lorentz invariant expression for the anomalous contribution to the generating function since that would imply that with a judicious choice of counterterms one may get rid of the anomaly altogether. However, when the background sources have an isometry direction then it is possible to obtain an explicit local expression for $W_{anom}$. In hydrostatic configurations we have such an isometry direction by definition, namely time. The local expression may be constructed as follows. Consider the anomaly polynomial $\fP$ of a $2n$-dimensional theory. (We fix our conventions for $\fP$ in Appendix~\ref{A:inflow}, where we also provide a review of the anomaly inflow mechanism.) This polynomial is a formal $2n+2$ form which is a polynomial in the Riemann tensor two-form $\form{R}^{\mu}{}_{\nu} = \frac{1}{2}R^{\mu}{}_{\nu\rho\sigma}dx^{\rho}\wedge dx^{\sigma}$, where $R\ab{\mu}{\nu\rho\sigma}$ is the Riemann curvature tensor, and field strength $\form{F} = \frac{1}{2}F_{\mu\nu}dx^{\mu}\wedge dx^{\nu}$. Both $\fR\ab{\mu}{\nu}$ and $\fF$ are constructed from the Christoffel connection $\form{\Gamma}^{\mu}{}_{\nu}={\Gamma}^{\mu}{}_{\nu\rho} dx^{\rho}$ and gauge connection $\form{A} = A_{\mu}dx^{\mu}$. We will adopt a notation where form fields are given by boldface characters. From the anomaly polynomial one may construct a Chern Simons form $\form{I}_{CS}$ via $\fP = d \form{I}_{CS}$. This Chern-Simons form is a polynomial in the connections and field strengths. From the connections $\form{A}$ and $\form{\Gamma}$ we can construct the hatted connections
\beq
\label{E:hatConnection}
	\hat{\form{A}} = \form{A}+\mu \form{u}
	\qquad 
	\hat{\form{\Gamma}}\ab{\mu}{\nu} = \form{\Gamma}\ab{\mu}{\nu} + (\mu_R)\ab{\mu}{\nu}\form{u}\,.
\eeq
where we have now defined $(\mu_R)^{\mu}{}_{\nu} = TD_{\nu} \left( \frac{u^{\mu}}{T} \right)$ and $\form{u} = u_{\mu}dx^{\mu}$, along with the covariant derivative $D_{\mu}$. These hatted connections give rise to hatted field strengths $\hat{\form{R}}^{\mu}{}_{\nu}$ and $\hat{\form{F}}$. From the hatted connections one can construct a hatted Chern-Simons form $\hat{\form{I}}_{CS}$ defined via $\hat{\form{I}}_{CS} = \form{I}_{CS}(\hat{\form{A}},\,\hat{\form{\Gamma}},\,\hat{\form{F}},\,\hat{\form{R}}) $. The physical reasoning behind the construction of these hatted connections which may seem somewhat mysterious at this point will be elaborated on in Section \ref{S:hydrostatics}. 

To construct $W_{anom}$, consider the $2n+2$ form $\form{u} \wedge \left( \form{I}_{CS} - \hat{\form{I}}_{CS} \right)$. As we will argue in the main text this $2n+2$ form is a polynomial in the vorticity two-form $\form{\omega} = \frac{1}{2}\omega_{\mu\nu}dx^{\mu}\wedge dx^{\nu}$ (with $\omega_{\mu\nu}=\frac{P_{\mu\rho}P_{\nu\sigma}}{2} \left( D^{\rho}u^{\sigma}-D^{\sigma}u^{\rho}\right)$ and $P_{\mu\nu} = g_{\mu\nu} + u_{\mu}u_{\nu}$ a projection matrix) which vanishes when we set $\form{\omega}=0$. In equations,  $\form{u} \wedge \left( \form{I}_{CS} - \hat{\form{I}}_{CS} \right) = \sum_{k=0}^n \form{c}_k \wedge (2\fomega)^{k+1}$, where $\form{c}_k$ is a $2(n-k)$ form. Our claim is that $W_{anom}$ is given by the integral of the $2n$ form $-\sum_{k=0}^n \form{c}_k \wedge (2\fomega)^k$. More formally, we write
\beq
\label{E:Wanom}
	W_{anom} = - \int \frac{\form{u}}{2\form{\omega}}\wedge \left( \form{I}_{CS}-\hat{\form{I}}_{CS}\right)\,.
\eeq

With $W_{anom}$ at hand one may now compute the anomalous contribution to the consistent stress tensor and current
\begin{equation}
	T_{anom}^{\mu\nu} = \frac{2}{\sqrt{-g}} \frac{\delta W_{anom}}{\delta g_{\mu\nu}}\,,
	\qquad
	J_{anom}^{\mu} = \frac{1}{\sqrt{-g}} \frac{\delta W_{anom}} {\delta A_{\mu}}\,,
\end{equation}
or any correlation function thereof.
The non-gauge-invariance of $W_{anom}$ implies that the anomalous current and stress tensor also fail to be gauge-invariant. As argued by Bardeen and Zumino~\cite{Bardeen:1984pm}, one may always construct a covariant stress tensor and current by adding appropriate compensating currents $T^{\mu\nu}_{BZ}$ and $J^{\mu}_{BZ}$ which are polynomials in the connections and field strengths,
\begin{equation}
	T^{\mu\nu}_{\mathcal{P}} = T_{anom}^{\mu\nu} + T^{\mu\nu}_{BZ} 
	\qquad
	J^{\mu}_{\mathcal{P}} = J_{anom}^{\mu} + J^{\mu}_{BZ}\,.
\end{equation}
We derive explicit and succinct expressions for the BZ polynomials and anomalous Ward identities in terms of the Chern-Simons form $\ICS$ and anomaly polynomial $\fP$ in Appendix \ref{A:inflow} and~\ref{A:conCovAnom}.

A second construction which we elaborate on in this paper allows us to obtain the covariant anomaly-induced stress tensor and current, $T^{\mu\nu}_{\mathcal{P}}$ and $J^{\mu}_{\mathcal{P}}$ without carrying out the explicit variation of $W_{anom}$. More explicitly we claim that given an anomaly polynomial $\fP$, one can construct the formal $2n+1$ form
\begin{equation}
\label{E:VPIntro}
	\VP=\frac{\fu}{2\fomega}\wedge \left( \form{\mathcal{P}}-\hat{\form{\mathcal{P}}}\right)\,,
\end{equation}
from which we take derivatives to obtain $T^{\mu\nu}_{\mathcal{P}}$ and $J^{\mu}_{\mathcal{P}}$ as
\begin{align}
\label{E:1ptFns}
\hodge\form{J}_{\fP} & = \frac{\partial\form{V}_{\mathcal{P}}}{\partial \form{B}}\,,
&\hodge\form{q}_{\fP} &= \frac{\partial\form{V}_{\mathcal{P}}}{\partial (2\form{\omega})}\,,
& \hodge (\form{\spp}_{\fP})\ab{\mu}{\nu} & = \frac{\partial \form{V}_{\mathcal{P}}}{\partial (\form{B}_R)\ab{\nu}{\mu}}\,.
\end{align}
In writing~\eqref{E:1ptFns} we have represented the flavor current $J_{\mathcal{P}}^{\mu}$, heat current $q_{\mathcal{P}}^{\mu}$, and $(\spp_{\mathcal{P}})\ab{\rho\mu}{\nu}$ in terms of their Hodge duals (we review our conventions for the Hodge star operator in Appendix \ref{ss:forms}), and have defined the magnetic flavor field and magnetic curvatures
\beq
	B_{\mu\nu} \equiv P_{\mu\rho}P_{\nu\sigma}F^{\rho\sigma}\,,
	\qquad
	(B_R)^{\mu}_{\phantom{\mu}\nu\rho\sigma} \equiv P_{\rho}{}^{\alpha}P_{\sigma}{}^{\beta}R^{\mu}{}_{\nu\alpha\beta}\,,
\eeq
where $P_{\mu\nu} \equiv g_{\mu\nu}+u_{\mu}u_{\nu}$ is the transverse projector to the velocity vector. The anomaly-induced covariant stress tensor is given by
\beq
\label{E:1ptTP}
T^{\mu\nu}_{\mathcal{P}} = u^{\mu}q_{\mathcal{P}}^{\nu}+u^{\nu}q_{\mathcal{P}}^{\mu} + D_{\rho}\left[ \spp^{\mu[\nu\rho]}_{\mathcal{P}}+\spp^{\nu[\mu\rho]}_{\mathcal{P}} - \spp^{\rho(\mu\nu)}_{\mathcal{P}}\right]\,,
\eeq
where the curved and straight brackets indicate symmetrization or anti-symmetrization,
\beq
A^{(\mu\nu)} =\frac{1}{2}(A^{\mu\nu}+A^{\nu\mu})\,, \qquad A^{[\mu\nu]}=\frac{1}{2}(A^{\mu\nu}-A^{\nu\mu})\,.
\eeq
For pure $U(1)$ anomalies, it was previously observed in~\cite{Loganayagam:2011mu} that one can generate the anomaly-induced currents using a functional $\VP$ in exactly the same way as in~\eqref{E:1ptFns}.

In order to make our work self-contained, we begin with a sequence of pedagogical sections (Sections~\ref{S:minimalhydrostatics}-\ref{S:abelAnomTrans}). In Section~\ref{S:minimalhydrostatics} we bring the reader up to speed on hydrostatics. We then re-derive known results regarding abelian anomalies using a novel framework which we will later apply to more  general anomalies. In Section~\ref{S:hydrostatics} we give a more extensive and modern discussion of hydrostatics including its covariant formulation and the role of hatted connections. A proof of \eqref{E:Wanom} and \eqref{E:1ptFns} for arbitrary anomalies, including gravitational ones, is presented in Section~\ref{S:allanomalies}. We conclude by discussing our results as well as prospects for future work in Section~\ref{S:conclusion}. Many of the technical details have been relegated to the Appendices.

\section{A minimalist's introduction to hydrostatics} \label{S:minimalhydrostatics}

Consider the hydrodynamic equations for a fluid placed in a slowly varying time-independent background. The background consists of a gauge field $A_{\mu}$ and metric $g_{\mu\nu}$ which we refer to as external sources. A hydrostatic configuration is a time-independent solution of these equations which is a local functional of these sources. We take it as a fundamental postulate that such a configuration exists. 

The simplest example of hydrostatic equilibrium is a fluid configuration in flat space where all hydrodynamical fields are constant, i.e., thermodynamic equilibrium. When a fluid is placed on a generic time-independent background,  we expect it to relax to a hydrostatic configuration at late times. Note that in a generic background spatial gradients of the hydrodynamic fields will not vanish. For instance, a hydrostatic configuration in the presence of a time-independent electric field has charge density gradients.

The  energy-momentum and charge distributions in hydrostatics are most efficiently summarised by  a  local generating functional $W_{QFT}[g_{\mu\nu},A_\mu]$  of the time-independent sources . This `hydrostatic' generating function is closely related to a Euclidean partition function $W_E$ \cite{Banerjee:2012iz,Jensen:2012jh}.

Let $K^\mu$ be the time-like Killing vector of the time independent background. We will choose a coordinate system where $K^{\mu}\partial_{\mu}=\partial_t$ (While our current formulation involves choosing a particular time direction, in Section \ref{S:hydrostatics} we will consider a covariant formulation of hydrostatics.) The Killing vector $K^{\mu}$ can be used to define a Euclidean partition function $Z_E$. This is done by Wick-rotating the time direction, compactifying it with coordinate periodicity $\beta$, and imposing thermal boundary conditions around the resulting thermal circle. The logarithm of $Z_E$ will give us the Euclidean generating functional $W_E$ for the connected correlation functions of the theory,
\beq
W_E = -i \ln {Z}_{E}\,.
\eeq
For slowly varying backgrounds, we can ``un-Wick rotate"  $W_E$  to get the hydrostatic generating  function $W_{QFT}$, defined earlier.

The  procedure outlined above then fixes the hydrostatic temperature, chemical potential and velocity profile in terms of the background sources to be
\begin{align}
\begin{split}
\label{E:hydrostatSol}
	T^{-1} &=  \beta \sqrt{-g_{00}} \,,\\ 
	\frac{\mu}{T} &=  \beta A_0 \,, \\
	u^{\mu} &= \frac{\delta^{\mu}_{0}}{\sqrt{-g_{00}}}\,,
\end{split}
\end{align}
up to a change of hydrodynamic frame. All other hydrodynamic quantities are constructed out of  the hydrodynamic variables in \eqref{E:hydrostatSol} together with the sources $A_{\mu}$ and $g_{\mu\nu}$. The expansion, acceleration vector, vorticity tensor, shear tensor, and electric and magnetic fields are respectively given by
\begin{align}
\label{E:defEB}
	\notag
	\vartheta &= \nabla_{\mu}u^{\mu}\,, & 
	a_{\mu} &= u^{\nu}\nabla_{\nu}u_{\mu}\,, \\
	\omega_{\mu\nu}&=\frac{P_{\mu\rho}P_{\nu\sigma}}{2} \left( \nabla^{\rho}u^{\sigma}-\nabla^{\sigma}u^{\rho}\right)\,, & 
	\sigma_{\mu\nu} &=\frac{P_{\mu\rho}P_{\nu\sigma}}{2} \left( \nabla^{\rho}u^{\sigma}+\nabla^{\sigma}u^{\rho}\right) - \frac{\vartheta}{d-1}P_{\mu\nu}\,, 
	\\
	\notag
	\qquad E_{\mu}&=F_{\mu\nu}u^{\nu}\,,& B_{\mu\nu}&=P_{\mu\rho}P_{\nu\sigma}F^{\rho\sigma}\,,
\end{align}
where $P_{\mu\nu} = g_{\mu\nu} + u_{\mu}u_{\nu}$ and $d$ is the number of space-time dimensions.
In form notation we can write
\begin{align}
\begin{split}
\label{E:formnotation}
	d\form{A} &= \form{u} \wedge \form{E} + \form{B} \\
	d\form{u} & = -\form{u}\wedge \form{a} + 2 \form{\omega}\,,
\end{split}
\end{align}

The fact that the system is in equilibrium, i.e., Equations \eqref{E:hydrostatSol} together with the time-independence of all fields, imposes various interrelations between the fields which imply that the field configuration is dissipationless \cite{Jensen:2012jh}. We have
\beq
\label{E:chemEquil}
	\nabla_{\mu} \left(\frac{u_{\nu}}{T} \right) + \nabla_{\nu} \left(\frac{u_{\mu}}{T} \right) = 0\,, 
	\qquad
	\nabla_{\mu}T + a_{\mu}  T= 0\,,
	\qquad 
	\nabla_\mu \mu +a_{\mu} \mu = E_{\mu}\,.
\eeq
These equations imply that the system is in thermal and chemical equilibrium and, moreover, that the shear and expansion vanish, $\sigma_{\mu\nu} = 0$ and $\vartheta=0$.

In this work we will often use hatted connections, e.g., $\hat{\form{A}} = \form{A} + \mu \form{u}$ (see Equation~\eqref{E:hatConnection}) and the corresponding electric and magnetic field $d\hat{\form{A}} = \fu \wedge \hat{\fE} + \hat{\fB}$. The virtue of $\hat{\form{F}} = d\hat{\form{A}}$ is that in hydrostatic equilibrium, the field strength $\hat{\form{F}}$ is transverse to the velocity field
\beq
\label{E:hatF}
\hat{\fF}=d\hat{\fA} = \fB+2\fomega \mu + \fu\wedge \left( \fE-(d+\fa)\mu\right)=\fB+2\fomega\mu\,,
\eeq
where we have used~\eqref{E:chemEquil}.

We return our attention to the hydrostatic generating functional $W_{QFT}$.
In the absence of gauge and gravitational anomalies the generating function, $W_{QFT}$ will be constructed from the most general gauge invariant and coordinate reparametrization-invariant combination of the fields $T$, $\mu$ and $u^{\mu}$ (as given in \eqref{E:hydrostatSol}) and the background fields $g_{\mu\nu}$ and $A_{\mu}$. It is often useful to organize the possible contributions to $W_{QFT}$ in a derivative expansion. We refer the reader to \cite{Banerjee:2012iz,Jensen:2012jh,Jensen:2012kj} for further details.

For a theory with anomalies, the Wess-Zumino consistency conditions demands that $W_{QFT}$ must exhibit a particular anomalous variation under gauge and coordinate transformations~\cite{Wess:1971yu}. As a result $W_{QFT}$ takes the form of a gauge and diffeomorphism-invariant term plus an extra, additive, and local contribution which we denote by $W_{anom}$. This additive term is not gauge and diffeomorphism-invariant and is constructed in such a way to correctly reproduce the anomalous variation of $W_{QFT}$. An explicit construction of $W_{anom}$ for arbitrary anomalies in arbitrary dimensions is one of the main results of this paper.

\section{Partition functions for theories with abelian anomalies} \label{S:abelAnomW}

Consider the hydrostatic generating functional $W_{QFT}$ for a $2n$-dimensional theory with a $U(1)$ anomaly.
We may decompose $W_{QFT}$ into a gauge-invariant contribution and an anomalous contribution,
\begin{equation}
\label{E:WQFT}
	W_{QFT}=W_{anom}+W_{gauge-invariant}\,.
\end{equation}
The separation in \eqref{E:WQFT} is somewhat arbitrary since there is an equivalence class of expressions for $W_{anom}$ under the addition of local gauge-invariant terms. In this work we advocate for a particular representative for $W_{anom}$
given in \eqref{E:Wanom}
\beq
\label{E:WanomU1}
W_{anom} = - \int \form{W}_{CS} = - \int \frac{\fu}{2\fomega}\wedge \left[ \form{I}_{CS} - \hat{\form{I}}_{CS}\right]\,,
\eeq
where $\form{I}_{CS}=c_{_A} \fA\wedge \fF^n$ is the Chern-Simons term associated with the abelian anomaly, and $\hat{\form{I}}_{CS}=c_{_A}\hat{\form{A}} \wedge\hat{\form{F}}^n$ is the Chern-Simons form evaluated for the hatted connection $\hat{\form{A}}=\fA+\mu \fu$. The appearance of the hatted connection might seem a bit mysterious at this point. In section \ref{S:hydrostatics} we attempt to elucidate its origin.

As we will now argue $W_{anom}$ correctly reproduces the anomalous gauge variation of the hydrostatic theory and therefore also reproduces existing results in the literature \cite{Banerjee:2012iz,Jain:2012rh,Banerjee:2012cr}. Consider the anomaly inflow mechanism of Callan and Harvey~\cite{Callan:1984sa}; we place our $2n$ dimensional field theory on the boundary of a $2n+1$ dimensional space-time $\mathcal{M}$. We denote the generating functional of the $2n$-dimensional theory as $W_{QFT}$. On $\mathcal{M}$ and its boundary $\partial\mathcal{M}$ we can define a covariant generating functional
\beq
\label{E:WcovDef}
W_{cov}[A,g] = W_{QFT}[A,g] + \int_{\mathcal{M}}\form{I}_{CS}[A]\,.
\eeq
The generating function $W_{cov}$ is gauge invariant while the Chern-Simons term $\form{I}_{CS}$ is gauge invariant up to boundary terms. The charge associated with this gauge invariance is conserved in $\mathcal{M}$ but may be deposited on the boundary $\partial\mathcal{M}$ rendering $W_{QFT}$ anomalous. Alternately, the reader familiar with Hall insulators may regard the second term on the right hand side of \eqref{E:WcovDef} as the action of a $2n+1$-dimensional Hall insulator. In the presence of background electromagnetic fields, Hall insulators carry edge currents on their boundary. In this instance the edge current is the (covariant) charge current of the field theory on $\partial\mathcal{M}$.

Under a gauge variation $\delta_{\Lambda}\fA=d\Lambda$ the Chern-Simons form varies as
\beq
\delta_{\Lambda} \form{I}_{CS} = c_{_A} \delta_{\Lambda} \fA \wedge \fF^n = d\left[ c_{_A} \Lambda \fF^n\right] = d\left[ \Lambda \frac{\partial \form{I}_{CS}}{\partial\fA}\right]\,,
\eeq
where in the last equality we have written the boundary term in a way that will be easier to generalize to more complicated non-abelian and gravitational anomalies. The gauge-invariance of $W_{cov}$ in \eqref{E:WcovDef} then gives
\beq
\label{E:deltaWinflow}
\delta_{\Lambda} W_{QFT} = - \int \Lambda\frac{\partial\form{I}_{CS}}{\partial\fA}\,.
\eeq
Meanwhile, the anomalous variation of $\form{W}_{CS}$ gives
\begin{align}
\begin{split}
\label{E:deltaWCSabelian}
\delta_{\Lambda}\form{W}_{CS}& = \delta_{\Lambda}\fA\wedge \frac{\partial\form{W}_{CS}}{\partial\fA} =d\Lambda \wedge\frac{\partial\form{W}_{CS}}{\partial\fA}= d\left(\Lambda\frac{\partial\form{W}_{CS}}{\partial\fA}\right) - \Lambda d\left(\frac{\partial\form{W}_{CS}}{\partial\fA}\right)
\\
& =d\left(\Lambda\frac{\partial\form{W}_{CS}}{\partial\fA}\right) + \Lambda d\left[\frac{\fu}{2\fomega}\right]\wedge \left( \frac{\partial\form{I}_{CS}}{\partial\fA}-\frac{\partial\hat{\form{I}}_{CS}}{\partial\hat{\fA}}\right) - \Lambda \frac{\fu}{2\fomega}\wedge d\left( \frac{\partial\form{I}_{CS}}{\partial\fA}-\frac{\partial\hat{\form{I}}_{CS}}{\partial\hat{\fA}}\right)
\\
& = d\left( \Lambda \frac{\partial\form{W}_{CS}}{\partial\fA}\right) + \Lambda \left( \frac{\partial\form{I}_{CS}}{\partial\fA}-\frac{\partial\hat{\form{I}}_{CS}}{\partial\hat{\fA}}\right)\,.
\end{split}
\end{align}
In going from the second line to the third we have used that
\beq
d\left(\frac{\partial\form{I}_{CS}}{\partial\fA}\right)=c_{_A} d\fF^n = 0\,,
\eeq
and similarly for the derivative of the hatted Chern-Simons form. We also used the formal identity
\beq
\label{E:magicFormula}
d\left[ \frac{\fu}{2\fomega}\right] = 1\,,
\eeq
valid when acting on a polynomial of at least degree one in $\fomega$, which we will now prove. Consider  $d\fu = 2\fomega - \fu\wedge \fa$ where $\fa = \acc_\mu dx^\mu$ is the acceleration 1-form. Then,
\begin{equation}
\begin{split}
0 = \fu\wedge d^2\fu = \fu\wedge d(2\fomega) - \fu\wedge \fa\wedge (2\fomega) =  \fu\wedge d(2\fomega)  + d\fu\wedge (2\fomega) -(2\fomega)^2\,,
\end{split}
\end{equation}
so that
\begin{equation}
\begin{split}
\label{E:ProofMagic}
d\brk{\frac{\fu}{2\fomega}}= \frac{u\wedge d(2\fomega)  + d\fu\wedge (2\fomega) } { (2\fomega)^2 } =1\,.
\end{split}
\end{equation}
By~\eqref{E:Wanom},~\eqref{E:deltaWCSabelian} becomes
\beq
\delta_{\Lambda}W_{anom} = - \int \delta_{\Lambda} \form{W}_{CS} = - \int \Lambda \left( \frac{\partial\form{I}_{CS}}{\delta \fA}-\frac{\partial\hat{\form{I}}_{CS}}{\partial\hat{\fA}}\right)\,.
\eeq
The second term in the integrand is
\beq
\frac{\partial \hat{\form{I}}_{CS}}{\partial\hat{\fA}}=c_{_A} \hat{\fF}^n\,.
\eeq
In hydrostatic equilibrium $\hat{\fF}=\fB+2\fomega\mu $ is a purely spatial form (see~\eqref{E:hatF}).  
It then follows that the $2n$-form $\hat{\fF}^n$ vanishes in $2n$-dimensions since it has no leg along the time direction, so that the gauge variation of $W_{anom}$ is given by
\beq
\delta_{\Lambda} W_{anom} = - \int \Lambda \frac{\partial \form{I}_{CS}}{\partial\fA}\,,
\eeq
which is the desired result~\eqref{E:deltaWinflow}.

We have shown that the expression~\eqref{E:Wanom} from the Introduction reproduces the correct anomalous variation of the hydrostatic $W_{QFT}$ for abelian anomalies. The proof that $W_{anom}$ reproduces the correct anomalous variation of $W_{QFT}$ can be extended to gravitational and non-abelian anomalies. In Section \ref{S:hydrostatics} we introduce, among other things, the non-abelian and spin chemical potential ($\mu_R$) which are the non-abelian and gravitational counterparts of the chemical potential $\mu$ used in this section. 
Using these chemical potentials one can define corresponding hatted connections, $\hat{A}_{\mu}$ and $\hat{\Gamma}\ab{\mu}{\nu\rho}$, which allow us to construct $W_{anom}$ for general anomalies.

\section{Abelian anomaly-induced transport} 
\label{S:abelAnomTrans}

We turn our attention from the anomalous gauge variation of $W_{QFT}$ to the study of the anomaly induced energy-momentum and $U(1)$ flavor currents.
The consistent current and stress tensor can be computed by varying the generating function $W_{QFT}$,
\beq
\delta W_{QFT} = \int d^{2n}x \sqrt{-g} \left[ \delta A_{\mu} J^{\mu} + \frac{1}{2}\delta g_{\mu\nu}T^{\mu\nu}\right]\,.
\eeq
In the previous section we  found an explicit expression for $W_{anom}$ which reproduces the anomalous gauge variation of $W_{QFT}$.
We proceed to compute the anomaly-induced charge current and stress tensor, i.e., the current and stress tensor that follow from varying $W_{anom}$,
\beq
\delta W_{anom} = \int d^{2n}x \sqrt{-g}\left[ \delta A_{\mu}J^{\mu}_{anom} + \frac{1}{2}\delta g_{\mu\nu}T_{anom}^{\mu\nu}\right]\,.
\eeq
It is possible (and not too difficult) to compute the anomalous contribution to the consistent current and stress tensor, $J^{\mu}_{anom}$ and $T_{anom}^{\mu\nu}$, by explicitly varying $\form{W}_{CS}$. We do this in Appendix \ref{A:deltaVPWCS}.

By construction, the consistent current $J^{\mu}$ varies under gauge transformations \cite{Bardeen:1984pm}.\footnote{
The non gauge invariance of $J^{\mu}$ follows from $ \delta_\Lambda \delta W_{QFT} =  \delta  \delta_\Lambda W_{QFT}$ implying that  $\int d^{2n}x \sqrt{-g} \,\delta A_{\mu} \delta_\Lambda J^{\mu}  = -\int \Lambda \frac{\partial\ICS}{\partial\fA}$ + (Boundary terms) where we have used~\eqref{E:deltaWinflow}. We refer the reader to Appendix \ref{A:inflow} for a more careful derivation.
}
However, it is possible to add to the consistent current a polynomial in $A_{\mu}$ and $F_{\mu\nu}$, the Bardeen-Zumino (BZ) polynomial $J_{BZ}^{\mu}$, such that $J_{cov}^{\mu} = J^{\mu} + J_{BZ}^{\mu}$, the covariant current, is invariant under gauge transformations. Rather than varying $\form{W}_{CS}$ and computing the Bardeen Zumino currents to obtain the covariant currents we elect to take a more straightforward approach; the covariant current can be computed by varying $W_{cov}$ defined in \eqref{E:WcovDef} directly,
\beq
\delta W_{cov} = \int d^{2n}x \sqrt{-g} \left[ \delta A_{\mu} J_{cov}^{\mu} + \frac{1}{2}\delta g_{\mu\nu} T_{cov}^{\mu\nu}\right] + (\text{bulk variation})\,.
\eeq
In what follows we carry out this variation.

Before varying $W_{cov}$, it is useful to decompose the bulk Chern-Simons form $\ICS$ in a specific way. Using $d \brk{\frac{\fu}{2\fomega}} =1$, we have
\beq
\label{E:dWCS}
\ICS-\hat{\form{I}}_{CS} = d\left[ \frac{\fu}{2\fomega}\wedge \left( \ICS-\hat{\form{I}}_{CS}\right)\right] + \frac{\fu}{2\fomega}\wedge \left( d\ICS-d\hat{\form{I}}_{CS}\right) = d\form{W}_{CS} + \frac{\fu}{2\fomega}\wedge \left( \form{\mathcal{P}}-\hat{\form{\mathcal{P}}}\right)\,,
\eeq
where we have used  $d\ICS=\fP$, $\fP(\fF)=c_{_A}\fF^{n+1}$ is the anomaly polynomial, and $\hat{\fP}=\fP(\hat{\fF})$. To avoid cluttering~\eqref{E:dWCS} we have used $\fP$ and $\hat{\fP}$ in place of $\fP(\fF)$ and $\fP(\hat{\fF})$. In the remainder of this Section we will continue to use these conventions, i.e., $\fP=\fP(\fF)$ and $\hat{\fP}=\fP(\hat{\fF})$. Equation~\eqref{E:dWCS} can be rewritten in the form
\beq
\label{E:ICSdecomp}
\ICS-\hat{\form{I}}_{CS}= \VP + d\form{W}_{CS}\,,
\eeq
where we have defined
\beq
\label{E:VPabelian}
\VP=\frac{\fu}{2\fomega}\wedge \left( \form{\mathcal{P}}-\hat{\form{\mathcal{P}}}\right)\,.
\eeq
From~\eqref{E:VPabelian}, $\VP$ is a gauge-invariant $2n+1$-form constructed out of $\form{u}$, $\form{\omega}$, $\form{F}$ and $\hat{\form{F}}$. Using~\eqref{E:defEB} we may decompose the field strength $\fF$ into an electric part and a magnetic part,
\beq
\fF = \fu\wedge \fE + \fB\,,
\eeq
allowing us to write $\VP$ in the form
\beq
\label{E:VPalt}
\VP = \frac{\fu}{2\fomega}\wedge \left( \form{\mathcal{P}}(\fB) - \form{\mathcal{P}}(\fB+2\fomega\mu)\right)\,.
\eeq
Thus we may regard $\VP$ as a functional of $\fu,\fomega, \fB$, and $\mu$. Both of these representations of $\VP$,~\eqref{E:VPabelian} and~\eqref{E:VPalt}, will prove useful. In the time-independent gauge which we are working in, both $\hat{\fA}$ and $\hat{\fF}$ are spatial forms. Consequently the $2n+1$-form $\hat{\form{I}}_{CS}=c_{_A}\hat{\fA}\wedge \hat{\fF}^n$ has no leg along the time-direction and therefore vanishes in $2n+1$ dimensions. Removing $\hat{\form{I}}_{CS}$ from~\eqref{E:ICSdecomp} and using~\eqref{E:WQFT} and~\eqref{E:WcovDef} we obtain
\beq
\label{E:WcovVP}
	W_{cov} = W_{gauge-invariant} + \int_{\mathcal{M}} \VP\,.
\eeq
The decomposition~\eqref{E:ICSdecomp} thereby leads to a rewriting of $W_{cov}$ in terms of manifestly gauge-invariant objects. Equation \eqref{E:WcovVP} implies that the covariant current and stress tensor will get contributions from both $W_{gauge-invariant}$ and $\VP$. 

The variation of $\VP$ leads to bulk and boundary currents, viz.,
\beq
\delta \int_{\mathcal{M}} \VP = \int d^{2n}x \sqrt{-g} \left[ \delta A_{\mu} J^{\mu}_{\mathcal{P}} + \frac{1}{2}\delta g_{\mu\nu} T^{\mu\nu}_{\mathcal{P}}\right] + (\text{bulk currents})\,,
\eeq
where we have denoted the contribution to the boundary stress tensor and current by $T_{\mathcal{P}}^{\mu\nu}$ and $J_{\mathcal{P}}^{\mu}$. Let us regard $\VP$ as a functional of $\fu,\fomega,\fB$, and $\mu$ as in~\eqref{E:VPalt}. In varying $\VP$ we need to convert the variations of $\fomega$ and $\fB$ to variations of $\fu$ and $\fA$ via an integration by parts,
\begin{equation}
\begin{split}
(d\delta \fu)\wedge \fu &= \brk{\delta(2\fomega)-\delta \fu \wedge \fa} \wedge \fu\,, \\
(d\delta \fA)\wedge \fu Ê&= \brk{\delta \fB+\delta \fu \wedge \fE} \wedge \fu\,. \\
\end{split}
\end{equation} 
The boundary variation of $\VP$ arises entirely from this integration by parts. We find 
\begin{align}
\begin{split}
\label{E:deltaVPabel}
\delta \VP &= Êd\left[ \delta\fA\wedge \left(\frac{\partial \VP}{\partial\fB}\right)_{\fu,\fomega,\mu} + \delta\fu\wedge \left(\frac{\partial\VP}{\partial(2\fomega)}\right)_{\fu,\fB,\mu}\right] + (\text{bulk contributions})\,.
\end{split}
\end{align}
From~\eqref{E:deltaVPabel} we find the covariant anomaly-induced flavor current and heat current
\beq
\label{E:JTAbelian}
	\hodge \form{J_{\mathcal{P}}} = \frac{\partial\VP}{\partial\fB}\,, \qquad \hodge \form{q_{\mathcal{P}}} = \frac{\partial\VP}{\partial(2\fomega)}\,,
\eeq
where $\hodge$ is the Hodge star operator on the boundary. Our conventions for the Hodge star operator may be found in Appendix~\ref{ss:forms}. Converting variations of $\fu$ to variations of the metric using \eqref{E:hydrostatSol} gives
\beq
\delta u_{\mu} q^{\mu} = \frac{1}{2}\delta g_{\mu\nu} (u^{\mu} q^{\nu} + u^{\nu}q^{\mu}) + \frac{1}{2} \delta g_{\mu\nu}u^{\mu}u^{\nu}u_{\rho}q^{\rho}\,,
\eeq
from which we obtain
\beq
\label{E:defqP}
	T^{\mu\nu}_{\mathcal{P}}=u^{\mu}q_{\mathcal{P}}^{\nu}+u^{\nu}q_{\mathcal{P}}^{\mu}\,.
\eeq

Let us see how this machinery works in detail by recomputing the $U(1)^3$ anomaly-induced transport in four dimensions~\cite{Son:2009tf}. For a theory with anomaly polynomial $\fP = c_{_A} \fF^3$, we have
\beq
\VP =c_{_A}\frac{\fu}{2\fomega}\wedge\left[ \fB^3-(\fB+2\fomega\mu)^3\right]=-\mu c_{_A}\fu \wedge \left[  3 \fB^2+3\mu(2\fomega) \fB + \mu^2 (2\fomega)^2\right]\,,
\eeq
which leads to the currents
\begin{align}
\begin{split}
\label{E:4dcurrents1}
	\hodge \form{J}_{\mathcal{P}}&=\frac{\partial\VP}{\partial\fB} = -6 c_{_A}\mu\, \fu\wedge \fB - 3 c_{_A}\mu^2\fu\wedge(2\fomega)\,,
	\\
	\hodge \form{q}_{\mathcal{P}} & = \frac{\partial\VP}{\partial(2\fomega)} = -3 c_{_A}\mu^2 \fu\wedge \fB -2 c_{_A}\mu^3\fu\wedge(2\fomega)\,.
\end{split}
\end{align}
Hodge dualizing the currents in~\eqref{E:4dcurrents1} leads to
\begin{align}
\begin{split}
\label{E:4dcurrents2}
	J_{\mathcal{P}}^{\mu} & = -6 c_{_A} \mu \,\epsilon^{\mu\nu\rho\sigma}u_{\nu}\partial_{\rho}A_{\sigma} - 3 c_{_A} \mu^2 \epsilon^{\mu\nu\rho\sigma} u_{\nu} \partial_{\rho}u_{\sigma}\,,
	\\
	q_{\mathcal{P}}^{\mu} & = - 3 c_{_A} \mu^2 \epsilon^{\mu\nu\rho\sigma} u_{\nu}\partial_{\rho}A_{\sigma}-2 c_{_A}\mu^3 \epsilon^{\mu\nu\rho\sigma}u_{\nu}\partial_{\rho}u_{\sigma}\,.
\end{split}
\end{align}
In~\eqref{E:4dconstitutive} we have summarized the hydrodynamic constitutive relations to first order in derivatives. We relate these results to the existing literature in Section~\ref{S:conclusion}.

Our general result~\eqref{E:JTAbelian} and~\eqref{E:defqP} is valid for $U(1)$ anomalies in even spacetime dimensions. It matches the anomaly-induced current and stress tensor computed using entropy arguments~\cite{Kharzeev:2011ds,Loganayagam:2011mu} or using the hydrostatic generating functional~\cite{Banerjee:2012cr}. Further, in \cite{Loganayagam:2011mu} it was observed that one can generate the anomaly induced currents using a generating function $\VP$ a' la \eqref{E:JTAbelian}. The current section allows one to interpret $\VP$ as the bulk contribution to the covariant generating function.

The main results of this section are equations~\eqref{E:JTAbelian} and~\eqref{E:defqP}, which describe the anomaly-induced flavor current and stress tensor. In the remainder of this work we will develop the technical machinery required to generalize~\eqref{E:JTAbelian} and~\eqref{E:defqP} to non-abelian and gravitational anomalies. We begin with a somewhat detailed and covariant exposition of hydrostatics before discussing that generalization.

\section{More on hydrostatics} \label{S:hydrostatics}

In the hydrodynamic limit, states of the system are characterized by hydrodynamic fields whose kinematic behavior is determined via energy-momentum conservation and charge conservation. In what follows we will consider a specialized subset of configurations which we call hydrostatic configurations that will be defined below. 

\subsection{Generalities}

Consider a field theory with a global symmetry group $G$ and algebra $\mathfrak{g}$. Following our convention earlier in the text, we refer to this symmetry as a ``flavor'' symmetry and to the corresponding symmetry current as a ``flavor'' current. Hydrodynamics can be thought of as a long wavelength approximation of a state of this theory which is close to thermal equilibrium. The hydrodynamic variables describing the evolution of the fluid are a velocity field $u^{\mu}$, a local temperature $T$, and a local chemical potential $\mu$ for the flavor charge. We place our fluids in a non-trivial but slowly varying background, described by a metric $g_{\mu\nu}$ and an external gauge field $A_{\mu}$ which couples to the flavor current. The chemical potential $\mu$ and external gauge field $A_{\mu}$ may be regarded as matrices in flavor space. We use a `$\cdot$' to denote a trace over flavor indices. In what follows we choose an anti-Hermitian basis for the generators $T_A$ of the adjoint representation of $\mathfrak{g}$, so that we notate e.g. the $\mathfrak{g}$-valued chemical potential $\mu$ as $\mu \equiv - i \mu^A (T_A)$.\footnote{In a background with nonzero chemical potential, the flavor symmetry is effectively broken to the subgroup which commutes with $\mu$. As a result the hydrodynamic limit is encoded in the response of the stress-energy tensor and the symmetry currents of the unbroken subgroup. For a typical chemical potential the unbroken subgroup is the Cartan subgroup, and so it is common practice to write thermodynamics with a non-abelian flavor symmetry $G$ in terms of thermodynamics with a number of $U(1)$ symmetries (see e.g.~\cite{Neiman:2010zi}). However this cannot be consistently done in a hydrostatic or hydrodynamic state. In this work we will be interested in the full structure of the global symmetry and so we retain the (potentially) non-abelian nature of the flavor symmetry.} From $A_{\mu}$ we construct the background field strength
\beq
F_{\mu\nu} = \partial_{\mu} A_{\nu}+ A_{\mu}A_{\nu} - (\mu\leftrightarrow \nu)\,,
\eeq
where matrix multiplication is implicit. Under a gauge transformation parameterized by $\Lambda$, the fields $\mu$, $A_{\mu}$ and $F_{\mu\nu}$ transform according to
\beq
\delta_{\Lambda} \mu = [\mu,\Lambda]\,, \qquad \delta_{\Lambda} A_{\mu} = \partial_{\mu} \Lambda + [A_{\mu},\Lambda] = D_{\mu} \Lambda\,, \qquad \delta_{\Lambda} F_{\mu\nu} = [F_{\mu\nu},\Lambda]\,.
\eeq

In coupling the fluid to the metric, we must also specify the connection. For simplicity we study fluids coupled to the Christoffel connection
\begin{equation}
\begin{split}
\label{E:christoffel}
\Gamma^\lambda{}_{\mu\nu} = \frac{1}{2}g^{\lambda\alpha}\brk{ \partial_\mu g_{\nu\alpha} + \partial_\nu g_{\mu\alpha} - \partial_\alpha g_{\mu\nu} }\,.
\end{split}
\end{equation}
We construct the Riemann curvature from the connection via 
\beq
R^\sigma{}_{\lambda\mu\nu} \equiv \partial_\mu \Gamma^\sigma{}_{\lambda\nu} + \Gamma^\sigma{}_{\alpha\mu} \Gamma^\alpha{}_{\lambda\nu} -
\prn{\mu \leftrightarrow \nu }\,.
\eeq
Using the Christoffel and gauge connections we extend the definition of $D_{\mu}$ so that it denotes a flavor and spacetime covariant derivative. For instance, consider a tensor $V\ab{\mu}{\nu}$ which transforms in the adjoint representation of the flavor symmetry. Then $D_{\mu}$ acts on $V\ab{\mu}{\nu}$ as
\beq
D_{\mu} V\ab{\nu}{\rho} = \partial_{\mu}V\ab{\nu}{\rho} + [A_{\mu},V\ab{\nu}{\rho}] + \Gamma\ab{\nu}{\sigma\mu}V\ab{\sigma}{\rho} - \Gamma\ab{\sigma}{\rho\mu}V\ab{\nu}{\sigma}\,,
\eeq
and similarly when the tensor has more indices.

Consider an infinitesimal coordinate transformation $\xi^{\mu}$ and gauge transformation $\Lambda$; we collectively notate the transformation as $\chi = \{\xi^{\mu},\Lambda\}$ and the variation as $\delta_{\chi}$. Under this transformation, a covariant tensor $\theta\ab{\mu}{\nu}$ in the adjoint representation of $\mathfrak{g}$ varies as
\beq
\label{E:deltaChi}
	\delta_{\chi}\theta\ab{\mu}{\nu} = \lieD_{\xi} \theta\ab{\mu}{\nu} + [\theta\ab{\mu}{\nu},\Lambda]\,.
\eeq
For later use, we find it useful to rewrite \eqref{E:deltaChi} in terms of arbitrary connections $\tilde{A}_{\mu}$, $\tilde{\Gamma}\ab{\mu}{\nu\rho}$ (not necessarily $A_{\mu}$ or the Christoffel connection) and their associated covariant derivative $\tilde{D}_{\mu}$. After some algebra one finds
\beq
	\delta_{\chi}\theta\ab{\mu}{\nu} = \xi^{\rho} \tilde{D}_{\rho} \theta\ab{\mu}{\nu} + \tilde{D}_{\nu} \xi^{\rho} \theta\ab{\mu}{\rho} - \tilde{D}_{\rho} \xi^{\mu} \theta\ab{\rho}{\nu} + \tilde{T}\ab{\sigma}{\rho\nu}\xi^{\rho}\theta\ab{\mu}{\sigma} - \tilde{T}\ab{\mu}{\rho\sigma}\xi^{\rho}\theta\ab{\sigma}{\nu} + [\theta\ab{\mu}{\nu},\xi^{\rho}\tilde{A}_{\rho}+\Lambda]\,,
\eeq
where we have defined the torsion $\tilde{T}\ab{\mu}{\nu\rho} = \tilde{\Gamma}\ab{\mu}{\rho\nu} - \tilde{\Gamma}\ab{\mu}{\nu\rho}$. We can use this result to deduce the transformation properties of $\xi^{\mu}$ and $\Lambda$ under a gauge and coordinate transformation as follows. We demand that $\delta_{\chi_1}\theta\ab{\mu}{\nu}$, being a covariant tensor transforming in the adjoint representation of $\mathfrak{g}$, varies under a second transformation $\chi_2$ as
\beq
	\delta_{\chi_2}(\delta_{\chi_1} \theta\ab{\mu}{\nu}) = \lieD_{\xi_2} \left(\delta_{\chi_1}\theta\ab{\mu}{\nu}\right) + [\delta_{\chi_1}\theta\ab{\mu}{\nu},\Lambda_2]\,.
\eeq
After some algebra one finds
\begin{align}
\begin{split}
	\delta_{\chi_2} \xi_1^{\mu} & = \lieD_{\xi_2} \xi_1^{\mu} = \xi_2^{\nu}\partial_{\nu} \xi_1^{\mu} - \xi_1^{\nu}\partial_{\nu}\xi_2^{\mu} = - \delta_{\chi_1}\xi_2^{\mu}\,,
	\\
	\delta_{\chi_2} \Lambda_1 & = \lieD_{\xi_2}\Lambda_1 + [\Lambda_1,\Lambda_2] - \xi_1^{\mu}\partial_{\mu}\Lambda_2 = \xi_2^{\mu}\partial_{\mu}\Lambda_1 - \xi_1^{\mu}\partial_{\mu}\Lambda_2 + [\Lambda_1,\Lambda_2] = - \delta_{\chi_1}\Lambda_2\,.
\end{split}
\end{align}
This motivates the definitions
\begin{align}
\begin{split}
	\xi_{[12]}^{\mu} &\equiv \xi_1^{\nu}\partial_{\nu}\xi_2^{\mu} - \xi_2^{\nu}\partial_{\nu}\xi_1^{\mu} \,,
	\\
	\Lambda_{[12]} & \equiv \xi_1^{\mu}\partial_{\mu}\Lambda_2 - \xi_2^{\mu}\partial_{\mu}\Lambda_1 - [\Lambda_1,\Lambda_2]\,,
\end{split}
\end{align}
so that we can write the algebra obeyed by coordinate and flavor gauge transformations as $[\delta_{\chi_1},\delta_{\chi_2}] = \delta_{\chi_{[12]}}$.

\subsection{A covariant formulation of hydrostatic equilibrium}

Suppose our field theory is coupled to a background metric $g_{\mu\nu}$ and gauge field $A_{\mu}$, both of which are invariant under the action of a time-like Killing vector $K^{\mu}$ and gauge transformation $\Lambda_K$,
\beq
\label{E:symmetry}
	\delta_K g_{\mu\nu} = \lieD_K g_{\mu\nu} = 0\,,
	\qquad
	\delta_K A_{\mu} = \lieD_K A_{\mu} + D_{\mu} \Lambda_K  = 0\,.
\eeq
Our main postulate, as phrased in section \ref{S:minimalhydrostatics} is that there exists a solution to the conservation equations of our field theory which respect the symmetry generated by $K$ and is a local function of the sources. We call such a configuration a hydrostatic state. In Section~\ref{S:minimalhydrostatics} we introduced hydrostatic equilibria in a particular gauge and coordinate choice where the background fields were explicitly time-independent (the ``transverse gauge'') i.e., $K^{\mu} = (1,0,\ldots,0)$. While it is  convenient to carry out computations in the transverse gauge, it is often useful to use a covariant formation of hydrostatics, especially when dealing with gravitational anomalies.

In analogy with \eqref{E:hydrostatSol} we define the temperature, $T$, velocity field $u^{\mu}$ and chemical potential $\mu$ via
\begin{align}
\begin{split}
\label{E:covHydroVar}
T & = \frac{1}{\beta \sqrt{-K^2}}\,, \qquad
u^{\mu} = \frac{K^{\mu}}{\sqrt{-K^2}}\,, \qquad \mu = \frac{K^{\mu}A_{\mu}+\Lambda_K}{\sqrt{-K^2}}\equiv \beta T \bar{\mu}\,,
\end{split}
\end{align}
where $\beta$ is the parametric length of the time circle. Since $T$, $u^{\mu}$ and $\mu$ are constructed from the background fields and symmetry generators, they are invariant  under the action of the symmetry. In particular, the chemical potential transforms covariantly due to
\begin{align}
\nonumber
	\delta_{\chi} (K^{\mu}A_{\mu}+\Lambda_K) & = A_{\mu}\delta_{\chi}K^{\mu} + K^{\mu}\delta_{\chi}A_{\mu} + \delta_{\chi}\Lambda_K\,,
	\\
	\nonumber
	& = A_{\mu}\lieD_{\xi}K^{\mu} +(K^{\mu}\lieD_{\xi} A_{\mu}+K^{\mu} D_{\mu}\Lambda)+\left(\lieD_{\xi}\Lambda_K+  [\Lambda_K,\Lambda] - K^{\mu}\partial_{\mu}\Lambda\right)\,,
	\\
	& = \lieD_{\xi}\bar{\mu} +[\bar{\mu},\Lambda]\,,
\end{align}
We also define the matrix-valued spin chemical potential $(\muR)\ab{\mu}{\nu}$ which in terms of $K^{\mu}$ becomes
\beq
\label{E:muR}
(\muR)\ab{\mu}{\nu} = T D_{\nu}\left( \frac{u^{\mu}}{T}\right) = \frac{D_{\nu}K^{\mu}}{\sqrt{-K^2}}\,.
\eeq
The spin chemical potential is the equivalent of the flavor chemical potential when constructing gravitational anomalies.\footnote{To understand this equivalence, we note that using the Mathisson-Papapetrou-Dixon formulation of the equations of motion in the presence of point torques the natural thermodynamic conjugate of the spin-current (which appears later in this Section in~\eqref{E:deltaWwithSpin}) is the chemical potential $\mu_R$ defined above. See Appendix~\ref{A:spin} for further discussion.} We elaborate on this point later in this Section (for example see~\eqref{E:deltaGammaCon} and~\eqref{E:spinEquil}) and in Appendix \ref{A:spin}. 

A particular realization of the transverse gauge which we have worked with in Section \ref{S:minimalhydrostatics} can be constructed as follows. Let us take the metric and gauge field to be of the form
\begin{equation}
\begin{split}
	\label{E:transversegaugeexplicit}
	g &= -e^{2\mathfrak{s}}(dt+\mathfrak{a})^2 + \mathfrak{p}_{ij} dx^i dx^j\,,\\
	\fA &= A_0 (dt+\mathfrak{a}) + \mathfrak{A}_i dx^i\,,
\end{split}
\end{equation}
where $\mathfrak{a} \equiv \mathfrak{a}_i dx^i$. All the metric and gauge-field components are functions of $x^i$ but are assumed to be independent of $t$. It is straightforward to verify that the metric and gauge field above satisfy the Killing conditions \eqref{E:symmetry} with  $K^\mu\partial_\mu =  \partial_t$ and $\Lambda_K=0$. The particular gauge \eqref{E:transversegaugeexplicit} was introduced in \cite{Banerjee:2012iz} and its significance in writing the hydrostatic generating functional and hydrodynamics was studied in detail there.

Using~\eqref{E:covHydroVar} the explicit expressions for the local temperature, fluid velocity, and chemical potential are 
\beq
T = \frac{e^{-\mathfrak{s}}}{\beta}\,, \qquad u^{\mu}\partial_{\mu} = e^{-\mathfrak{s}}\partial_t\,, \qquad u_{\mu}dx^{\mu} = - e^{\mathfrak{s}}(dt+\mathfrak{a})\,, \qquad \mu = e^{-\mathfrak{s}}A_0\,.
\eeq
We also note in passing that $\mathfrak{p}_{ij}$ is equivalent to the projection matrix $P_{\mu\nu}$, 
\begin{equation*}
P_{\mu\nu}dx^{\mu}dx^{\nu} = (g_{\mu\nu} + u_{\mu}u_{\nu})dx^{\mu}dx^{\nu} = - e^{2\mathfrak{s}}(dt+\mathfrak{a})^2 + \mathfrak{p}_{ij}dx^idx^j + e^{2\mathfrak{s}}(dt+\mathfrak{a})^2 = \mathfrak{p}_{ij} dx^i dx^j\,,
\end{equation*}
The physical interpretation of $\mathfrak{A}_i$ will be described shortly.

\subsection{The electric-magnetic decomposition}
\label{S:EMdecomp}

Motivated by the results in Section \ref{S:abelAnomTrans} we use the local fluid velocity $u^{\mu}$ to decompose the various covariant tensors into ``electric'' and ``magnetic'' parts. For instance, we decompose the background flavor field strength $F_{\mu\nu}$ into an electric flavor field $\El_\mu \equiv  F_{\mu\nu} u^\nu$ and a flavor magnetic field which is transverse to $u^{\mu}$,
\begin{equation}
\begin{split}
B_{\mu\nu} &\equiv F_{\mu\nu} - \brk{ u_\mu  \El_\nu -  u_\nu  \El_\mu }
\end{split}
\end{equation}
It is easily checked that $ B_{\mu\nu} u^\nu=0$ and hence the magnetic field can be thought of as
the transverse part of the field strength. We refer to this as an ``electro-magnetic'' decomposition insofar as $E_{\mu}$ and $B_{\mu\nu}$ describe the electric and magnetic fields in the local fluid rest frame. We decompose the exterior derivative of the fluid velocity $\partial_{\mu}u_{\nu}-\partial_{\nu}u_{\mu}$ in a similar way: the acceleration $a_{\mu}$ and vorticity $\omega_{\mu\nu}$ are given by
\begin{align}
\begin{split}
\label{E:defaomega}
a_{\mu} &\equiv -u^{\nu}(\partial_{\mu}u_{\nu} - \partial_{\nu} u_{\mu})\,,
\\
2\omega_{\mu\nu} & \equiv (\partial_{\mu}u_{\nu}-\partial_{\nu}u_{\mu}) + \brk{ u_{\mu}a_{\nu}-u_{\nu}a_{\mu}   }\,.
\end{split}
\end{align}
The definitions~\eqref{E:defaomega} definitions coincide with the usual ones~\eqref{E:defEB} in the absence of torsion. We see that the acceleration and vorticity are the fluid analogues of a local electric and magnetic field respectively.

Our next step is to mimic this construction in gravity. As it turns out, it is most convenient to regard the Riemann tensor as a matrix-valued antisymmetric two-tensor. That is, we treat the first two indices of the Riemann tensor as matrix indices and the last two indices as spacetime indices. We then decompose the last two indices into electric and magnetic parts just as we did for $F_{\mu\nu}$: the electric and magnetic parts part of the Riemann tensor are
\begin{align}
\begin{split}
	\prn{\ER}^\lambda{}_{\sigma\mu} &\equiv  R^\lambda{}_{\sigma\mu\nu} u^\nu\,,
\\
	\prn{\BR}^\lambda{}_{\sigma\mu\nu} &\equiv R^\lambda{}_{\sigma\mu\nu} 
- \brk{ u_\mu  \prn{\ER}^\lambda{}_{\sigma\nu} -  u_\nu \prn{\ER}^\lambda{}_{\sigma\mu} }\,.
\end{split}
\end{align}
This magnetic Riemann tensor is transverse in its last two indices $\prn{\BR}^\lambda{}_{\sigma\mu\nu} u^\nu =0$.\footnote{Unfortunately, there are many other notions of electric-magnetic decomposition of gravitational tensors which are prevalent in this context. The first one, which often goes by the name of `gravito-magnetism' involves a decomposition of the connection $\fGamma$ while the second one, more familiar in general relativity , is the electric-magnetic decomposition of the Weyl tensor (a closely related decomposition is the so called Bel decomposition of the Riemann tensor).  We will be using {none} of those notions in this paper and the reader interested in comparisons with other literature is warned to be mindful of these distinctions. As will become clear later on, the electric-magnetic decomposition we describe in this section is the one relevant to questions about transport - in particular, this is the most convenient decomposition to study gravitational anomalies at finite temperature. } 

As we mentioned in Section~\ref{S:minimalhydrostatics}, the quantities in~\eqref{E:covHydroVar} and~\eqref{E:muR} obey certain differential interrelations. These may be obtained from the condition~\eqref{E:symmetry} that $\{K^{\mu},\Lambda_K\}$ generates a symmetry of the background fields. To see this, write
\begin{subequations}
\begin{align}
\begin{split}
	\delta_K g_{\mu\nu} & = D_{\mu}K_{\nu} + D_{\nu}K_{\mu} 
	\\
	 & = \frac{2}{\beta T}\left( \sigma_{\mu\nu} + \frac{P_{\mu\nu}}{d-1}\vartheta\right) - \frac{1}{\beta T^2}\left[ u_{\mu}\left(D_{\nu} + a_{\nu}\right)T+ u_{\nu}\left( D_{\mu}+a_{\mu}\right)T\right] = 0\,,
\end{split}
\end{align}
together with
\begin{align}
\begin{split}
\label{E:deltaACon}
	\delta_K A_{\mu} & = K^{\nu}\partial_{\nu}A_{\mu} + \partial_{\mu}K^{\nu} A_{\nu} + D_{\mu}\Lambda_K  =  - F_{\mu\nu}K^{\nu} + D_{\mu}(K^{\nu}A_{\nu}+\Lambda_K)
	\\
	& = \frac{1}{\beta T^2}\left[T \left( - E_{\mu} + (D_{\mu}+a_{\mu})\mu\right) - \mu (D_{\mu} + a_{\mu})T\right] \,,
\end{split}
\end{align}
\end{subequations}
and
\begin{align}
\begin{split}
\label{E:deltaGammaCon}
	\delta_K \Gamma\ab{\mu}{\nu\rho} & = \lieD_K \Gamma\ab{\mu}{\nu\rho}  + \partial_{\nu}\partial_{\rho}K^{\mu} = -R\ab{\mu}{\nu\rho\sigma}K^{\sigma} + D_{\rho}D_{\nu}K^{\mu}  
	\\
	& = \frac{1}{\beta T^2}\left[ T\left( - (E_R)\ab{\mu}{\nu\rho} + (D_{\rho}+a_{\rho})(\muR)\ab{\mu}{\nu}\right) + (\muR)\ab{\mu}{\nu}(D_{\rho}+a_{\rho})T\right]\,.
\end{split}
\end{align}
Assuming that $K$ generates a symmetry, we then have
\beq
\label{E:chemEquil2}
D_{\mu}\left( \frac{u_{\nu}}{T}\right) + D_{\nu}\left(\frac{u_{\mu}}{T}\right) = 0\,,
\qquad
(D_{\mu}+a_{\mu}) T = 0\,,
\qquad
(D_{\mu}+a_{\mu})\mu  = E_{\mu}\,,
\eeq
which reproduces the conditions~\eqref{E:chemEquil} described earlier together with
\beq
\label{E:spinEquil}
(D_{\rho}+a_{\rho})(\muR)\ab{\mu}{\nu} = (E_R)\ab{\mu}{\nu\rho}\,,
\eeq
in close analogy with the equation of chemical equilibrium $(D_{\mu}+a_{\mu})\mu=E_{\mu}$.  Note that if we have covariant fields $\{T,u^{\mu},\mu\}$ that satisfy these equations, then we can define symmetry data $\{K^{\mu},\Lambda_K\}$ from them. As a result, solutions to~\eqref{E:chemEquil2} are in one-to-one correspondence with geometries possessing a timelike symmetry.

\subsection{Hatted connections}

Many of the quantities above can be more easily manipulated when written as differential forms. We follow the notation of Section~\ref{S:minimalhydrostatics} and notate form fields with a boldface font. We begin by writing the velocity co-vector as a one-form, $\fu = u_{\mu} dx^{\mu}$. From its exterior derivative $d\fu$ we obtain the acceleration and vorticity as
\beq
\fa =a_{\mu}dx^{\mu} =  \iota_{u} d\fu\,, \qquad 2\fomega = \omega_{\mu\nu}dx^{\mu} \wedge dx^{\nu} =  d\fu +\fu\wedge \fa\,,
\eeq
where in a slight abuse of notation we refer to $\iota_{u}$ as the interior product along the vector $ u^{\mu}\partial_{\mu}$. The background gauge field may be written as a one-form $\fA = A_{\mu}dx^{\mu}$, from which the field strength is defined as
\beq
\fF = \frac{1}{2}F_{\mu\nu}dx^{\mu}\wedge dx^{\nu} = d\fA + \fA \wedge \fA\,,
\eeq
where matrix multiplication is implied. The electric and magnetic flavor fields are
\beq
\fE = E_{\mu}dx^{\mu} = - \iota_u \fF\,, \qquad \fB = \frac{1}{2}B_{\mu\nu} dx^{\mu} \wedge dx^{\nu} = \fF - \fu \wedge \fE\,.
\eeq

To treat the gravitational quantities efficiently we write the Christoffel connection as a matrix-valued one-form
\beq
\form{\Gamma}\ab{\mu}{\nu} \equiv \Gamma\ab{\mu}{\nu\rho}dx^{\rho}\,,
\eeq
whose non-abelian field strength is the Riemann curvature (regarded as a matrix-valued two-form)
\beq
\form{R}\ab{\mu}{\nu} = \frac{1}{2}R\ab{\mu}{\nu\rho\sigma}dx^{\rho}\wedge dx^{\sigma} = d\form{\Gamma}\ab{\mu}{\nu} + \form{\Gamma}\ab{\mu}{\rho}\wedge \form{\Gamma}\ab{\rho}{\nu}\,.
\eeq
The electric and magnetic curvatures then take the same form as the electric and magnetic flavor fields,
\begin{align}
\begin{split}
(\fE_R)\ab{\mu}{\nu} &= (E_R)\ab{\mu}{\nu\rho}dx^{\rho} =  - \iota_u \form{R}\ab{\mu}{\nu}\,,
\\
(\fB_R)\ab{\mu}{\nu} &= \frac{1}{2}(B_R)\ab{\mu}{\nu\rho\sigma}dx^{\rho}\wedge dx^{\sigma} = \fR\ab{\mu}{\nu} - \fu\wedge (\fE_R)\ab{\mu}{\nu}\,.
\end{split}
\end{align}
Finally, we can define the exterior covariant derivative $D$. In this work we require the action of $D$ on $p$-forms $\form{V}$ which transform in the adjoint representation of $\mathfrak{g}$,
\beq
D\form{V} = d\form{V} + \fA \wedge \form{V} -(-1)^p \form{V} \wedge \fA\,,
\eeq
as well as on matrix valued $p$-forms $\form{V}\ab{\mu}{\nu}$ transforming in the adjoint of $\mathfrak{g}$,
\beq
D\form{V}\ab{\mu}{\nu} = d\form{V}\ab{\mu}{\nu} + \fA \wedge \form{V}\ab{\mu}{\nu} - (-1)^p \form{V}\ab{\mu}{\nu} \wedge \fA + \form{\Gamma}\ab{\mu}{\rho}\wedge \form{V}\ab{\rho}{\nu} - (-1)^p \form{V}\ab{\mu}{\rho}\wedge \form{\Gamma}\ab{\rho}{\nu}\,.
\eeq

We are now in a position to introduce {hatted connections}~\eqref{E:hatConnection}, which play a critical role in this work. In terms of forms, the hatted connections are
\begin{align}
\begin{split}
\label{E:hatConnectionsAgain}
\hat{\fA} & = \fA + \mu \fu= \fA + (K^{\nu}A_{\nu}+\Lambda_K)\form{K}\,, 
\\
\hat{\form{\Gamma}}\ab{\mu}{\nu} & = \form{\Gamma}\ab{\mu}{\nu} + (\muR)\ab{\mu}{\nu}\fu = \form{\Gamma}\ab{\mu}{\nu} + (D_{\nu}K^{\mu})\form{K}\,.
\end{split}
\end{align}
The hatted connections, and the quantities constructed from them exhibit a number of useful properties. Consider the hatted field strengths,
\begin{align}
\begin{split}
\hat{\form{F}} &= d\hat{\form{A}} + \hat{\fA}\wedge \hat{\fA}\,,
\\
\hat{\form{R}}\ab{\mu}{\nu} & = d\hat{\form{\Gamma}}\ab{\mu}{\nu} + \hat{\form{\Gamma}}\ab{\mu}{\rho}\wedge \hat{\form{\Gamma}}\ab{\rho}{\nu}\,,
\end{split}
\end{align}
which may be decomposed into electric and magnetic parts. Following the previous subsection, we have
\begin{align}
\begin{split}
\hat{\fF} & = \fu \wedge \hat{\fE} + \hat{\fB}\,,
\\
\hat{\form{R}}\ab{\mu}{\nu} &= \fu \wedge (\hat{\fE}_R)\ab{\mu}{\nu} + (\hat{\fB}_R)\ab{\mu}{\nu}
\end{split}
\end{align}
with
\begin{align}
\begin{split}
\hat{\fE} &= \fE - (D+\fa)\mu\,,
\\
\hat{\fB} &= \fB + 2\fomega \mu\,,
\\
(\hat{\fE}_R)\ab{\mu}{\nu} &  = (\fE_R)\ab{\mu}{\nu} - (D+\fa)(\muR)\ab{\mu}{\nu}\,,
\\
(\hat{\fB}_R)\ab{\mu}{\nu} & = (\fB_R)\ab{\mu}{\nu} + 2\fomega (\muR)\ab{\mu}{\nu}\,.
\end{split}
\end{align}
However, upon using~\eqref{E:chemEquil2} and~\eqref{E:spinEquil} we see that the hatted electric fields vanish,
\beq
\hat{\fE}=0\,, \qquad (\hat{\fE}_R)\ab{\mu}{\nu}=0\,,
\eeq
so that the hatted field strengths are purely transverse,
\beq
\label{E:hatCurvaturesAreTransverse}
\hat{\fF} = \hat{\fB} = \fB + 2\fomega\mu\,, \qquad \hat{\fR}\ab{\mu}{\nu} = (\hat{\fB}_R)\ab{\mu}{\nu} = (\fB_R)\ab{\mu}{\nu} + 2\fomega (\muR)\ab{\mu}{\nu}\,.
\eeq
Physically, the hatted electric fields encode the violation of chemical and spin equilibrium.

While the hatted field strengths are transverse to the velocity field the hatted connections are generally not transverse to the velocity. The transverse parts of $\hat{A}_{\mu}$ and $\hat{\Gamma}\ab{\mu}{\nu\rho}$ are given by
\beq
\hat{A}_{\mu}K^{\mu} = \Lambda_K\, K^2\,, \qquad \hat{\Gamma}\ab{\mu}{\nu\rho}K^{\rho} =  \partial_{\nu}K^{\mu}\, K^2\,.
\eeq
Nevertheless, one can always switch to a transverse gauge where the hatted connections are transverse $(\{\Lambda_K =0,\, \partial_{\nu}K^{\mu} = 0\})$.

In the transverse gauge the hatted flavor connection is
\beq
\hat{\fA}= \fA +\mu \fu = \fA - A_0(dt+\mathfrak{a}) = \mathfrak{A}_i dx^i\,,
\eeq
In our covariant analysis we showed that the hatted field strength $\hat{\fF}$ is a transverse form. We can easily recover that result in transverse gauge, since $\hat{\fA}$ is independent of time and has no leg along the time direction. Consequently, the hatted field strength $\hat{\fF}$ also has no leg along the time direction, which is equivalent to the statement that $\hat{\fF}$ is transverse. Similarly since $\hat{\Gamma}\ab{\mu}{\nu\rho}$ is transverse and independent of time, the hatted curvature $\hat{\fR}\ab{\mu}{\nu}$ is also transverse.

\subsection{The Euclidean partition function}
\label{S:hydrostatEuc}

Often, it is useful to Wick rotate a hydrostatic configuration to Euclidean signature. To this end we define Euclidean time via $t = -i t_E$ and associate with every function of $t$ a function of $t_E$ such that these two functions are restrictions of a single analytic function in the lower half of the complex $t$-plane. This is easily accomplished in the transverse gauge where all the components of the metric and gauge-fields are $t$-independent and hence their analytic continuation is trivial. For instance,
\begin{equation}
\mathfrak{a}_E \equiv i \mathfrak{a}\,, \qquad  e^{\mathfrak{s}_{_E}}\equiv  - i e^{\mathfrak{s}} \,,\qquad
(A_0)_E \equiv -i A_0\,,  
\end{equation}
so that the analytically continued metric and gauge field become
\begin{equation}
 \begin{split}
 \label{E:eucBack}
g&= -e^{2\mathfrak{s}_{_E}}(dt_E+\mathfrak{a}_E )^2 + \mathfrak{p}_{ij} dx^i dx^j\\
 \fA &= (A_0)_E (dt_E +\mathfrak{a}_E) + \mathfrak{A}_i dx^i\,.
 \end{split}
\end{equation}
Similalry, under Wick rotation, the Killing vector becomes $K^\mu\partial_\mu =i \partial_{t_E} $; the Killing vector along the Euclidean time is actually $-iK^\mu$ in our notation. We have deliberately adopted a notation where the metric takes the same form before and after Wick rotating. The advantage of such a notation is that we can continue to use the Lorentzian expressions in the Euclidean theory except for the fact that the temporal components are taken to be imaginary.\footnote{Note that the Euclidean metric here is not really Riemannian but is actually complex when the metric is not static (when $\mathfrak{a}\neq 0$ ). So, the adjective `Euclidean', though commonly employed, is an abuse of terminology. We will however, following the common convention, continue to use the adjective `Euclidean' ignoring this fact.} In what follows, we will continue to think of the Euclidean theory in such a Lorentzian notation. almost all the expressions in the previous subsection then carry over to the corresponding Euclidean expressions. 

We now compactly the Euclidean time direction by making the periodic identification $t_E \sim t_E + \beta $ along the imaginary time. Put differently, we identify two points on the integral curves of the Killing  vector $K^\mu$ provided they are separated by affine distance $\beta$---two points along  the curves satisfying 
\[ 
	\frac{d x^\mu_E}{d\lambda} = -iK^\mu 
\]
are identified if they are separated by $\Delta \lambda = \beta$. We refer to the integral orbits of the  Killing vector as the thermal circles. It is interesting to note that in our new language, the geometry of the Euclidean spacetime is that of a fibre bundle, where the fibres are the thermal circles. The  base manifold of the fibre bundle is the transverse space described by the $x$-coordinates. As before, we identify the temperature, fluid velocity, and flavor chemical potential as in~\eqref{E:covHydroVar}, although in the transverse gauge we have $\Lambda_K=0$ and so the chemical potential becomes
\beq
\mu = \frac{A_{\mu}K^{\mu}}{\sqrt{-K^2}} = A_{\mu}u^{\mu}\,.
\eeq
Similarly the spin chemical potential in transverse gauge is given by
\beq
(\muR)\ab{\mu}{\nu} = \frac{D_{\nu}K^{\mu}}{\sqrt{-K^2}} = \frac{\Gamma\ab{\mu}{\nu\rho}K^{\rho}}{\sqrt{-K^2}}=\Gamma\ab{\mu}{\nu\rho}u^{\rho}\,,
\eeq
where we have used that the Christoffel connection is torsionless, $\Gamma\ab{\mu}{\nu\rho}=\Gamma\ab{\mu}{\rho\nu}$ and that $\partial_{\nu}K^{\mu}=0$ in transverse gauge. The Euclidean partition function is related to the thermal partition function via
\beq
Z_E = \text{tr} \, \exp(-\beta \mathcal{H})\,,
\eeq
with an appropriate definition of the Hamiltonian. See Appendix~\ref{A:Boltzmann} for further discussion.

We can now use the Euclidean partition function to compute hydrostatic expectation values of the stress tensor and current. The consistent flavor current and stress-energy tensor are defined by variation of the generating functional $W_{QFT} = -i \ln Z_E$ with respect to the gauge field and metric respectively. For theories with gravitational anomalies, we find it useful to carry out the variation with respect to the metric in two stages. We first consider the metric and connection as independent, keeping in mind that this separation is somewhat artificial. This gives
\beq
\label{E:deltaWwithSpin}
\delta W_{QFT} = \int d^{2n}x \sqrt{-g} \left[ \delta A_{\mu}\cdot J^{\mu} + \frac{1}{2}\delta g_{\mu\nu} t^{\mu\nu} + \delta \Gamma\ab{\mu}{\nu\rho} \spp\ab{\rho\nu}{\mu}\right] + (\text{boundary terms})\,,
\eeq
where, as usual, we have notated a trace over flavor indices with a `$\cdot$'. The tensor $\spp\ab{\rho\nu}{\mu}$ is the ``spin current'' which we have alluded to earlier. It will frequently be useful to regard it as a matrix-valued one-form,
\beq
\form{\spp}\ab{\mu}{\nu} = \spp_{\rho}{}^{\mu}{}_{\nu} dx^{\rho}\,.
\eeq
The variation of the connection n terms of the variation of the metric is given by
\beq
\delta\Gamma\ab{\mu}{\nu\rho} = \frac{1}{2}\left[ D_{\nu}\delta g\ab{\mu}{\rho} + D_{\rho} \delta g\ab{\mu}{\nu} - D^{\mu} \delta g_{\nu\rho}\right]\,,
\eeq
Thus, integrating~\eqref{E:deltaWwithSpin} by parts we find
\beq
\delta W_{QFT} = \int d^{2n}x \sqrt{-g} \left[ \delta A_{\mu}\cdot J^{\mu} + \frac{1}{2}\delta g_{\mu\nu} T^{\mu\nu}\right] + (\text{boundary terms})\,,
\eeq
where the stress tensor $T^{\mu\nu}$ is given by
\beq
\label{E:stressIntoTandSpin}
T^{\mu\nu} = t^{\mu\nu} + D_{\rho}\left[ \spp^{\mu[\nu\rho]} + \spp^{\nu[\mu\rho]} - \spp^{\rho(\mu\nu)}\right]\,.
\eeq
In the last equation the round (square) brackets indicate (anti-)symmetrization
\beq
A^{(\mu\nu)} = \frac{1}{2}\left( A^{\mu\nu}+A^{\nu\mu}\right)\,, \qquad A^{[\mu\nu]} = \frac{1}{2}\left( A^{\mu\nu}-A^{\nu\mu}\right)\,.
\eeq
This decomposition of the stress-energy tensor is naturally related to the Mathisson-Papapetrou-Dixon equations which treat point torques in a gravitational setting~\cite{Mathisson:1937zz,Papapetrou:1951pa,Dixon:1970zza} which review in Appendix~\ref{A:spin}.

When the background fields are slowly varying (over length scales longer than the static screening length), the equilibrium state is {hydrostatic}. As discussed in \cite{Banerjee:2012iz,Jensen:2012jh}, since the hydrostatic configuration is slowly varying one can argue that all correlation functions of the theory may be expanded in a power series in derivatives  of the sources. Thus, in the absence of anomalies the generating function for connected correlators of a hydrostatic configuration can be constructed as a local gauge and diffeomorphism invariant  functional of the hydrostatic fields $\{T,u^{\mu},\mu\}$, the sources $g_{\mu\nu}$ and $A^{\mu}$, and covariant derivatives thereof. From the point of view of hydrodynamics, the fields $\{T,u^{\mu},\mu\}$ comprise a time-independent solution of the hydrodynamic equations of motion in a particular choice of hydrodynamic frame, known as the thermodynamic frame~\cite{Jensen:2012jh}. In the presence of anomalies this generating functional needs to be appropriately modified. This is the content of the next section.

\section{Non-abelian, gravitational and mixed anomalies} \label{S:allanomalies}

In the previous Section we have developed some technology which will allow us to put gravitational anomalies on a similar footing as $U(1)$ anomalies in our analysis of Sections~\ref{S:abelAnomW} and~\ref{S:abelAnomTrans}. We now go on to consider anomalies of all stripes. Our goal in this Section is to obtain a simple expression for the generating functional of equilibrium covariant currents $W_{cov}$, which we then vary to obtain the anomaly-induced response. Some of the formal manipulations that will be carried out in this Section can be understood from the anomaly inflow mechanism and do not depend on the existence of a time-independent equilibrium. In order to avoid confusion we will often emphasize those equalities which are valid only in equilibrium.

Let $W_{QFT}$ be the generating functional of our quantum field theory. According to the anomaly inflow mechanism~\cite{Callan:1984sa}, the non gauge and reparametrization invariance of $W_{QFT}$ can be captured by thinking of the manifold on which our $2n$ dimensional quantum theory lives on as a boundary of a higher, $2n+1$, dimensional manifold $\mathcal{M}$. The anomalies of the field theory are encoded in a Chern-Simons form $\form{I}_{CS}[\fA,\form{\Gamma}]$ which is the generating functional on $\mathcal{M}$. Using $\form{I}_{CS}$, we define the generating functional of the $2n+1$-dimensional theory to be
\eqref{E:WcovDef},
\beq
W_{cov}[A,g] = W_{QFT}[A,g] + \int_{\mathcal{M}}\form{I}_{CS}[\fA,\form{\Gamma}]\,,
\eeq
The generating functional $W_{cov}$ is gauge and diffeomorphism-invariant. As a result the total flavor charge and energy-momentum currents are conserved, but they may flow from the bulk into the boundary $\partial\mathcal{M}$, which from the perspective of $W_{QFT}$ leads to an anomaly.

Mimicking the construction in Sections~\ref{S:abelAnomW} and~\ref{S:abelAnomTrans}, we use the hatted connections~\eqref{E:hatConnectionsAgain} to write
\begin{align}
\begin{split}
\label{E:theDecomposition}
	\form{I}_{CS}-\hat{\form{I}}_{CS} 
	&= d\left[ \frac{\fu}{2\fomega}\wedge \left( \form{I}_{CS}-\hat{\form{I}}_{CS}\right)\right] + \frac{\fu}{2\fomega} \wedge d\left( \form{I}_{CS}-\hat{\form{I}}_{CS}\right) \\
	& = d\WCS + \VP\,,
\end{split}
\end{align}
with
\begin{equation}
	\WCS = \frac{\fu}{2\fomega}\wedge \left(\form{I}_{CS}-\hat{\form{I}}_{CS}\right)\,, \qquad \VP = \frac{\fu}{2\fomega}\wedge \left( \fP-\hat{\fP}\right)\,,
\end{equation}
where we have used $d\left[\frac{\fu}{2\fomega}\right]=1$ (see \eqref{E:ProofMagic}) along with $d\ICS = \fP$ and $d\hat{\form{I}}_{CS}=\hat{\fP}$ (for $\fP=\fP[\fF,\fR]$ the anomaly polynomial of the theory). We remind the reader that $\hat{\form{I}}_{CS}$ and $\hat{\fP}$ are the Chern-Simons form and anomaly polynomial evaluated for the hatted connections. Using~\eqref{E:theDecomposition} the covariant generating functional is then given by
\beq
\label{E:WcovWithVW}
W_{cov}[A,g] = W_{QFT}[A,g] + \int_{\partial\mathcal{M}} \WCS + \int_{\mathcal{M}} \left( \VP+\hat{\form{I}}_{CS}\right)\,.
\eeq

We now exploit the fact that, in hydrostatic equilibrium \emph{and} in a transverse gauge, all of the hatted connections and curvatures are transverse to $\fu$. As a result, the $2n+1$-dimensional form $\hat{\form{I}}_{CS}$ does not have a leg along the time direction and so its integral over the $2n+1$-dimensional manifold $\mathcal{M}$ vanishes. Thus, in equilibrium,
\beq
W_{cov}[A,g] =W_{QFT}[A,g] + \int_{\partial\mathcal{M}}\WCS + \int_{\mathcal{M}} \VP\,.
\eeq
In contrast to the $2n+1$-form $\VP$, $\WCS$ explicitly depends on the connections $\fA$ and $\form{\Gamma}\ab{\mu}{\nu}$ and is neither gauge-invariant nor diffeomorphism-covariant. Separating $W_{QFT}$ into an anomalous and gauge invariant contribution $W_{QFT} = W_{gauge-invariant}+W_{anom}$, the gauge invariance of $W_{cov}$ implies that 
\begin{equation}
\label{E:WanomisWCS}
	W_{anom} = - \int \form{W}_{CS}.
\end{equation}
Equation~\eqref{E:WanomisWCS} proves the claim made in the Introduction regarding the existence and form of a local expression for $W_{anom}$. With this choice of representative 
\beq
\label{E:WcovEquil}
W_{cov}[A,g]= W_{gauge-invariant}[A,g] + \int_{\mathcal{M}}\VP\,.
\eeq
We have obtained \eqref{E:WcovEquil} in the transverse gauge. However, since $W_{cov}$  is gauge and diffeomorphism invariant and since both terms on the right hand side of \eqref{E:WcovEquil} are also gauge and diffeomorphism invariant, then the expression \eqref{E:WcovEquil} is valid in any gauge and coordinate choice (provided the system is in equilibrium).

In the rest of this Section we will vary $W_{cov}$ to obtain the equilibrium anomaly-induced transport. When the equilibrium state is hydrostatic, this response is the part of the hydrodynamic constitutive relations due to the anomalies. The reader interested in obtaining the consistent currents generated by varying $W_{anom}$ is referred to Appendix \ref{A:deltaVPWCS}. According to~\eqref{E:WcovEquil}, the covariant anomaly-induced currents follow from the variation of $\int \VP$
\begin{align}
\begin{split}
\delta \int_{\mathcal{M}}\VP =
& \int d^{2n}x \sqrt{-g} \left[ \delta A_{\mu} \cdot J_{\mathcal{P}}^{\mu} + \frac{1}{2}\delta g_{\mu\nu} t_{\mathcal{P}}^{\mu\nu} + \delta \Gamma\ab{\mu}{\nu\rho} (\spp_{\mathcal{P}})\ab{\rho\nu}{\mu}\right] + (\text{bulk terms})\,,
\end{split}
\end{align}
such that
\beq
T_{\mathcal{P}}^{\mu\nu}=u^{\mu}q_{\mathcal{P}}^{\nu}+u^{\nu}q_{\mathcal{P}}^{\mu} + D_{\rho}\left( \spp_{\mathcal{P}}^{\mu[\nu\rho]}+\spp_{\mathcal{P}}^{\nu[\mu\rho]}-\spp_{\mathcal{P}}^{\rho(\mu\nu)}\right)\,.
\eeq

To obtain explicit expressions for  $J_{\mathcal{P}}^{\mu}, t_{\mathcal{P}}^{\mu\nu},$ and $\spp_{\mathcal{P}}^{\mu\nu\rho}$, it is helpful to rewrite $\VP$ in the form
\beq
\label{E:VPdecomposition}
\VP=\frac{\fu}{2\fomega}\wedge \left( \fP-\hat{\fP}\right) = \frac{\fu}{2\fomega}\wedge \left( \fP[\fB,\fB_R] - \fP[\fB + 2\fomega \mu, \fB_R+2\fomega\mu_R]\right)\,.
\eeq
The second equality follows by arguing that only the transverse parts of $\fP$ and $\hat{\fP}$ contribute to $\VP$, and that the hatted curvatures are given by~\eqref{E:hatCurvaturesAreTransverse} to be the transverse forms $\hat{\fB}=\fB +2\fomega\mu$ and $\hat{\fB}_R = \fB_R + 2\fomega \mu_R$. In rewriting $\VP$ in the form \eqref{E:VPdecomposition} it becomes transparent that $\VP$ may be considered as a function of $\fB,\mu,\fB_R,\mu_R,\fu$, and $\fomega$ (moreover, it should also be clear that the expression inside the brackets vanishes at zero vorticity, so that we may consistently act on it with the operator $\fu/(2\fomega)$.) Therefore, under a general variation, $\VP$ varies via the chain rule as
\begin{align}
\begin{split}
	\delta \VP  =& \delta \fB\wedge \cdot \frac{\partial \VP}{\partial \fB}+ \delta \mu \cdot \frac{\partial \VP}{\partial\mu} + \delta (\fB_R)\ab{\mu}{\nu}\wedge\frac{\partial\VP}{\delta (\fB_R)\ab{\nu}{\mu}} + \delta (\mu_R)\ab{\mu}{\nu}\frac{\partial\VP}{\partial(\mu_R)\ab{\nu}{\mu}}
	\\
	&+\delta(2 \fomega) \wedge \frac{\partial \VP}{\partial(2\fomega)} + \delta \fu\wedge \frac{\partial\VP}{\partial\fu}\,.
\end{split}
\end{align}
We are interested in the contribution of the variation to boundary terms. To this end let us write the variations of $\fomega, \fB$, and $\fB_R$ in terms of derivatives of $\delta\fu,\delta\fA$, and $\delta\form{\Gamma}$,
\begin{align}
\begin{split}
\label{E:deltawedgeu}
	(d\delta\fu)\wedge\fu & = (\delta(2\fomega)-\delta\fu\wedge\fa)\wedge\fu\,,
	\\
	(D\delta \fA)\wedge \fu &=( \delta\fB+\delta\fu\wedge \fE) \wedge \fu\,,
	\\
	(D\delta\form{\Gamma}\ab{\mu}{\nu}) \wedge \fu & = (\delta (\fB_R)\ab{\mu}{\nu} + \delta\fu\wedge (\fE_R)\ab{\mu}{\nu})\wedge \fu\,.
\end{split}
\end{align}
Since $\partial\VP/\partial\fB$, $\partial\VP/\partial(2\fomega)$, and $\partial\VP/\partial (\fB_R)\ab{\nu}{\mu}$ each have a leg along $\fu$, we find
\beq
\label{E:deltaVPboundary}
\delta \VP = d\left[ \delta \fA\wedge \cdot  \frac{\partial \VP}{\partial \fB}+\delta\fu\wedge \frac{\partial\VP}{\partial(2\fomega)}+\partial \form{\Gamma}\ab{\mu}{\nu}\wedge \frac{\partial\VP}{\partial(\fB_R)\ab{\nu}{\mu}}\right] + (\text{bulk terms})\,.
\eeq
Comparing \eqref{E:deltaVPboundary} with~\eqref{E:deltaVP}, we find
\beq
\label{E:allAnomTrans}
	\hodge \form{J}_{\fP} = \frac{\partial\VP}{\partial \fB}\,, 
	\qquad 
	\hodge\form{q}_{\fP} = \frac{\partial\VP}{\partial(2\fomega)}\,, 
	\qquad 
	\hodge (\form{\spp}_{\fP})\ab{\mu}{\nu} = \frac{\partial\VP}{\partial (\fB_R)\ab{\nu}{\mu}}\,.
\eeq
In Appendix  \ref{A:deltaVPWCS} we show explicitly that the bulk terms correctly reproduce the bulk flavor and spin currents that correspond to the Chern-Simons form $\form{I}_{CS}$.

Up to this point, we have presented a functional form for $W_{anom}$ and have shown that it correctly reproduces the anomalous variation of $W_{QFT}$. However, the curious reader may be somewhat puzzled as to the why a local $W_{anom}$ should exist and to the origin of the hatted connections that were so crucial in $W_{anom}$'s construction. In the remainder of this Section we will discuss the physical origin of $W_{anom}$ and the hatted connections as well as the relation between our construction and transgression formulae.

On general grounds one does not expect to be able to capture the anomaly by a local term in the generating function. The key ingredient which allows for such a term in our setup is that in the transverse gauge, the generating functional for a hydrostatic state is a local functional on the $2n-1$-dimensional spatial slice~\cite{Banerjee:2012iz,Jensen:2012jh}. This essentially follows from analytically continuing to the Euclidean theory and dimensionally reducing on the thermal circle. For a theory with a finite static screening length, such a dimensional reduction generates an effective $2n-1$-dimensional theory with a mass gap. Hydrostatic response may then be described by a local generating function on the spatial slice. When the $2n$-dimensional theory has anomalies, the theory on the spatial slice will necessarily be anomalous under gauge and/or coordinate transformations. However, in odd dimensions any local gauge or coordinate variation of $W_{QFT}$ may be removed by the addition of a suitable local counterterm. Thus we are guaranteed that a local $W_{anom}$ exists in hydrostatic equilibrium.

We now turn to the relation between $W_{anom}$, the hatted connections and transgression formulae.  To understand this relation it is useful to back up a step and consider a general (non-equlibrated) configuration of background fields. That is, we no longer demand that the background fields are invariant under the action of a timelike symmetry $K$. Consider {two} sets of background fields $\{\fA_1,g_1\}$ and $\{\fA_2,g_2\}$, from which we construct the corresponding field strengths $\{\fF_1,\fR_1\}$ and $\{\fF_2,\fR_2\}$ and so the anomaly polynomials and Chern-Simons forms evaluated on the ``1'' or ``2'' connections. To save space, we notate these as
\beq
\fP_i \equiv \fP[\fF_i,\fR_i]\,, \qquad \form{I}_i \equiv \ICS[\fA_i,\fF_i;\form{\Gamma}_i,\fR_i]\,.
\eeq
It is a classic result that one may construct a gauge and coordinate-invariant functional $\form{V}_{12}\equiv \form{V}_{12}[\fF_i,\fR_i]$ such that
\beq
\label{E:V12}
\fP_1-\fP_2 = d\form{V}_{12}\,,
\eeq
where $\form{V}_{12}$ may be given an integral expression in terms of a flow in the space of connections from $\{\fA_2,\form{\Gamma}_2\}$ to $\{\fA_1,\form{\Gamma}_1\}$. Similarly, the difference of Chern-Simons forms is given by
\beq
\label{E:VandW12}
\form{I}_1 - \form{I}_2 = \form{V}_{12} + d\form{W}_{12}\,,
\eeq
where $\form{W}_{12}$ explicitly depends on both sets of connections and may be expressed in terms of a double integral in the space of connections. We reproduce the construction of $\form{V}_{12}$ and $\form{W}_{12}$ in Appendix~\ref{A:transgression}. These results are collectively known as ``transgression formulae'' of the first and second kind respectively and are useful when studying the relation between anomalies and algebraic topology. For instance, when the ``1'' and ``2'' connections differ by a gauge transformation, the integral of $\form{V}_{12}$ calculates the phase picked up by $W_{cov}$, which must be $2\pi i$ times an integer so that the theory is invariant. In such an instance $\form{V}_{12}$ computes a topological invariant of the bundle in which the connections live.

By~\eqref{E:VandW12}, the difference of Chern-Simons terms $\form{I}_1-\form{I}_2$ varies under a gauge or coordinate transformation by a boundary term given by the variation of $\form{W}_{12}$. As a result, the integral of $\form{W}_{12}$ {almost} gives a $2n$-dimensional functional which reproduces the anomalies of $W_{QFT}$. Indeed, if we denote the variation of $W_{QFT}$ under a gauge/coordinate transformation $\delta_{\lambda}$ as
\beq
\delta_{\lambda} W_{QFT}[\fA,g] = \int \form{G}_{\lambda}[\fA,\form{\Gamma}]\,,
\eeq
then we have
\beq
\delta_{\lambda} \int \form{W}_{12} = - \int \left(\form{G}_{\lambda}[\fA_1,\form{\Gamma}] - \form{G}_{\lambda}[\fA_2,\form{\Gamma}_2]\right)\,.
\eeq
That is, the anomalous variation of $-\int \form{W}_{12}$ reproduces the variation of $W_{QFT}$ were it coupled to the ``1'' background, minus the variation it would exhibit if it were coupled to the ``2'' background. 

Let us return to hydrostatic equilibrium. In our construction, we also introduced two sets of connections: the physical ones $\{\fA,\form{\Gamma}\ab{\mu}{\nu}\}$ which we coupled to our field theory, and the hatted connections. We relate our hatted and unhatted connections to transgression upon the identification
\beq
\label{E:transgressTohats}
\fA_1 = \fA\,, \qquad \fA_2 = \hat{\fA}
\eeq
and similarly for the gravitational connection. Crucially, in the transverse gauge the hatted connections are transverse $\hat{A}_{\mu}u^{\mu}=0$ and $\hat{\Gamma}\ab{\mu}{\nu\rho}u^{\rho}$. Additionally, $\form{G}_{\lambda}$ is a $2n$-form given by a sum of wedge products of the connections with themselves and the field strengths. As a result $\hat{\form{G}}_{\lambda} = \form{G}_{\lambda}[\hat{\fA},\hat{\form{\Gamma}}]$ vanishes in transverse gauge, in which case the anomalous variation of $W_{QFT}$ when coupled to the background $\{\fA,\form{\Gamma}\}$ is reproduced by the {local} functional $-\int \form{W}_{12}$, just like $-\int \WCS$. Indeed, one can show that $\WCS$ is precisely $\form{W}_{12}$ under the identification~\eqref{E:transgressTohats}. Similarly, we have $\VP=\form{V}_{12}$.

In summary, the mechanism behind our construction of $W_{anom}$ and $\VP$ is the transgression machinery of e.g.~\cite{Becchi:1975nq,Stora:1983ct,Zumino:1983rz}, applied to the physical connections $\{\fA,\form{\Gamma}\}$ to which we coupled our field theory and the hatted connections~\eqref{E:hatConnectionsAgain} built from them. The hatted connections are special because, in hydrostatic equilibrium, one can go to a gauge where they are completely transverse. In that case the boundary term $\form{W}_{12}$ in the decomposition~\eqref{E:VandW12} provides a representative for $W_{anom}$, which is, of course, the one given in~\eqref{E:Wanom}.

\section{The relation to hydrodynamics} \label{S:conclusion}

Our results~\eqref{E:WanomisWCS} and~\eqref{E:allAnomTrans} describe anomaly-induced response in equilibrium. In this Section we make contact with recent developments in fluid mechanics and the macroscopic manifestation of anomalies in hydrodynamics. Among other things, the following section provides an example where the computational simplicity of the formalism established in this paper is made clear. It also sets the stage for a companion paper~\cite{companion} which will summarize a much more subtle form of anomaly-induced response.

In Section~\ref{S:intro} we briefly summarized hydrodynamic theory. When studying a many body system at distances which are much larger than the typical mean free path, the effective degrees of freedom are a local temperature $T$, a local rest frame characterized by the timelike vector $u^{\mu}$ satisfying $u^2=-1$, and a local chemical potential $\mu$. Taking these fields as well as the background metric $g_{\mu\nu}$ and gauge field $A_{\mu}$ to be slowly varying, one expands the energy-momentum tensor $T_{cov}^{\mu\nu}$ and current $J_{cov}^{\mu}$ in a gradients of the hydrodynamic and background fields. The relation between the hydrodynamic fields and background sources and the conserved currents are called the constitutive relations. 

The hydrodynamic variables are then determined by treating the Ward identities for $T_{cov}^{\mu\nu}$ and $J_{cov}^{\mu}$ as equations of motion.  To construct the Ward identities we must identify the anomalies of the theory. In four dimensions there are two types of anomalies, pure flavor anomalies, and mixed flavor-gravitational anomalies. For simplicity, consider a theory with a single $U(1)$ flavor symmetry which has both types of anomalies. These are encoded in the anomaly polynomial
\beq
\label{E:4dAnomPoly}
\fP = c_{_A} \fF\wedge \fF\wedge \fF + c_m \fF \wedge \form{R}\ab{\mu}{\nu} \wedge \form{R}\ab{\nu}{\mu}\,,
\eeq
where the coefficients $c_{_A}$ and $c_m$ describe the strength of the flavor and mixed anomalies respectively. In a theory with a functional integral description with chiral fermions they are given by
\beq
c_{_A} = - \frac{1}{3!(2\pi)^2}\sum_{species} \chi_i q_i^3\,, \qquad c_m = -\frac{1}{4!(8\pi)^2}\sum_{species} \chi_i q_i\,,
\eeq
where $\chi_i=\pm 1$ indicates the fermion chirality (in our conventions right-handed fermions have $\chi_i=1$) and $q_i$ denotes the fermion $U(1)$ charge. In a four-dimensional theory with the anomaly polynomial~\eqref{E:4dAnomPoly}, the anomalous Ward identities are
\begin{align}
\begin{split}
\label{E:4dWard}
D_{\mu} J^{\mu}_{cov} &= \frac{1}{4}\epsilon^{\mu\nu\rho\sigma}\left[ 3c_{_A} F_{\mu\nu}F_{\rho\sigma} + c_m R\ab{\alpha}{\beta\mu\nu}R\ab{\beta}{\alpha\rho\sigma}\right]\,,
\\
D_{\nu}T^{\mu\nu}_{cov} & = F\ab{\mu}{\nu}J^{\nu}_{cov} + \frac{c_m}{2}D_{\nu}\left[\epsilon^{\rho\sigma\alpha\beta}F_{\rho\sigma}R\ab{\mu\nu}{\alpha\beta}\right]\,.
\end{split}
\end{align}

The constitutive relations for the current and stress tensor have been worked out in detail (see e.g~\cite{Jensen:2012kj}). Let us decompose them into irreducible representations of the rotational invariance which fixes $u^{\mu}$,
\beq
\label{E:4dhydroDecomp}
J^{\mu}_{cov} = \mathcal{N} u^{\mu} + \nu^{\mu}\,, \qquad T^{\mu\nu}_{cov} = \mathcal{E} u^{\mu}u^{\nu} + \mathcal{P} P^{\mu\nu} + u^{\mu}q^{\nu}+u^{\nu}q^{\mu} + \tau^{\mu\nu}\,,
\eeq
where
\beq
P^{\mu\nu} = g^{\mu\nu}+u^{\mu}u^{\nu}\,, \qquad \nu_{\mu}u^{\mu} = q_{\mu}u^{\mu} = \tau_{\mu\nu}u^{\nu} = \tau_{\mu\nu}g^{\mu\nu}=0\,.
\eeq
To first order in derivatives, the constitutive relations are parameterized by
\begin{subequations}
\label{E:4dconstitutive}
\begin{align}
\mathcal{P}&  = P-\zeta D_{\mu}u^{\mu}\,, & \mathcal{E} &= - P + T\frac{\partial P}{\partial T} + \mu \frac{\partial P}{\partial \mu}\,, & \mathcal{N} = \frac{\partial P}{\partial\mu}\,,
\end{align}
and
\begin{align}
\begin{split}
\nu^{\mu} & = \sigma \left(E^{\mu}-TP^{\mu\nu}D_{\nu}\left(\frac{\mu}{T}\right)\right)+\chi_E E^{\mu} + \chi_T P^{\mu\nu}D_{\nu} T + \xi_1 B^{\mu} + \xi_2 \omega^{\mu} \,,
\\
q^{\mu} & = \xi_2 B^{\mu} + \xi_3\omega^{\mu}\,,
\\
\tau^{\mu\nu} &= -\eta \sigma^{\mu\nu}\,,
\end{split} 
\end{align}
where we have defined
\begin{align}
B^{\mu} &= \frac{1}{2} \epsilon^{\mu\nu\rho\sigma} u_{\nu}F_{\rho\sigma}\,, & \omega^{\mu} &=\epsilon^{\mu\nu\rho\sigma}u_{\nu}\partial_{\rho}u_{\sigma}\,,
\\
E_{\mu} & = F_{\mu\nu}u^{\nu}\,, & \sigma^{\mu\nu} &= P^{\mu\rho}P^{\nu\sigma} (D_{\rho}u_{\sigma}+D_{\sigma}u_{\rho}) - \frac{2}{3}P^{\mu\nu}D_{\rho}u^{\rho}\,.
\end{align}
\end{subequations}
The quantity $P$ is the pressure, $\eta$ the shear viscosity, $\sigma$ the conductivity, and $\zeta$ the bulk viscosity.

Equation~\eqref{E:4dconstitutive} is the most general one compatible with the symmetries of the system as well with the equations of motion of ideal hydrodynamics. However, further constraints arise when we demand the existence of an entropy current~\cite{landau_fluid_1987,Bhattacharya:2011tra}. That is, we demand the existence of a current whose divergence is positive for fluid flows which solve the Ward identities~\eqref{E:4dWard} (and in a thermodynamic equilibrium in the absence of external sources it reduces to the entropy density times the velocity field). Solving this constraint leads to a set of relations between the parameters appearing in~\eqref{E:4dconstitutive}.\footnote{In presenting the solution~\eqref{E:equalityCon}, we are writing the constitutive relations in a particular hydrodynamic frame (see~\cite{Loganayagam:2011mu,Jensen:2012jy}) in which we readily make contact with hydrostatic equilibrium.}
The equality type relations are given by
\begin{align}
\begin{split}
\label{E:equalityCon}
	\chi_E & = \chi_T = 0\,,
	\\
	\xi_1 &= - 6 c_{_A}\mu\,,
	\\
	\xi_2 &= - 3 c_{_A} \mu^2 + \tilde{c} T^2\,,
	\\
	\xi_3 &= - 2 c_{_A} \mu^3 + 2\tilde{c} \mu T^2\,,
\end{split}
\end{align}
where $\tilde{c}$ is a constant and we have dropped CPT-violating terms~\cite{Neiman:2010zi,Banerjee:2012iz,Jensen:2012jy}. The inequality-type constraints on the remaining parameters are
\beq
\eta \geq 0\,, \qquad \sigma \geq 0\,, \qquad \zeta \geq 0\,.
\eeq

Before proceeding, we note that the equality-type constraints for $\{\xi_1,\xi_2,\xi_3\}$ in~\eqref{E:equalityCon} are unusual from the point of view of the entropy current. They are in stark contrast with equality-type relations as they usually appear in hydrodynamics, which allow for hydrostatic response parameterized by unspecified functions of state. As an example, consider $(2+1)$-dimensional parity-violating fluids at first order in derivatives~\cite{Jensen:2011xb}. Six new response coefficients are allowed in such a system, four of which may be measured in equilibrium (in the notation of~\cite{Jensen:2011xb} they are $\{\tilde{\chi}_E,\tilde{\chi}_T,\tilde{\chi}_B,\tilde{\chi}_{\Omega}\}$); those four response coefficients obey two equality-type relations, whose solution is given by two arbitrary functions of state. In this sense, the equality-type constraints for $\{\xi_1,\xi_2,\xi_3\}$ are qualitatively different from standard equality-type constraints as they ordinarily appear in hydrodynamics. Correspondingly, the constants $c_{_A}$ and $\tilde{c}$ appear in the hydrostatic generating functional in a unique way relative to other response coefficients. The special role played by $c_{_A}$ shouldn't be a surprise: $c_{_A}$ is an anomaly coefficient and so occupies a special role in both the conservation equations of hydrodynamics and in $W_{anom}$. The role of $\tilde{c}$ is also special, as we now discuss.

The equality constraints~\eqref{E:equalityCon} may also be determined by the properties of hydrostatic states. In hydrostatic equilibrium, the expansion and shear tensors vanish, $D_{\mu}u^{\mu} = 0$ and $\sigma_{\mu\nu}=0$ as does the Einstein term $E^{\mu} - T P^{\mu\nu}D_{\nu}\left(\frac{\mu}{T}\right) = 0$. As a result the remaining terms in~\eqref{E:4dconstitutive} are allowed in equilibrium, and so the corresponding response parameters $\chi_E,\chi_T,\xi_1,\xi_2$, and $\xi_3$ are computed by the hydrostatic generating functional $W_{QFT}$. Writing down the most general CPT-preserving $W_{QFT}$ to one-derivative order, one finds~\cite{Jensen:2012kj}
\beq
\label{E:WQFT4d}
W_{QFT} = \int d^4x \sqrt{-g} \left[ P(T,\mu) + \tilde{c} A_{\mu} T^2 \omega^{\mu} \right] + W_{anom} + O(\partial^2)\,.
\eeq
The current $T^2\omega^{\mu}$ is conserved in equilibrium, and as a result the term proportional to $\tilde{c}$ is gauge-invariant provided that $\tilde{c}$ is {constant}. In fact, in transverse gauge this term becomes a Chern-Simons term on the spatial slice (see~\cite{Jensen:2012kj} for details). In this way, the somewhat unusual presence of the constant $\tilde{c}$ in the response~\eqref{E:equalityCon} may be understood naturally from the point of view of $W_{QFT}$: $\tilde{c}$ is just a Chern-Simons coefficient on the spatial slice.

To obtain $W_{anom}$ we follow the prescription of Section \ref{S:allanomalies}. We first identify the Chern-Simons term associated with the anomaly polynomial \eqref{E:4dAnomPoly}. This can be done via transgression formula as described in Appendix \ref{A:transgression}; As it turns out, there are various equivalent Chern-Simons forms that one may use to describe the $U(1)^3$ and mixed anomalies. One can choose the Chern-Simons form in such a way that $\WCS$ is diffeomorphism-invariant but not gauge-invariant, giving
\beq
\form{I}_{CS} = c_{_A} \fA\wedge \fF\wedge \fF + c_m \fA \wedge \form{R}\ab{\mu}{\nu}\wedge \form{R}\ab{\nu}{\mu}\,.
\eeq
Defining a trace over matrix-valued forms to be
\beq
\text{tr}(A_1\hdots A_m) = (A_1)\ab{\mu_1}{\mu_2}(A_2)\ab{\mu_2}{\mu_3}\hdots (A_m)\ab{\mu_m}{\mu_1}
\eeq
we see that the corresponding forms $\VP$ and $\WCS$ are  given by
\begin{align}
\begin{split}
	\VP 
	&= \frac{\fu}{2\fomega}\wedge \left( \fP[\fB,\fB_R] - \fP[\fB+2\fomega\mu,\fB_R+2\fomega\mu_R]\right) 
\\
	&=- \fu\wedge \left[ c_{_A} (3 \mu \fB\wedge \fB + 6 \mu^2\fB \wedge \fomega + 4 \mu^3\fomega\wedge \fomega)\right.
\\
	&\hspace{.6in}\left.+c_m \left( 2(\fB+2\fomega\mu)\wedge \text{tr}(\mu_R\fB_R+\mu_R^2\fomega)+\mu\,\text{tr}(\fB_R\wedge \fB_R) \right)\right]\,,
\end{split}
\end{align}
and
\begin{align}
\begin{split}
\label{E:4dWCS}
	\WCS
	& = \frac{\fu}{2\fomega}\wedge \left( \form{I}_{CS}- \hat{\form{I}}_{CS}\right)
\\
	& =  -2\fu\wedge \fA\wedge \left[ c_{_A} \mu(\fB+\mu\fomega) + c_m \text{tr}(\mu_R \fB_R+\mu_R^2\fomega)\right]\,.
\end{split}
\end{align}
Note the similarity between the flavor and mixed anomaly terms in $\WCS$. From $\WCS$ we construct the anomalous contribution to the generating functional, $W_{anom}$
\begin{subequations}
\label{E:Wanom4d}
\beq
W_{anom} = - \int \WCS = \int d^4x \sqrt{-g} A_{\mu}(c_{_A} j_{_A}^{\mu} + c_m j_m^{\mu})\,,
\eeq
where
\begin{align}
\label{E:jA}
	j_{_A}^{\mu} 
	&= -2 \epsilon^{\mu\nu\rho\sigma} \mu u_{\nu} \left( \partial_{\rho}A_{\sigma} + \frac{\mu}{2}\partial_{\rho}u_{\sigma}\right)\,, 
	\\
\label{E:jm}
	j_m^{\mu}
	& = - \epsilon^{\mu\nu\rho\sigma} u_{\nu} \left( (\mu_R)\ab{\alpha}{\beta}R\ab{\beta}{\alpha\rho\sigma} + (\mu_R)\ab{\alpha}{\beta}(\mu_R)\ab{\beta}{\alpha} \partial_{\rho}u_{\sigma}\right)\,.
\end{align}
\end{subequations}
Our representatives for the anomalous contribution to the generating function specified by equations \eqref{E:Wanom4d} agrees with results obtained previously in the literature. The contribution from the $U(1)^3$ anomaly is identical to the one found in~\cite{Banerjee:2012iz,Jensen:2012kj} while the contribution of the mixed anomaly agrees with that found in \cite{Jensen:2012kj} up to a gauge and coordinate-invariant expression. Denoting the representative for the anomalous part of $W_{QFT}$ in~\cite{Jensen:2012kj} as $W_A$, we find 
\beq
\label{E:deltaWanom}
W_{anom} - W_A =4c_m  \int d^4x \sqrt{-g}\,\omega_{\mu\nu} a^{\mu}B^{\nu}+(\text{boundary terms}) \,.
\eeq

Varying the generating functional \eqref{E:WQFT4d} and adding appropriate Bardeen-Zumino terms, one obtains $T_{cov}^{\mu\nu}$ and $J_{cov}^{\mu}$ in terms of $P$, $T$, $\mu$, $\tilde{c}_{4d}$, and $c_{_A}$. As described in the text, a simpler way of obtaining the covariant currents is to vary $\VP$ as in~\eqref{E:allAnomTrans}, which leads directly to
\begin{align}
\nonumber
	\hodge \form{J}_{\fP}
	& = \frac{\partial\VP}{\partial\fB} = - \fu\wedge \left[  c_{_A}(6\mu\fB+3\mu^2 (2\fomega)) + c_m \,\text{tr}(2\mu_R\fB_R+\mu_R^2(2\fomega))\right]\,,
	\\
	\nonumber
	\hodge \form{q}_{\fP} 
	& = \frac{\partial\VP}{\partial (2\fomega)} = -\fu\wedge \left[ c_{_A}(3\mu^2\fB+2\mu^3 (2\fomega))+ c_m \left( 2\mu\,\text{tr}(\mu_R\fB_R+ \mu_R^2 (2\fomega))+\text{tr}(\mu_R)^2\fB  \right) \right]\,,
	\\
	\hodge (\form{\spp}_{\fP})\ab{\mu}{\nu} 
	& = \frac{\partial\VP}{\partial (\fB_R)\ab{\nu}{\mu}} = -2c_m\fu\wedge \left[ (\mu_R)\ab{\mu}{\nu} (\fB+\mu (2\fomega))+ \mu(\form{B}_R)\ab{\mu}{\nu}\right]\,.
\end{align}
Dualizing the three-forms $\fu\wedge \fB,\fu\wedge (\fB_R)\ab{\mu}{\nu}$, and $\fu\wedge (2\fomega)$ to the pseudovectors
\beq
b^{\mu} \equiv \half \epsilon^{\mu\nu\rho\sigma}u_{\nu}F_{\rho\sigma}\,, \qquad (b_R)\ab{\mu\alpha}{\beta} \equiv  \half \epsilon^{\mu\nu\rho\sigma}u_{\nu}R\ab{\alpha}{\beta\rho\sigma}\,, \qquad w^{\mu} \equiv \epsilon^{\mu\nu\rho\sigma}u_{\nu}\partial_{\rho}u_{\sigma}\,,
\eeq
the covariant currents are given by the expressions 
\begin{align}
\label{E:4dCurrentsTotal}
\nonumber
	J^{\mu}_{\mathcal{P}} 
	& = -6 c_{_A} \mu \,b^{\mu} - 2c_m (\mu_R)\ab{\alpha}{\beta} (b_R)\ab{\mu\beta}{\alpha} - \left(3c_{_A} \mu^2 + c_m \,\text{tr}(\mu_R^2)\right)w^{\mu}\,,
	\\
	\nonumber
	q^{\mu}_{\mathcal{P}}
	& = -\left(3c_{_A}\mu^2 +  c_m \text{tr}(\mu_R^2)\right)b^{\mu} - 2c_m \mu(\mu_R)\ab{\alpha}{\beta}(b_R)\ab{\mu\beta}{\alpha} - 2\left(c_A \mu^3 + c_m \mu \,\text{tr}(\mu_R^2)\right)w^{\mu}\,,
	\\
	(\spp_{\mathcal{P}})\ab{\mu\alpha}{\beta}
	& = -2 c_m\left((\mu_R)\ab{\alpha}{\beta}b^{\mu}+\mu (b_R)\ab{\mu\alpha}{\beta} + \mu (\mu_R)\ab{\alpha}{\beta} w^{\mu}\right)\,.
\end{align}
The contribution of the $U(1)^3$ anomaly to the stress tensor and currents agree with those in the literature~\cite{Son:2009tf} while the mixed anomaly-induced currents differ from those computed in Appendix B of~\cite{Jensen:2012kj} by the variations of the right hand side of~\eqref{E:deltaWanom}.

Upon matching \eqref{E:4dCurrentsTotal} to the hydrodynamic constitutive relations \eqref{E:4dconstitutive} , one finds precisely the equality-type conditions~\eqref{E:equalityCon}. In particular, the terms proportional to the $U(1)^3$ anomaly coefficient $c_{_A}$ are computed by the corresponding terms in the anomaly-induced currents in~\eqref{E:4dCurrentsTotal}. If we were to extend this analysis to include terms in $W_{QFT}$ with up to three derivatives, we would find a plethora of higher derivative contributions to $T_{cov}^{\mu\nu}$ and $J_{cov}^{\mu}$. These would include the mixed anomaly-induced currents proportional to $c_m$ in~\eqref{E:4dCurrentsTotal}, which give rise to three-derivative terms in the covariant current and stress tensor.

Note that by construction $\tilde{c}$ is not encoded in our choice of representative for $W_{anom}$ or $\VP$. In that sense it is not obviously related to the anomalies of the underlying theory. However, arguments that go beyond hydrodynamics~\cite{Jensen:2012kj,Golkar:2012kb} have verified a conjecture based on calculations at weak~\cite{Landsteiner:2011cp} and strong coupling~\cite{Landsteiner:2011iq} that $\tilde{c}$ is related to the mixed anomaly coefficient as
\beq
\tilde{c} = - 8 \pi^2 c_m\,.
\eeq
We see that anomalies appear in $W_{QFT}$ in two very different ways: (i.) through our representative $W_{anom}$ encoding the anomalous variation of $W_{QFT}$, and (ii.) through the Chern-Simons-like term in $W_{QFT}$~\eqref{E:WQFT4d} proportional to $\tilde{c}$ (and its analogues in other dimensions). 

We elaborate on the case of four-dimensional theories because it serves as a template for the story in general dimension. Motivated by the four-dimensional results, we are led to decompose $W_{QFT}$ into three parts as
\beq
\label{E:WseparatedTransAnom}
W_{QFT} = W_{gauge-invariant} + W_{trans} + W_{anom}\,,
\eeq
where $W_{trans}$ is defined to be the sum of the Chern-Simons-like terms in $W_{QFT}$. In four dimensions this is just the term proportional to $\tilde{c}$ in~\eqref{E:WQFT4d}. This uniquely specifies $W_{trans}$ up to boundary terms. In general, we also choose the representative $W_{anom}$ so that it and $W_{gauge-invariant}$ have vanishing Chern-Simons-like coefficients; one can check that our represenative~\eqref{E:Wanom} for $W_{anom}$ does just this. 

Upon variation of $W_{anom}$ and $W_{trans}$, we obtain the anomaly-induced transport: the terms in the current and stress tensor which are fixed by anomalies. We further distinguish the response due to $W_{anom}$ and $W_{trans}$ by terming the former {rational} anomaly-induced transport, and the latter {transcendental} anomaly-induced transport. We call the latter trascendental due to the relative factor of $\pi^2$ between $\tilde{c}$ and $c_m$. One might worry that such a division is contrived, but from~\eqref{E:equalityCon} we see that it is physically well-motivated: the rational anomaly-induced transport (proportional to $c_{_A}$) is temperature-independent, while and the transcendental response (proportional to $\tilde{c})$ is proportional to powers of the temperature.

To summarize, the methods of this paper may be used to easily compute the anomalous part $W_{anom}$ of the hydrostatic generating functional $W_{QFT}$. Varying $W_{anom}$ leads to the {rational} anomaly-induced transport. Transport associated with rational terms may also be determined by demanding the existence of an entropy current with positive divergence and extracting the terms in the consequent constitutive relations which are explicitly proportional to anomaly coefficients.\footnote{The construction of an entropy current becomes prohibitively difficult as one goes to higher orders in the gradient expansion or includes gravitational anomalies. Nevertheless, we will show in a companion paper~\cite{companion} that the anomaly-induced transport computed in this work solves the entropy constraint for arbitrary anomalies.} However,  $W_{trans}$ and so the {transcendental} anomaly-induced transport is presently uncomputed. We calculate it in an upcoming paper~\cite{companion} by suitably generalizing the methods of~\cite{Jensen:2012kj}.

\acknowledgments

We would like to thank T.~Azeyanagi, K.~Balasubramanian, J.~Bhattacharya, S. Bhattacharyya, C.~Cordova, S.~Jain, Z.~Komargodski, V.~Kumar, P.~Kovtun, H.~Liu, S.~Minwalla, R.~Myers, G.~Ng, M.~Rangamani, A.~Ritz, M.~Rodriquez, S.~Sachdev, and D.~Son for useful conversations and correspondence. RL would like to thank various colleagues at the Harvard Society of Fellows for interesting discussions. KJ also thanks the Perimeter Institute for Theoretical Physics for their hospitality while a portion of this work was completed. KJ was supported in part by NSCERC, Canada and by the National Science Foundation under grant PHY-0969739. RL was supported by the Harvard Society of Fellows through a junior fellowship. AY is a Landau fellow, supported in part by the Taub foundation as well as the ISF under grand number $495/11$, the BSF under grant number $2014350$ the European commission FP7, under IRG $908049$ and the GIF under grant number $1156/2011$.

\begin{appendix}

\section{Ward identities in the absence of anomalies}
\label{A:ward}

In this Appendix, we will state some of the basic results regarding various currents and their associated conservation equations as derived from a generating functional. The results are standard and the reader is encouraged to skim through this subsection paying special attention to how we define spin currents, as it will be useful later.

Let $e^{iW_{QFT}[g,A]}$ denote the partition function of a  quantum field theory living in $d$ spacetime dimensions coupled to a background metric $g_{\mu\nu}$. We will take the metric to be in Lorentzian signature, though later on, we will Wick-rotate this metric in order to get the thermal partition function. In addition, we will assume that the background has profiles for various non-abelian flavor gauge fields (i.e., sources for flavor currents) jointly denoted by $A_\mu$.

The diffeomorphism and flavor gauge invariance of this generating functional leads to conservation equations for the stress tensor and the flavor current. Let us outline how this relation works in a non-anomalous theory and we will then carefully adopt it to anomalous theories in Appendix \ref{A:conCovAnom}. 

By varying the connected generating function $W_{QFT}[g,A]$ with respect to the flavor gauge field and metric we obtain the current and stress tensor,
\begin{equation}\label{eq:deltaWdef}
\begin{split}
\delta W_{QFT}  &\equiv  \int  d^dx\sqrt{-g} \Bigl\{ \delta A_{\mu}\cdot J^\mu 
+ \half  \delta g_{\mu\nu} T^{\mu\nu}\Bigr\} + (\text{boundary terms})\,. 
\end{split}
\end{equation}
Throughout the Appendices and in the main text we find it useful to carry out the variation of the metric in a two stage process. We first treat the connection and metric as separate entities, under which the generating functional varies as 
\beq
\delta W_{QFT} = \int d^{d}x \sqrt{-g} \left[ \delta A_{\mu} \cdot J^{\mu} + \frac{1}{2}\delta g_{\mu\nu} t^{\mu\nu}+ \delta \Gamma\ab{\mu}{\nu\rho} \spp\ab{\rho\nu}{\mu}\right] +(\text{boundary terms})\,.
\eeq
Then, to get $T^{\mu\nu}$ we rewrite $\delta \Gamma\ab{\mu}{\nu\rho}$ in terms of $\delta g_{\mu\nu}$ and integrate by parts, 
\beq
T^{\mu\nu} = t^{\mu\nu} + D_{\rho}\left( \spp^{\mu[\nu\rho]} + \spp^{\nu[\mu\rho]}-\spp^{\rho(\mu\nu)}\right)\,.
\eeq
Here circular (square) brackets indicate (anti-)symmetrization,
\beq
A^{(\mu\nu)} = \frac{1}{2}\left( A^{\mu\nu}+A^{\nu\mu}\right)\,, \qquad A^{[\mu\nu]} = \frac{1}{2}\left( A^{\mu\nu}-A^{\nu\mu}\right)\,.
\eeq

The conservation of the stress tensor and current follow from the diffeomorphism and gauge invariance of $W_{QFT}$. Indeed, let us denote the variation under an infinitesimal gauge transformation $\Lambda$ and coordinate transformation $\xi^{\mu}$ by $\delta_{\chi}$, i.e.,
\begin{align}
\begin{split}
\label{E:deltaAg}
	\delta_{\chi} A_{\mu} &= \pounds_{\xi} A_{\mu} + D_{\mu}\Lambda = \partial_{\mu} \left(\Lambda+\xi^{\sigma}A_{\sigma}\right) + \left[A_{\mu},\,\Lambda+\xi^{\sigma}A_{\sigma}\right]+\xi^{\sigma}F_{\sigma\mu} \\
	\delta_{\chi}g_{\mu\nu} &= \pounds_{\xi} g_{\mu\nu} = D_{\mu}\xi_{\nu} + D_{\nu}\xi_{\mu}\,.
\end{split}
\end{align}
Then, using \eqref{E:deltaAg} and integrating by parts, we find that
\begin{equation}\label{eq:WpreNoether}
J^\mu \cdot \diffF A_\mu 
+ \half  T^{\mu\nu}\diffF g_{\mu\nu} 
= D_\mu N_{\chi}^\mu -  \prn{\Lambda +\xi^\nu A_\nu}  \cdot D_{\mu} J^{\mu} - \xi_{\mu}\Bigl\{ D_{\nu} T^{\mu\nu}
- F^\mu{}_\nu \cdot  J^\nu \   \Bigr\} 
\end{equation}
with
\begin{equation}\label{eq:NoetherXi}
\begin{split}
N_\chi^\mu  &\equiv \prn{\Lambda +\xi^\alpha A_\alpha} \cdot J^\mu+ \xi_\alpha T^{\alpha\mu}   \,.
\end{split}
\end{equation}
Thus, 
\begin{equation}\label{eq:deltaLW}
\begin{split}
\diffF W 
	 &= -\int d^dx \sqrt{-g} \left( 
	 	\prn{\Lambda +\xi^\alpha A_\alpha}  \cdot D_{\mu} J^{\mu} 
 		+ \xi_{\mu}\left( D_{\nu} T^{\mu\nu} - F^\mu{}_\nu \cdot  J^\nu \right) 
		\right)
	+  \left(\substack{\hbox{boundary} \\ \hbox{ terms}} \right)\,.
\end{split}
\end{equation}
The diffeomorphism/flavor gauge invariance of $W$, $\diffF W =0 $, directly implies the conservation equations for the flavor current $J^{\mu}$ and stress tensor $T^{\mu\nu}$, 
\begin{equation}\label{eq:WConservation}
\begin{split}
 D_{\mu} J^{\mu} &=0 \ ,\quad
 D_{\nu} T^{\mu\nu}- F^\mu{}_\nu \cdot  J^\nu = 0 \ ,\quad
 -T^{\mu\nu}+T^{\nu\mu} =0\ ,
\end{split}
\end{equation}
where in the last line we have added in the statement that $T^{\mu\nu}$ is symmetric (which is
equivalent to the conservation of angular momentum). Thus, we conclude that the flavor currents
and angular momentum are covariantly conserved and the energy-momentum is covariantly 
conserved except for the energy-momentum injected via Lorentz force.

Further, substituting \eqref{eq:WConservation} into \eqref{eq:WpreNoether}, we get the Noether identity 
\begin{equation}\label{eq:WNoether}
\begin{split}
D_\mu N_\chi^\mu &= J^\mu \cdot \diffF A_\mu 
+ \half  T^{\mu\nu}\diffF g_{\mu\nu}   \\
\end{split}
\end{equation}
This implies that whenever we place the quantum field theory on a symmetric background with $\diffF A_\mu=0$ and $\diffF g_{\mu\nu}=0$, there is a Noether current $N_\chi^\mu$ which is conserved. Note that the Ward identities in~\eqref{eq:WConservation} are related to, but conceptually distinct from the Noether conservation law~\eqref{eq:WNoether} that arises when \emph{the background sources} are invariant under diffeomorphism/flavor transformations, i.e., when there exists a $\{\xi^\mu,\Lambda\}$ such that $\{\diffF A_\mu =0,\diffF g_{\mu\nu}=0\}$. As we will discuss later, this Noether conservation can hold sometimes even when the conservation laws above get modified by anomalies.

\section{Anomaly inflow}
\label{A:inflow}

The anomaly inflow mechanism of Callan and Harvey \cite{Callan:1984sa} plays a pivotal role in our construction of functions $\form{W}_{CS}$ and $\form{V}_{\mathcal{P}}$ described in detail in Section \ref{S:allanomalies}. In this Appendix, after reviewing the anomaly inflow mechanism we obtain various compact expressions for Hall and Bardeen-Zumino currents associated with anomalies. (For a nice discussion of anomaly inflow in the context of condensed matter physics, see e.g.~\cite{Stone:2012ud}.)

As discussed in Section \ref{S:allanomalies}, the non gauge and reparametrization-invariance of the generating functional $W_{QFT}$ in $2n$ dimensions can be encoded in a $2n+1$ dimensional Chern-Simons form. Indeed, if we think of the $2n$ dimensional manifold on which the anomalous quantum field theory lives as the boundary of a $2n+1$ dimensional manifold $\mathcal{M}$, then the non gauge- and (or) diffeomorphism-invariance of $W_{QFT}$ amounts to the statement that the covariant generating functional $W_{cov}$ defined via
\eqref{E:WcovDef},
\beq
W_{cov} = W_{QFT}+ W_{Hall}\,,
\eeq
with
\begin{equation}
	W_{Hall} = \int_{\mathcal{M}}\form{I}_{CS}[\fA,\form{\Gamma}]
\end{equation}
is gauge and (or) diffeomorphism invariant. Thus,
\begin{equation}
	\delta_{\chi}W_{QFT} = - \delta_{\chi} W_{Hall}\,.
\end{equation}
We remind the reader that bold-face characters label $p$-forms and refer her/him to Section \ref{S:hydrostatics} for the definitions of the connection one-forms $\form{A}$ and $\form{\Gamma}$. In labeling the Chern-Simons action with a ``$Hall$'' subscript, we indicate that this bulk action is reminiscent of the action of a Hall insulator. The reader who is unfamiliar with Hall systems may safely ignore this association. 

The currents $J^{\mu}$ and $T^{\mu\nu}$ obtained by varying $W_{QFT}$ would have been conserved if it were not for the non-gauge and (or) diffeomorphism invariance of $W_{QFT}$. Following the literature, we refer to these currents as consistent currents since $W_{QFT}$ satisfies the Wess-Zumino consistency condition \cite{Wess:1971yu}. 

We refer to the currents on $\mathcal{M}$ which are obtained by varying $W_{Hall}$ as Hall currents and denote them by $T_{H}^{\mu\nu}$ and $J_{H}^{\mu}$. Since $W_{Hall}$ is gauge invariant up to boundary terms the Hall currents are conserved. However, the currents obtained from the boundary variation of $W_{Hall}$ are neither conserved nor are they covariant. We will refer to these currents as Bardeen-Zumino currents (which we will also refer to as Bardeen-Zumino polynomials), $T_{BZ}^{\mu\nu}$ and $J_{BZ}^{\mu}$. More formally, we write the variation of $W_{Hall}$ as
\begin{align}
\begin{split}
\delta W_{Hall} = 
&\int d^{2n+1}x \sqrt{-G} \left[ \delta A_M \cdot J_H^M + \delta \Gamma\ab{M}{NP}(\spp_H)\ab{PN}{M}\right]
\\
& +\int d^{2n}x\sqrt{-g} \left[ \delta A_{\mu} \cdot J_{BZ}^{\mu} + \delta \Gamma\ab{\mu}{\nu\rho}(\spp_{BZ})\ab{\rho\nu}{\mu}\right]\,,
\end{split}
\end{align}
where we have extended the boundary metric $g$ on $\partial \mathcal{M}$ to a metric $G$ on $\mathcal{M}$. Note that since $W_{Hall}$ only depends on the metric $G$ through the connection $\Gamma$, there is only a flavor current $J_H$ and a spin current $\spp_H$. 
In terms of forms we can rewrite the variation of $W_{Hall}$ as
\begin{align*}
\delta W_{Hall} = \int_{\mathcal{M}} \left[\delta \form{A}\wedge \cdot \hodge \form{J}_H +\delta \form{\Gamma}\ab{b}{a}\wedge \hodge(\form{\spp}_H)\ab{a}{b}\right] + \int_{\partial\mathcal{M}} \left[\delta \form{A}\wedge \cdot \hodge\form{J}_{BZ} + \delta \form{\Gamma}\ab{\nu}{\mu}\wedge \hodge (\form{\spp}_{BZ})\ab{\mu}{\nu}\right]\,.
\end{align*}

\begin{figure}[hbt]
\begin{center}
\includegraphics[width=9 cm]{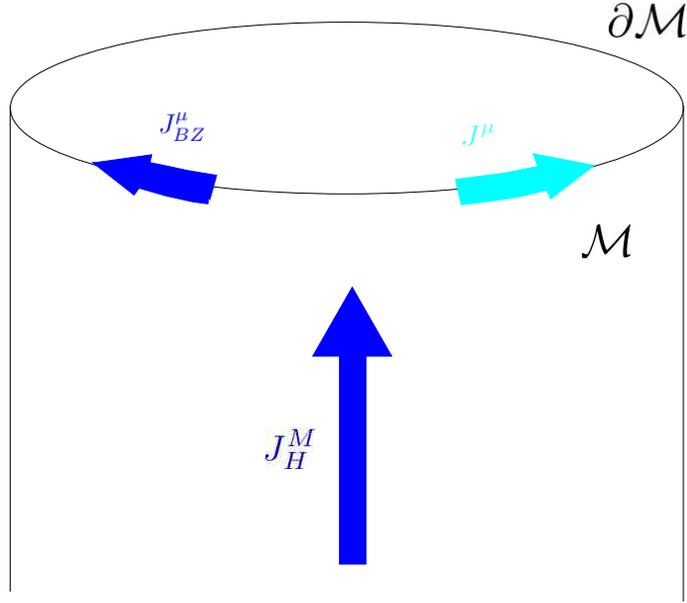}
\caption{\label{F:bulk boundary} Schematic diagram of the inflow mechanism where the manifold $\mathcal{M}$ is depicted as a semi-infinite cylinder and $\partial \mathcal{M}$ as its boundary. The current $J_{Hall}^M$ defined on the manifold $\mathcal{M}$ is conserved but transfers charge to the boundary theory on $\partial\mathcal{M}$ rendering it anomalous. The anomalous boundary current gets a contribution from the Bardeen-Zumino term $J_{BZ}^{\mu}$ associated with the flow of bulk charge and a consistent current associated with the theory defined on $\partial \mathcal{M}$.}
\end{center}
\end{figure}

We also define covariant currents as boundary variations of $W_{cov}$, $J_{cov}^{\mu}$ and $T_{cov}^{\mu\nu}$,
\begin{align}
\delta W_{cov} =& \int d^{2n}x \sqrt{-g} \left[ \delta A_{\mu} \cdot (J^{\mu}+J_{BZ}^{\mu}) + \frac{1}{2}\delta g_{\mu\nu}(T^{\mu\nu} + T_{BZ}^{\mu\nu})\right]
\\
\nonumber
& + \int d^{2n+1}x \sqrt{-G} \left[ \delta A_M \cdot J_H^M + \frac{1}{2}\delta G_{MN}T_H^{MN}\right] + (\text{boundary terms on }\partial\mathcal{M})\,.
\end{align}
From \eqref{E:WcovDef} we find that the covariant currents are the sum of the consistent currents and Bardeen Zumino currents, 
\beq
J_{cov}^{\mu} = J^{\mu} + J_{BZ}^{\mu}\,, \qquad T_{cov}^{\mu\nu} = T^{\mu\nu} + T_{BZ}^{\mu\nu}\,.
\eeq
 Because $W_{cov}$ is both gauge and diffeomorphism-invariant, $J_{cov}^{\mu}$ and $T_{cov}^{\mu\nu}$ are indeed gauge and diffeomorphism-covariant as their name advertises. However, they are not conserved. See Figure \ref{F:bulk boundary}.

It is possible to obtain explicit expressions for the Hall and Bardeen-Zumino currents. Consider the Chern-Simons form $\form{I}_{CS}$ which depends on the connections $\form{A}$ and $\form{\Gamma}$ and on the field strengths $\form{F}$ and $\form{R}$. Under a variation of the connections $\{\fA,\form{\Gamma}\ab{a}{b}\}$ it varies as
\begin{equation}
\label{E:deltaICS}
	\delta \form{I}_{CS} = \delta \form{A} \cdot \frac{\partial \form{I}_{CS}}{\partial \form{A} }
					    + \delta \form{F} \cdot \frac{\partial \form{I}_{CS}}{\partial \form{F}}
					    + \delta \form{\Gamma}^{a}{}_b \frac{\partial \form{I}_{CS}}{\partial \form{\Gamma}^a{}_{b}} 
					    + \delta \form{R}^{a}{}_b \frac{\partial \form{I}_{CS}}{\partial \form{R}^a{}_{b} }\,,
\end{equation}
where we have suppressed wedge products for brevity. Under a general variation of the connections $\delta\form{A}$ and $\delta\form{\Gamma}\ab{a}{b}$, the curvatures vary as
\beq
\delta \form{F} = D\delta\form{A}, \qquad \delta \form{R}\ab{a}{b} = D\form{\Gamma}\ab{a}{b},
\eeq
from which we obtain
\begin{align}
\begin{split}
\label{E:deltaICS2}
	\delta \form{I}_{CS} = &\delta \form{A} \cdot \left( \frac{\partial \form{I}_{CS}}{\partial \form{A}} + D \left(\frac{\partial \form{I}_{CS}}{\partial \form{F}} \right) \right)
				       	    + \delta \form{\Gamma}^{a}{}_b \left(\frac{\partial \form{I}_{CS}}{\partial \form{\Gamma}^a{}_b} + D \left( \frac{\partial \form{I}_{CS}}{\partial \form{R}^a{}_b}\right) \right)   
					    \\
					    &\qquad + d \left(\delta \form{A} \cdot \frac{\partial \form{I}_{CS}}{\partial \form{F}} + \delta \form{\Gamma}^{\mu}{}_{\nu} \frac{\partial \form{I}_{CS}}{\partial \form{R}^{\mu}{}_{\nu}}\right) \,.
\end{split}
\end{align}				
In \eqref{E:deltaICS2} we have defined the action of $D$ on the non-covariant quantities $\partial \form{I}_{CS}/\partial\form{F}$ and $\partial\form{I}_{CS}/\partial \form{R}\ab{a}{b}$ as if they were covariant forms, namely
\begin{align}
\begin{split}
	D \left(\frac{\partial \form{I}_{CS}}{\partial \form{F}}\right) 
	&= d\left(\frac{\partial \form{I}_{CS}}{\partial \form{F}} \right)+ \form{A} \frac{\partial \form{I}_{CS}}{\partial \form{F}} + \frac{\partial\form{I}_{CS}}{\partial \form{F}}\form{A} \,, 
	\\
	D\left(\frac{\partial \form{I}_{CS}}{\partial\form{R}\ab{a}{b}}\right)
	&=d\left(\frac{\partial\form{I}_{CS}}{\partial\form{R}\ab{a}{b}}\right)+\form{\Gamma}\ab{b}{c}\frac{\partial\form{I}_{CS}}{\partial \form{R}\ab{a}{c}}+\frac{\partial \form{I}_{CS}}{\partial\form{R}\ab{c}{b}}\form{\Gamma}\ab{c}{a}\,.
\end{split}
\end{align}
Taking an exterior derivative of both sides and using $d\delta \form{I}_{CS} = \delta \fP$ we find that
\begin{multline}
	\delta\fP= \delta\form{F} \cdot \left( \frac{\partial \form{I}_{CS}}{\partial \form{A}} + D \left( \frac{\partial \form{I}_{CS}}{\partial \form{F}} \right) \right) + \delta \form{R}^a{}_b \left(  \frac{\partial \form{I}_{CS}}{\partial \form{\Gamma}^a{}_b} + D \left( \frac{\partial \form{I}_{CS}}{\partial \form{R}^a{}_b} \right) \right) \\
	- \delta \form{A} \cdot D\left(\frac{\partial \form{I}_{CS}}{\partial \form{A}} + D \left(\frac{\partial \form{I}_{CS}}{\partial \form{F}} \right) \right) 	- \delta \form{\Gamma}^a{}_b  D\left(\frac{\partial \form{I}_{CS}}{\partial \form{\Gamma}^a{}_b} + D \left(\frac{\partial \form{I}_{CS}}{\partial \form{R}^a{}_b} \right) \right)\,.
\end{multline}
Since $\fP$ does not depend explicitly on $\form{A}$ or $\form{\Gamma}$ we conclude that
\begin{equation}
	D\left(\frac{\partial \form{I}_{CS}}{\partial \form{A}} + D \left(\frac{\partial \form{I}_{CS}}{\partial \form{F}} \right) \right)=0\,,
	\qquad
	D\left(\frac{\partial \form{I}_{CS}}{\partial \form{\Gamma}^a{}_b} + D \left(\frac{\partial \form{I}_{CS}}{\partial \form{R}^a{}_b} \right) \right) = 0\,,
\end{equation}
and
\begin{equation}
\label{E:PandICS}
	\frac{\partial \fP}{\partial \form{F}} = \frac{\partial \form{I}_{CS}}{\partial \form{A}} + D \left( \frac{\partial \form{I}_{CS}}{\partial \form{F}} \right)\,,
	\qquad
	\frac{\partial \fP}{\partial \form{R}^a{}_b} = \frac{\partial \form{I}_{CS}}{\partial \form{\Gamma}^a{}_b} + D \left( \frac{\partial \form{I}_{CS}}{\partial \form{R}^a{}_b} \right)\,.
\end{equation}
Combining these results leads to the useful identities
\beq
\label{E:conservedHall}
D\left(\frac{\partial\fP}{\partial\form{F}}\right)=0\,,
\qquad
D\left(\frac{\partial \fP}{\partial\form{R}\ab{a}{b}}\right)=0\,.
\eeq
Thus,
\begin{equation}
	\delta \form{I}_{CS} = \delta \form{A} \cdot \frac{\partial \fP}{\partial \form{F}} + \delta \form{\Gamma}^a{}_b \frac{\partial \fP}{\partial \form{R}^a{}_b} + d \left(\delta \form{A} \cdot \frac{\partial \form{I}_{CS}}{\partial \form{F}} + \delta \form{\Gamma}^{\mu}{}_{\nu} \frac{\partial \form{I}_{CS}}{\partial \form{R}^{\mu}{}_{\nu}} \right)
\end{equation}
from which
\begin{equation}\label{E:HallBZ}
\begin{split}
 \hodge \form{J}_H  &= \prn{\frac{\partial \fP}{\partial \fF}}_{\fR}\ ,\hspace{.6in}
(\hodge \form{\spp}_H)^b{}_a = \prn{\frac{\partial \fP}{\partial \fR^a{}_b }}_{\fF}  ,\\	
 \hodge \form{J}_{BZ} &=  \prn{\frac{\partial \ICS }{\partial \fF }}_{\fA,\fGamma,\fR} \ ,\quad	
 (\hodge \form{\spp}_{BZ})^\nu{}_\mu =  \prn{\frac{\partial \ICS }{\partial \fR^\mu{}_\nu } }_{\fGamma,\fA,\fF}
\end{split}
\end{equation}
follows. In \eqref{E:HallBZ} we have kep the variables which are kept fixed in the subscript for completeness. 
The identities~\eqref{E:conservedHall} then amount to the fact that the Hall flavor and spin currents are covariantly conserved,
\begin{equation}\label{eq:HallProp1}
\begin{split}
D_a J_H^a =0 \ ,\quad D_c \spp_{H}^{cab} =0 \ ,\quad  
\spp_{H}^{cab} =-\spp_H^{cba} \ ,\quad  (\spp_H)^a{}_{ac}=0\,,
\end{split}
\end{equation}
where the last property follows from applying the first Bianchi identity. 

In order to go from the spin currents to the Hall and Bardeen-Zumino stress-energy tensors, we convert the variation of the Levi-Civita connections $\delta\form{\Gamma}\ab{b}{a}$ and $\delta\form{\Gamma}\ab{\nu}{\mu}$ to variations of the metric. This leads to a Hall stress tensor
\beq
T_H^{MN} = D_P\left(\spp_H^{M[NP]} + \spp_H^{N[MP]} - \spp_H^{P(MN)}\right)\,.
\eeq
Going back to the variation of $W_{Hall}$ and integrating the variations of $\Gamma$ by parts, we can write
\beq
\label{E:deltaWhall}
\delta W_{Hall} = \int d^{2n+1}x \sqrt{-G} \left[ \delta A_M \cdot J_H^M + \frac{1}{2}\delta G_{MN}T_H^{MN}\right] + \int d^{2n} x \sqrt{-g} \left[ \delta A_{\mu}\cdot J_{BZ}^{\mu} + \frac{1}{2}\delta g_{\mu\nu} T_{BZ}^{\mu\nu}\right]\,.
\eeq
which defines the Bardeen-Zumino (BZ) polynomials $J_{BZ}^{\mu}$ and $T_{BZ}^{\mu\nu}$. The BZ polynomial for the stress tensor is related to the $\spp_{BZ}$ and the Hall spin current as
\begin{align}
\begin{split}
\label{E:BZStress}
T_{BZ}^{\mu\nu} &= t_{BZ}^{\mu\nu} + D_{\rho}\left(\spp_{BZ}^{\mu[\nu\rho]} + \spp_{BZ}^{\nu[\mu\rho]} -\spp_{BZ}^{\rho(\mu\nu)}\right)\,,
\\
t_{BZ}^{\mu\nu} & = -\left( \spp_H^{\mu[\nu\perp]} + \spp_H^{\nu[\mu\perp]} - \spp_H^{\perp(\mu\nu)}\right)\,,
\end{split}
\end{align}
where the $\perp$ direction is perpendicular to the boundary $\partial\mathcal{M}$. 
The $t_{BZ}^{\mu\nu}$ contribution to the Bardeen-Zumino stress tensor arises from integrating parts in the bulk. It is a covariant, purely extrinsic contribution, in the sense that it involves the curvature forms $\form{R}\ab{a}{\perp}$ and $\form{R}\ab{\perp}{a}$ on the hypersurface $\partial\mathcal{M}$ where our theory lives. Put differently, it provides information on how $\partial\mathcal{M}$ is embedded into $\mathcal{M}$. In what follows we consistently set these extrinsic terms to zero. In field theory terms, the anomalies of our theory only depend on the intrinsic $2n$-dimensional sources which we couple to the theory.\footnote{We point out that in topologically non-trivial phases with anomalous edge states, there may be anomalies associated with the extrinsic data of the edge state.} As a result, the BZ polynomial for the stress tensor may be understood as coming from a BZ polynomial for the spin current.

\section{Ward identities in the presence of anomalies}
\label{A:conCovAnom}

In this Appendix we will obtain the anomalous Ward identities for a general theory with anomaly polynomial $\fP$. We will obtain the Ward identities obeyed by the consistent currents as well as by the covariant currents. 

We begin with the most general variation of the generating functional $W_{QFT}$ for our theory given by equation \eqref{eq:deltaWdef} which we reproduce here for convenience.
\beq
\label{E:deltaWQFT}
\delta W_{QFT} = \int d^{2n}x\sqrt{-g} \left[ \delta A_{\mu}\cdot J^{\mu} + \frac{1}{2}\delta g_{\mu\nu}T^{\mu\nu}\right]\,.
\eeq
To obtain the Ward identities, we perform an infinitesimal gauge transformation $\Lambda$ and coordinate variation $x^{\mu}\to x^{\mu}+\xi^{\mu}$, which we collectively notate as $\delta_{\chi}$ (see \eqref{E:deltaAg}) and we obtain
\beq
\label{E:deltaWlambda}
\delta_{\chi} W_{QFT} = - \int d^{2n}x\sqrt{-g} \left[ \Lambda \cdot D_{\mu} J^{\mu}+ \xi_{\mu} \left( D_{\nu}T^{\mu\nu}-F\ab{\mu}{\nu}\cdot J^{\nu}+ A^{\mu}\cdot D_{\nu}J^{\nu}\right)\right]
\eeq
as in \eqref{eq:deltaLW}.
For non-anomalous theories we have $\delta_{\chi}W_{QFT}=0$ and so~\eqref{E:deltaWlambda} leads to the standard Ward identities for the current and stress tensor as in \eqref{eq:WConservation}. When our theory has anomalies the nonzero variation of $W_{QFT}$ is related to the nonzero variation of the Chern-Simons form in one higher dimension via
\beq
\label{E:deltachiWQFT}
\delta_{\chi} W_{QFT} = -\delta_{\chi} W_{Hall}\,,
\eeq
where $W_{Hall} = \int_{\mathcal{M}}\form{I}_{CS}$ was studied in Appendix~\ref{A:inflow}.

To continue we must determine the explicit gauge and diffeomorphism variation of the Chern-Simons form. Since the Chern-Simons form is defined in one higher dimension we must extend the gauge and coordinate variations on the boundary to variations in the bulk. The gauge and coordinate variations of the connections and curvatures may be efficiently written in terms of forms as
\begin{align}
\begin{split}
\label{E:deltaLambdaSources}
	\delta_{\chi}\form{A} &= d\Lambda + [\form{A},\Lambda] + \lieD_{\xi}\form{A}\,,
	\qquad
	\delta_{\chi}\form{F} = [\form{F},\Lambda] + \lieD_{\xi}\form{F}\,,
	\\
	\delta_{\chi}\form{\Gamma}\ab{a}{b} &= dv\ab{a}{b} + \lieD_{\xi}\form{\Gamma}\ab{a}{b}\,,
	\hspace{.57in}
	\delta_{\chi}\form{R}\ab{a}{b}  = \lieD_{\xi}\form{R}\ab{a}{b}\,,
\end{split}
\end{align}
where $\lieD_{\xi}$ is a Lie derivative along $\xi$, $\Lambda$ is the gauge transformation parameter and $v\ab{a}{b}$ is defined as the 0-form
\beq
v\ab{a}{b}=\partial_b\xi^a\,.
\eeq
The operators $\lieD_{\xi}$, and $[\cdot,\Lambda]$ (meaning the adjoint action of $\Lambda$ on a tensor in some representation of the flavor symmetry group) all satisfy linearity and the Leibniz rule and so act like derivatives on objects constructed out of differential forms. As a result we have
\begin{align}
\begin{split}
\label{E:derivations}
\lieD_{\xi}\form{I}_{CS} &=\lieD_{\xi}\form{A}\cdot \frac{\partial\form{I}_{CS}}{\partial\form{A}}+\lieD_{\xi}\form{\Gamma}\ab{a}{b}\frac{\partial\form{I}_{CS}}{\partial\form{\Gamma}\ab{a}{b}}+\lieD_{\xi}\form{F}\cdot \frac{\partial\form{I}_{CS}}{\partial\form{F}}+\lieD_{\xi}\form{R}\ab{a}{b}\frac{\partial\form{I}_{CS}}{\partial\form{R}\ab{a}{b}}\,,
\\
[\form{I}_{CS},\Lambda] &=[\form{A},\Lambda]\cdot \frac{\partial\form{I}_{CS}}{\partial\form{A}}+[\form{\Gamma}\ab{a}{b},\Lambda]\frac{\partial\form{I}_{CS}}{\partial\form{\Gamma}\ab{a}{b}}+[\form{F},\Lambda]\cdot \frac{\partial\form{I}_{CS}}{\partial\form{F}}+[\form{R}\ab{a}{b},\Lambda]\frac{\partial\form{I}_{CS}}{\partial\form{R}\ab{a}{b}}\,,
\end{split}
\end{align}
Since $\form{I}_{CS}$ is a flavor singlet it must satisfy $[\form{I}_{CS},\Lambda]=0$. The Lie derivative $\lieD_{\xi} \form{I}_{CS}$ is a total derivative since $\form{I}_{CS}$ is a top form, and so only contributes a boundary term which in fact vanishes.
Putting together~\eqref{E:deltaICS} (for a gauge and diffeomorphism variation), \eqref{E:deltaLambdaSources} and~\eqref{E:derivations} we find
\begin{align}
\begin{split}
\label{E:deltaLambdaICS}
\delta_{\chi} \form{I}_{CS} &= d\Lambda \cdot \frac{\partial\form{I}_{CS}}{\partial\form{A}}+dv\ab{a}{b}\frac{\partial\form{I}_{CS}}{\partial\form{\Gamma}\ab{a}{b}} + \lieD_{\xi}\form{I}_{CS}+ [\form{I}_{CS},\Lambda]
\\
& = -\Lambda \cdot d\left(\frac{\partial\form{I}_{CS}}{\partial\form{A}}\right)
- v\ab{a}{b} d\left(\frac{\partial\form{I}_{CS}}{\partial\form{\Gamma}\ab{a}{b}}\right)+d\left[ \Lambda \cdot \frac{\partial\form{I}_{CS}}{\partial\form{A}}+v\ab{\mu}{\nu}\frac{\partial\form{I}_{CS}}{\partial\form{\Gamma}\ab{\mu}{\nu}}\right] + \lieD_{\xi}\form{I}_{CS}\,.
 \end{split}
\end{align}
The Chern-Simons form is gauge and coordinate invariant up to a boundary term. It then follows that the bulk variations in the second line of~\eqref{E:deltaLambdaICS} must vanish,
\beq
\label{E:closedJT}
d\left(\frac{\partial\form{I}_{CS}}{\partial\form{A}}\right) = 0\,, 
\qquad
d\left(\frac{\partial\form{I}_{CS}}{\partial\form{\Gamma}\ab{a}{b}}\right)=0\,. 
\eeq

We are now in a position to derive the Ward identities for the consistent currents. Using~\eqref{E:deltaLambdaICS}, and \eqref{E:deltachiWQFT} the gauge and coordinate variation of $W_{QFT}$ is given by
\beq
\label{E:deltaLambdaWH1}
\delta_{\chi}W_{QFT} = -\int d^{2n}x \sqrt{-g} \left[ \Lambda \cdot \mathcal{J} + \partial_{\nu}\xi^{\mu}\mathcal{T}\ab{\nu}{\mu}\right]\,.
\eeq
where we have defined
\beq
\label{E:defConAnom}
\hodge\form{\mathcal{J}} = \frac{\partial\form{I}_{CS}}{\partial \form{A}}\,,
\qquad
\hodge\form{\mathcal{T}}\ab{\mu}{\nu} = \frac{\partial\form{I}_{CS}}{\partial \form{\Gamma}\ab{\nu}{\mu}}\,.
\eeq
Equation \eqref{E:closedJT} implies that the $2n$ forms $\hodge\form{\mathcal{J}}$ and $\hodge\form{\mathcal{T}}\ab{\mu}{\nu}$ are closed. 
Comparing \eqref{E:deltaLambdaWH1} with \eqref{E:deltaWlambda} leads to the consistent Ward identities
\begin{align}
\begin{split}
\label{E:conWard}
D_{\mu}J^{\mu} & = \mathcal{J}\,,
\\
D_{\nu}T^{\mu\nu} & = F\ab{\mu}{\nu}\cdot J^{\nu}-A^{\mu}\cdot \mathcal{J} - \frac{1}{\sqrt{-g}}g^{\mu\nu}\partial_{\rho}\left[\sqrt{-g}\mathcal{T}\ab{\rho}{\nu}\right]\,.
\end{split}
\end{align} 
where $\mathcal{J}$ and $\mathcal{T}\ab{\rho}{\nu}$ are given by \eqref{E:defConAnom}.

To compute the Ward identities obeyed by the covariant currents it is useful to first identify the Hall currents in~\eqref{E:deltaWhall} and then carry out the variation \eqref{E:deltaLambdaSources}. obtaining
\begin{align}
\begin{split}
\label{E:deltaLambdaWH2}
\delta_{\lambda} W_{Hall} = -\int d^{2n}x \sqrt{-g} &\left[  \xi_{\mu}(D_{\nu}T_{BZ}^{\mu\nu} -F\ab{\mu}{\nu}\cdot J_{BZ}^{\nu} + A^{\mu}\cdot(D_{\nu}J_{BZ}^{\nu}-J_H^{\perp})-T_H^{\perp\mu})\right.
\\
& \left.\,\,\,\,+\Lambda \cdot (D_{\mu}J_{BZ}^{\mu} - J_H^{\perp})\right]\,,
\end{split}
\end{align}
where $T_H^{\perp\mu}$ is a component of the Hall stress tensor
\beq
T_H^{\perp\mu} = D_{\nu}(\spp_H^{\perp[\mu\nu]} + \spp_H^{\mu[\perp\nu]} - \spp_H^{\nu(\perp\mu)}) = D_{\nu}\spp_H^{\perp[\mu\nu]}\,,
\eeq
and we have dropped extrinsic terms in the final expression. Equating~\eqref{E:deltaLambdaWH1} and~\eqref{E:deltaLambdaWH2} leads to the {Bardeen-Zumino anomaly equations}
\begin{align}
\begin{split}
D_{\mu}J_{BZ}^{\mu} & = J_H^{\perp}-\mathcal{J}\,,
\\
D_{\nu}T_{BZ}^{\mu\nu} & = F\ab{\mu}{\nu}\cdot J_{BZ}^{\nu}-A^{\mu}\cdot \mathcal{J} + \frac{1}{\sqrt{-g}}g^{\mu\nu}\partial_{\rho}\left[ \sqrt{-g}\mathcal{T}\ab{\rho}{\nu}\right] + D_{\nu} \spp_H^{\perp[\mu\nu]}\,,
\end{split}
\end{align}
where $J_{BZ}^{\mu}, J_H^{\perp}, T_{BZ}^{\mu\nu}$ and $\spp_H^{\perp[\mu\nu]}$ are given by \eqref{E:HallBZ} and~\eqref{E:BZStress}. Further, $\mathcal{J}$ and $\mathcal{T}\ab{\rho}{\nu}$ are given by \eqref{E:defConAnom}. It then follows that the covariant current and stress tensor obey the {covariant Ward identities}
\begin{align}
\begin{split}
\label{E:covWard}
D_{\mu}J_{cov}^{\mu} &= J_H^{\perp}\,,
\\
D_{\nu}T_{cov}^{\mu\nu} & = F\ab{\mu}{\nu}\cdot J_{cov}^{\nu} + D_{\nu} \spp_H^{\perp[\mu\nu]}\,,
\end{split}
\end{align}
where $J_H^{\perp}$ and $\spp_H^{\perp[\mu\nu]}$ are given by \eqref{E:HallBZ}. Note that the covariant anomalies are given by transverse components of the Hall current and stress-energy tensor---this is essentially a consequence of the anomaly inflow mechanism, wherein the total current and stress-energy is conserved, but may flow from the bulk to the boundary. 

Before closing this discussion, we point out that one may define a conserved covariant stress tensor
\beq
T_{conserved}^{\mu\nu} = T^{\mu\nu}_{cov} - \spp_H^{\perp[\mu\nu]}\,,
\eeq 
which is conserved by virtue of the Ward identity~\eqref{E:covWard}. However this stress tensor is clearly non-symmetric. This result corresponds to the fact that a diffeomorphism anomaly, which is manifested in the non-conservation of stress-energy, may be exchanged for a Lorentz anomaly, whereby the stress tensor has an antisymmetric part in the presence of background fields.

\section{Transgression formulae}
\label{A:transgression}

At the end of Section~\ref{S:allanomalies} we briefly mentioned the transgression technique and its relation to the anomaly polynomial and the Chern-Simons form. Historically, transgression was useful in completing the classification of anomalies in general dimension as well as for understanding their connection to topological invariants (see e.g.~\cite{Becchi:1975nq,Stora:1983ct,Zumino:1983rz}). In this Appendix we will rederive the transgression formulae of the first and second kind as well as relate them to the functionals $\VP$ and $\WCS$ studied in this work.

We begin by consider a flow of connections $\{ \fA(\tau),\fGamma(\tau)\}$ which vary with a real flow parameter $\tau$. The associated field strengths
\beq
\fF(\tau) = d\fA(\tau) + \fA(\tau)\wedge \fA(\tau)\,, \qquad \fR\ab{\mu}{\nu}(\tau) = d\fGamma\ab{\mu}{\nu}(\tau) + \fGamma\ab{\mu}{\rho}(\tau)\wedge \fGamma\ab{\rho}{\nu}(\tau)\,,
\eeq
are closed under the covariant derivative $D(\tau)$ constructed from these connections,
\beq
D(\tau)\fF(\tau) = 0\,, \qquad D(\tau)\fR\ab{\mu}{\nu}(\tau) = 0\,.
\eeq
The anomaly polynomial $\fP(\tau) = \fP(\fF(\tau),\fR(\tau))$ evaluated for these connections then satisfies
\beq
\label{E:DHalltau}
D(\tau) \left( \frac{\partial \fP(\tau)}{\partial \fF(\tau)}\right) = 0\,, \qquad D(\tau)\left(\frac{\partial \fP(\tau)}{\partial \fR\ab{\mu}{\nu}(\tau)} \right)= 0\,,
\eeq
on account of the fact that $\fP(\tau)$ is a polynomial of the field strengths $\fF(\tau)$ and $\fR\ab{\mu}{\nu}(\tau)$. We also have
\begin{align}
\begin{split}
\label{E:dFdtau}
	\partial_{\tau} \fF(\tau) & = d\partial_{\tau} \fA(\tau) + \fA(\tau) \wedge \partial_{\tau} \fA(\tau) + \partial_{\tau} \fA(\tau) \wedge \fA(\tau) = D(\tau) \partial_{\tau} \fA(\tau)\,,
	\\
	\partial_{\tau} \fR\ab{\mu}{\nu}(\tau) & = d\partial_{\tau} \fGamma\ab{\mu}{\nu}(\tau) + \fGamma\ab{\mu}{\rho}(\tau)\wedge \partial_{\tau} \fGamma\ab{\rho}{\nu}(\tau) + \partial_{\tau}\fGamma\ab{\mu}{\rho}(\tau)\wedge \fGamma\ab{\rho}{\nu}(\tau) = D(\tau)\partial_{\tau}\fGamma\ab{\mu}{\nu}(\tau)\,.
\end{split}
\end{align}
Note that $\partial_{\tau}\fA(\tau)$ and $\partial_{\tau}\fGamma\ab{\mu}{\nu}(\tau)$ are given by differences of connections. As a result these derivatives are gauge and diffeomorphism-covariant and so it makes sense to define the action of $D(\tau)$ on them. Collecting these results, by the chain rule, $\tau$-derivatives of the anomaly polynomial $\fP(\tau)$ are given by
\begin{align}
\begin{split}
\label{E:dPdtau}
\partial_{\tau} \fP(\tau) &=D(\tau) \partial_{\tau} \fA \wedge \cdot \frac{\partial\fP(\tau)}{\partial \fF(\tau)} + D(\tau) \partial_{\tau}\fGamma\ab{\mu}{\nu}(\tau)\wedge  \frac{\partial\fP(\tau)}{\partial\fR\ab{\mu}{\nu}(\tau)}
\\
& = d\left[ \partial_{\tau} \fA(\tau)\wedge \cdot  \frac{\partial\fP(\tau)}{\partial \fF}+\partial_{\tau}\fGamma\ab{\mu}{\nu}(\tau) \wedge \frac{\partial \fP(\tau)}{\partial \fR\ab{\mu}{\nu}(\tau)}\right]\,,
\end{split}
\end{align}
where in going from the first line to the second we have used~\eqref{E:DHalltau}. This immediately gives
\beq
\label{E:deltaPtau}
\fP(\tau_1) - \fP(\tau_2) = d \left[ \int_{\tau_2}^{\tau_1} d\tau \left(\partial_{\tau} \fA(\tau) \wedge \cdot \frac{\partial \fP(\tau)}{\partial \fF(\tau)} +\partial_{\tau}\fGamma\ab{\mu}{\nu}(\tau) \wedge  \frac{\partial \fP(\tau)}{\partial \fR\ab{\mu}{\nu}(\tau)} \right)\right]\,.
\eeq
This provides an integral expression for the difference of the anomaly polynomial evaluated for the connections at the endpoints of integration $\tau_2$ and $\tau_1$. Without loss of generality, we may consider a flow along the interval $\tau\in [0,1]$ which interpolates between the connections $\{\fA_2,\fGamma_2\}$ and $\{\fA_1,\fGamma_1\}$ as
\beq
\label{E:connectionsFlow}
\fA(\tau) = \fA_2 + \tau (\fA_1 - \fA_2) = \fA_2 + \tau \Delta \fA\,, \qquad \fGamma(\tau) = \fGamma_2 + \tau(\fGamma_1-\fGamma_2) = \fGamma_2 + \tau \Delta \fGamma\,.
\eeq
Defining the shorthand expressions
\begin{align}
\begin{split}
\label{E:Pshorthand}
\fP_{\fF}[\form{f}] &\equiv\form{f} \wedge \cdot \frac{\partial \fP(\tau)}{\partial \fF(\tau)} \,,
\\
\fP_{\fR}[\form{g}] & \equiv \form{g}\ab{\mu}{\nu}\wedge \frac{\partial \fP(\tau)}{\partial \fR\ab{\mu}{\nu}(\tau)} \,,
\end{split}
\end{align}
\eqref{E:deltaPtau} becomes
\begin{align}
\begin{split}
\label{E:transgressionFirst}
\fP_1 - \fP_2 &= d\left[ \int_0^1 d\tau \left( \fP_{\fF}[\Delta \fA] + \fP_{\fR}[\Delta \fGamma]\right)\right]
\\
& \equiv d\form{V}_{12}\,,
\end{split}
\end{align}
where we have defined $\fP_i$ to be the anomaly polynomial evaluated on the ``$i$'' connections. Equation~\eqref{E:transgressionFirst} is the transgression formula of the first kind, giving the difference of $\fP_i$'s in terms of an integral $\form{V}_{12}$ as we claimed in our discussion at the end of Section~\ref{S:allanomalies} near Equation~\eqref{E:V12}.

We can use this technology to construct a transgression formula for the Chern-Simons form $\fP = d\ICS$. We simply take the ``$1$'' connections to be the ones of interest, $\fA_1 = \fA, \fGamma_1 = \fGamma$, and take the ``$2$'' connections to vanish. Then~\eqref{E:transgressionFirst} gives
\beq
\label{E:ICStransgression}
\fP = d\left[ \int_0^1 d\tau \left( \fP_{\fF}[\fA] + \fP_{\fR}[\fGamma]\right)\right] \equiv d\ICS\,,
\eeq
which defines a ``canonical'' Chern-Simons form. However, unlike the anomaly polynomial, the Chern-Simons form $\ICS$ is better understood as a representative of an equivalence class in which we identify $\ICS \sim \ICS + d\form{H}$. These total derivatives are sometimes important. For instance, when an anomaly is mixed between a gauge symmetry and a global symmetry as in the $AVV$ anomalies of the Standard Model, just such a term $d\form{H}$ may be used to define the Chern-Simons form (and so $W_{QFT}$) in a way that is invariant under the gauge symmetry, but anomalous under the global symmetry. In that context $\form{H}$ is known as a Bardeen counterterm~\cite{Bardeen:1984pm}. The net result is that while~\eqref{E:ICStransgression} provides a canonical expression for $\ICS$, we find it to be more useful to consider more general Chern-Simons forms $\ICS$.

Having dispensed with these comments about $\ICS$, we proceed to derive the transgression formula of the second kind. As above, consider a flow of connections $\{\fA(\tau),\fGamma(\tau)\}$ as a function of a real flow parameter $\tau$. Denoting the Chern-Simons form evaluated for these connections as $\form{I}(\tau) = \ICS(\fA(\tau),\fGamma(\tau))$, we have by the chain rule and~\eqref{E:dFdtau}
\begin{align}
\begin{split}
\label{E:dIdtau}
\partial_{\tau} \form{I}(\tau) = 
& \partial_{\tau} \fA(\tau) \wedge \cdot \frac{\partial \form{I}(\tau)}{\partial\fA(\tau)} + D(\tau)\partial_{\tau}\fA(\tau)
 \wedge \cdot \frac{\partial \form{I}(\tau)}{\partial \fF(\tau)}
\\
& \qquad + \partial_{\tau}\fGamma\ab{\mu}{\nu}(\tau) \wedge \frac{\partial \form{I}(\tau)}{\partial\fGamma\ab{\mu}{\nu}(\tau)}  +D(\tau)\partial_{\tau}\fGamma\ab{\mu}{\nu}(\tau) \wedge \frac{\partial\form{I}(\tau)}{\partial\fR\ab{\mu}{\nu}(\tau)} \,.
\end{split}
\end{align}
To simplify these expressions we exploit the identities~\eqref{E:PandICS}, which in the present context are
\begin{align}
\begin{split}
\frac{\partial \form{I}(\tau)}{\partial\fA(\tau)}+D(\tau)\left( \frac{\partial\form{I}(\tau)}{\partial\fF(\tau)}\right) &= \frac{\partial \fP(\tau)}{\partial\fF(\tau)}\,, 
\\
\frac{\partial\form{I}(\tau)}{\partial\fGamma\ab{\mu}{\nu}(\tau)} + D(\tau)\left( \frac{\partial\form{I}(\tau)}{\partial\fR\ab{\mu}{\nu}(\tau)}\right) & = \frac{\partial\fP(\tau)}{\partial\fR\ab{\mu}{\nu}(\tau)}\,.
\end{split}
\end{align}
Then~\eqref{E:dIdtau} becomes
\beq
\label{E:dIdtau2}
\partial_{\tau}\form{I}(\tau) = \fP_{\fF}[\partial_{\tau}\fA(\tau)] + \fP_{\fR}[\partial_{\tau}\fGamma(\tau)] + d \left[ \partial_{\tau}\fA(\tau) \wedge \cdot \frac{\partial \form{I}(\tau)}{\partial \fF(\tau)} + \partial_{\tau}\fGamma\ab{\mu}{\nu}(\tau)\wedge \frac{\partial\form{I}(\tau)}{\partial\fR\ab{\mu}{\nu}(\tau)}\right]\,.
\eeq
Inspired by~\eqref{E:Pshorthand} we define
\begin{align}
\begin{split}
\form{I}_{\fF}[\form{f}] & \equiv \form{f}\wedge \cdot \frac{\partial \form{I}(\tau)}{\partial\fF(\tau)}\,,
\\
\form{I}_{\fR}[\form{g}] & \equiv \form{g}\ab{\mu}{\nu}\wedge \frac{\partial \form{I}(\tau)}{\partial \fR\ab{\mu}{\nu}(\tau)}\,.
\end{split}
\end{align}
We then integrate~\eqref{E:dIdtau} with respect to $\tau$ to give
\beq
\label{E:deltaI}
\form{I}(\tau_1)-\form{I}(\tau_2) = \int_{\tau_2}^{\tau_1} d\tau \left( \fP_{\fF}[\partial_{\tau}\fA(\tau)] + \fP_{\fR}[\partial_{\tau}\fGamma(\tau)]\right) + d\left[ \int_{\tau_2}^{\tau_1} d\tau \left( \form{I}_{\fF}[\partial_{\tau}\fA(\tau)] + \form{I}_{\fR}[\partial_{\tau}\fGamma(\tau)]\right)\right]
\eeq
Specializing to the flow~\eqref{E:connectionsFlow} along $\tau\in [0,1]$ between the ``$2$'' and ``$1$'' connections, the above difference of Chern-Simons forms becomes
\begin{align}
\begin{split}
\label{E:transgressionSecond}
\form{I}_1 - \form{I}_2 & = \int_0^1 d\tau \left( \fP_{\fF}[\Delta \fA] + \fP_{\fR}[\Delta \fGamma]\right) + d\left[ \int_0^1 d\tau \left( \form{I}_{\fF}[\Delta \fA] + \form{I}_{\fR}[\Delta \fGamma]\right)\right]
\\
& \equiv \form{V}_{12} + d\form{W}_{12}\,,
\end{split}
\end{align}
where we recognize the bulk term to be $\form{V}_{12}$ as earlier, and we define $\form{W}_{12}$ in the obvious way. This is the transgression formula of the second kind. In writing~\eqref{E:transgressionSecond}, we have explicitly demonstrated our claim at the end of Section~\ref{S:allanomalies} that the difference of Chern-Simons forms $\form{I}_1-\form{I}_2$ may be decomposed as in~\eqref{E:VandW12}. Incidentally, if we used the canonical Chern-Simons form~\eqref{E:ICStransgression}, then the integral for $\form{W}_{12}$ may be represented as a double integral over flows of connections.

We conclude this Appendix by relating the transgression formulae above to our $\VP$ and $\WCS$. To do so we relate our hatted and unhatted connections to those above by assigning
\begin{subequations}
\label{E:hatUnhatToTrans}
\beq
\{\fA_1,\fGamma_1\}= \{\fA,\fGamma\}\,, \qquad  \{\fA_2,\fGamma_2\}= \{ \hat{\fA},\hat{\fGamma}\}\,,
\eeq
so that the connections along the flow are given by~\eqref{E:connectionsFlow} to be
\beq
\fA(\tau) = \fA + (1-\tau)\mu \fu\,, \qquad \fGamma\ab{\mu}{\nu}(\tau) = \fGamma\ab{\mu}{\nu} + (1-\tau)(\muR)\ab{\mu}{\nu}\fu\,,
\eeq
and
\beq
\Delta \fA = - \mu \fu\,, \qquad \Delta \fGamma\ab{\mu}{\nu} = - (\muR)\ab{\mu}{\nu}\fu\,.
\eeq
\end{subequations}
The corresponding field strengths may be decomposed into electric and magnetic parts as in Subsection~\ref{S:EMdecomp} to give
\begin{align}
\begin{split}
\fF(\tau) &= \fu \wedge \fE(\tau) + \fB(\tau)
\\
& = \fu \wedge \left( \fE + (\tau-1)(D+\fa)\mu\right) + \left( \fB + (1-\tau)2\fomega\mu\right)\,,
\\
\fR\ab{\mu}{\nu}(\tau) & = \fu \wedge (\fER)\ab{\mu}{\nu}(\tau) + (\fBR)\ab{\mu}{\nu}(\tau)
\\
& =  \fu \wedge \left( (\fER)\ab{\mu}{\nu} + (\tau-1)(D+\fa)(\muR)\ab{\mu}{\nu}\right) + \left( (\fBR)\ab{\mu}{\nu} + (1-\tau)2 \fomega (\muR)\ab{\mu}{\nu}\right)\,,
\end{split}
\end{align}
where $D$ is the usual covariant derivative defined using the connections $\{\fA,\fGamma\}$. Since both $\Delta \fA$ and $\Delta \fGamma$ are longitudinal, the integral for $\form{V}_{12}$ in~\eqref{E:transgressionFirst} becomes
\begin{align}
\begin{split}
\label{E:V12simpler}
\form{V}_{12} = & - \fu \wedge \int_0^1 d\tau \left[ \mu\cdot \left.\frac{\partial\fP}{\partial \fF}\right|_{\fF = \fB(\tau), \fR = \fBR(\tau)} + (\muR)\ab{\mu}{\nu}\left.\frac{\partial\fP}{\partial\fR\ab{\mu}{\nu}}\right|_{\fF = \fB(\tau),\fR = \fBR(\tau)}\right]\,.
\end{split}
\end{align}
This expression may be simplified even further by viewing $\fP(\tau)$ as a functional of $\fB(\tau)$ and $\fBR(\tau)$ alone, i.e. $\fP(\tau) = \fP(\fF = \fB(\tau),\fR = \fBR(\tau))$. With this identification,~\eqref{E:V12simpler} then becomes
\beq
\form{V}_{12} = - \fu \wedge \int_0^1 d\tau\left[ \mu\cdot \frac{\partial }{\partial\fB(\tau)} + (\muR)\ab{\mu}{\nu}\frac{\partial}{\partial(\fBR)\ab{\mu}{\nu}(\tau)}\right]\fP(\fB(\tau),\fBR(\tau))\,.
\eeq
Now note that, when acting on functionals of $\fB(\tau)$ and $\fBR(\tau)$, the chain rule gives us
\beq
\frac{\partial}{\partial \tau}  = - (2\fomega)\wedge \left( \mu \cdot \frac{\partial}{\partial \fB(\tau)} + (\muR)\ab{\mu}{\nu}\frac{\partial}{\partial (\fBR)\ab{\mu}{\nu}(\tau)}\right)\,.
\eeq
This allows us to simplify~\eqref{E:V12simpler} enormously to become
\beq
\form{V}_{12} = \frac{\fu}{2\fomega}\wedge \int_0^1 d\tau \frac{\partial \fP(\tau)}{\partial \tau} = \frac{\fu}{2\fomega}\wedge \left( \fP-\hat{\fP}\right)=\VP\,,
\eeq
where we have used that $\fu \wedge \fP(\fB,\fBR)  = \fu \wedge \fP(\fF,\fR)$ and similarly for the hatted connections. Of course this is the expression for $\VP$ we quoted in the Introduction in~\eqref{E:VPIntro}.

We may simplify $\form{W}_{12}$ similarly. Under the identification~\eqref{E:hatUnhatToTrans}, the integral for it in~\eqref{E:transgressionSecond} becomes
\begin{align}
\begin{split}
\label{E:W12simpler}
\form{W}_{12} = &- \fu \wedge \int_0^1d\tau\left[ \mu\cdot \left.\frac{\partial\form{I}}{\partial\fF}\right|_{\fA = \fA(\tau),\fF = \fB(\tau),\fGamma = \fGamma(\tau),\fR = \fBR(\tau)}\right.
\\
& \qquad \qquad \qquad \left.+ (\muR)\ab{\mu}{\nu}\left.\frac{\partial\form{I}}{\partial\fR\ab{\mu}{\nu}}\right|_{\fA = \fA(\tau),\fF = \fB(\tau),\fGamma = \fGamma(\tau),\fR = \fBR(\tau)}\right]\,,
\end{split}
\end{align}
which may be simplified by viewing $\form{I}(\tau)$ as a functional of the connections $\{\fA(\tau),\fGamma(\tau)\}$ as well as $\fB(\tau)$ and $\fBR(\tau)$, i.e. $\form{I}(\tau) = \form{I}(\fA = \fA(\tau),\fF=\fB(\tau),\fGamma=\fGamma(\tau),\fR=\fBR(\tau))$. With this identitifation~\eqref{E:W12simpler} becomes
\beq
\form{W}_{12} = -\fu \wedge \int_0^1d\tau\left[ \mu\cdot \frac{\partial}{\partial\fB(\tau)} + (\muR)\ab{\mu}{\nu}\frac{\partial}{\partial (\fBR)\ab{\mu}{\nu}(\tau)}\right]\form{I}(\tau)\,.
\eeq
Now, when acting on functionals of the connections $\{\fA(\tau),\fGamma(\tau)\}$ and the magnetic field strengths $\fB(\tau)$ and $\fBR(\tau)$, the chain rule and~\eqref{E:hatUnhatToTrans} gives
\begin{align}
\begin{split}
\frac{\partial}{\partial\tau} =& - (2\fomega)\wedge \left( \mu \cdot \frac{\partial}{\partial \fB(\tau)} + (\muR)\ab{\mu}{\nu}\frac{\partial}{\partial (\fBR)\ab{\mu}{\nu}(\tau)}\right) 
\\
& \qquad \qquad - \fu \wedge \left( \mu \cdot \frac{\partial}{\partial \fA(\tau)} + (\muR)\ab{\mu}{\nu}\frac{\partial}{\partial\fGamma\ab{\mu}{\nu}(\tau)}\right)\,.
\end{split}
\end{align}
This allows us to enormously simplify the expression~\eqref{E:W12simpler} for $\form{W}_{12}$ as
\beq
\form{W}_{12} = \frac{\fu}{2\fomega}\wedge \int_0^1d\tau \frac{\partial\form{I}(\tau)}{\partial \tau} = \frac{\fu}{2\fomega}\wedge \left( \form{I}_{CS}-\hat{\form{I}}_{CS}\right)\,,
\eeq
where we have used that the terms involving derivatives with respect to $\fA(\tau)$ and $\fGamma(\tau)$ are proportional to $\fu\wedge\fu=0$ and that $\fu\wedge \form{I}(\tau_1) = \fu \wedge \form{I}_{CS}$, $\fu\wedge \form{I}(\tau_2) = \fu \wedge \hat{\form{I}}_{CS}$. This final expression is the one we quoted in Section~\ref{S:allanomalies} for $\WCS$ in~\eqref{E:theDecomposition}.

\section{Computing the variation of $V_{\mathcal{P}}$ and $W_{CS}$}
\label{A:deltaVPWCS}

In the main text we have argued that the covariant currents may be obtained by varying the $2n+1$-form $\VP$ and the consistent currents by varying the $2n$-form $\form{W}_{CS}$,
\begin{equation}
\label{E:reminderVPWCS}
	\VP = \frac{\fu}{2\fomega}\wedge \left( \fP-\hat{\fP}\right)\,,
	\qquad
	\WCS = \frac{\fu}{2\fomega}\wedge \left(\form{I}_{CS}-\hat{\form{I}}_{CS}\right)\,.
\end{equation} 
In this Appendix we give a detailed account of the variational procedure preserving many details which have been omitted from the main text. While straightforward, the computation is somewhat tedious. To assist the reader we begin by stating our final result upfront: we will show that under a general variation $\delta$, the variations of $\VP$ and $\WCS$ are given by 
\begin{align}
\begin{split}
\label{E:variationsmain}
\delta \VP &= d\brk{\delta \fu\wedge  \hodge \form{q_{\mathcal{P}}} +\delta \fA \wedge \cdot \hodge\form{J_{\mathcal{P}}}
 + \delta \fGamma^\alpha{}_\beta\wedge (\hodge \form{\spp_{\mathcal{P}}})^\beta{}_\alpha }  \\ 
&\qquad + \delta \fA\wedge  \cdot  \hodge \form{J}_H- \delta \hat{\fA}\wedge  \cdot \hodge  \hat{\form{J}}_H 
+ \delta \fGamma^\alpha{}_\beta\wedge  (\hodge \hat{\form{\spp}}_H)^\beta{}_\alpha - \delta \hat{\fGamma}^\alpha{}_\beta\wedge    (\hodge \hat{\form{\spp}}_H)^\beta{}_\alpha \,,  \\
-\delta \WCS 
&= -d\brk{\delta \fu \wedge \frac{\partial \WCS}{\partial (2\fomega)} +\delta \fA\wedge \cdot  \frac{\partial \WCS}{\partial \fB}
 + \delta \fGamma^\alpha{}_\beta\wedge  \frac{\partial \WCS}{\partial (\fBR)^\alpha{}_\beta} }  \\ 
&\qquad +  \delta \fu \wedge  \hodge \form{q_{\mathcal{P}}} + \delta \fA\wedge \cdot \brk{\hodge \form{J_{\mathcal{P}}}-\hodge \form{J}_{BZ}  } + \delta \hat{\fA}\wedge \cdot \hodge \hat{\fJ}_{BZ} \\
&\qquad + \delta \fGamma^\alpha{}_\beta \wedge  \brk{(\hodge \form{\spp_{\fP}})^\beta{}_\alpha-  (\hodge \form{\spp}_{BZ})^\beta{}_\alpha   }
+ \delta \hat{\fGamma}^\alpha{}_\beta\wedge    (\hodge \hat{\form{\spp}}_{BZ})^\beta{}_\alpha\,.  \\
\end{split}
\end{align}
In writing~\eqref{E:variationsmain} we have defined the $2n-1$-forms as in~\eqref{E:1ptFns}
\begin{equation}
\begin{split}
\label{E:anomcurrents}
\hodge \form{J_{\mathcal{P}}} &\equiv \frac{\partial\VP}{\partial \fB}\,, \qquad
\hodge \form{q_{\mathcal{P}}} \equiv \frac{\partial\VP}{\partial(2 \fomega)}\,, \qquad
\hodge \form{\spp_{\mathcal{P}}} \equiv \frac{\partial\VP}{\partial \fBR}\,,
 \end{split}
\end{equation}
which give the covariant anomaly-induced flavor, heat, and spin currents respectively. The currents with `$H$' and `$BZ$' subscripts denote the Hall and Bardeen-Zumino currents respectively, which we derived in~\eqref{E:HallBZ}. The Hall currents live in the $2n+1$-dimensional bulk $\mathcal{M}$, hence their bulk Hodge-duals are $2n$-forms; the boundary Bardeen-Zumino currents are valued on the boundary $\partial\mathcal{M}$ and so their boundary Hodge-duals are $2n-1$-forms. The hat denotes that the object is evaluated for the hatted connections~\eqref{E:hatConnection}.

To start, consider  the variation of a general $p$-form $\form{W}$ which has a leg along the velocity field $\fu$. If it may be written as $\form{W}=\fu\wedge \Sigma(2\fomega; \mu, \fA, \fB; \muR, \fGamma, \fBR)$ (and not in terms of say the derivatives of those variables), then by the chain rule its variation is
\begin{subequations}
\label{E:GeneralVariation}
\begin{equation}
\begin{split}
\delta \form{W} &= 
	\delta \fu  \wedge \frac{\partial \form{W}}{\partial \fu}  + \delta (2\fomega)  \wedge\frac{\partial \form{W}}{\partial (2\fomega)} 
	+ \delta \mu\wedge \cdot \frac{\partial \form{W}}{\partial \mu}   + \delta \fA\wedge \cdot \frac{\partial \form{W}}{\partial \fA}  
	+ \delta \fB\wedge \cdot \frac{\partial \form{W}}{\partial \fB}\,,  
	\\
	&\qquad + (\delta \muR)^\alpha{}_\beta \wedge  \frac{\partial \form{W}}{\partial \muR^\alpha{}_\beta}   
	+ \delta \fGamma^\alpha{}_\beta\wedge   \frac{\partial \form{W}}{\partial \fGamma^\alpha{}_\beta } 
	+  (\delta \fBR)^\alpha{}_\beta \wedge  \frac{\partial \form{W}}{\partial (\fBR)^\alpha{}_\beta} \,.
\end{split}
\end{equation}
Using~\eqref{E:deltawedgeu} we exchange variations of $\{\fomega,\fB,\fBR\}$ for variations of $\{\fu,\fA,\fGamma;\fa,\fE,\fER\}$ giving
\begin{align}
	\delta \form{W}  =& 
	d\brk{\delta \fu \wedge  \frac{\partial \form{W}}{\partial (2\fomega)} +\delta \fA\wedge \cdot  \frac{\partial \form{W}}{\partial \fB}
	 + \delta \fGamma^\alpha{}_\beta\wedge  \frac{\partial \form{W}}{\partial (\fBR)^\alpha{}_\beta} } 
	 \\
 	\nonumber
	&\,\,\, +  \delta \fu  \wedge\brk{ \frac{\partial \form{W}}{\partial \fu}  + D\prn{  \frac{\partial \form{W}}{\partial (2\fomega)} } + \fa \wedge \frac{\partial \form{W}}{\partial (2\fomega)}
-\fE\wedge \cdot \frac{\partial \form{W}}{\partial \fB} - (\fER)^\alpha{}_\beta \wedge  \frac{\partial \form{W}}{\partial (\fBR)^\alpha{}_\beta} } 
	\\
	\nonumber 
	&\,\,\, + \delta\mu\wedge \cdot \frac{\partial \form{W}}{\partial \mu}   + \delta \fA\wedge\cdot \brk{ \frac{\partial \form{W}}{\partial \fA}  + D\prn{\frac{\partial \form{W}}{\partial \fB}} } 
	\\
	\nonumber
	&+ \delta \muR^\alpha{}_\beta   \wedge\frac{\partial \form{W}}{\partial \muR^\alpha{}_\beta}   + \delta \fGamma^\alpha{}_\beta\wedge   \brk{ \frac{\partial \form{W}}{\partial \fGamma^\alpha{}_\beta } 
	+  D\prn{  \frac{\partial \form{W}}{\partial (\fBR)^\alpha{}_\beta} } } \,.
\end{align}
\end{subequations}
We can now use \eqref{E:GeneralVariation} to obtain an explicit expression for the variations of $\WCS$ and $\VP$ in terms of derivatives thereof. Before doing so, it is useful to note the identities:
\begin{align}\label{eq:muDeriv}
\frac{\partial \VP}{\partial \mu} &= -\fu\wedge \hodge  \hat{\form{J}}_H\,, &  \frac{\partial \VP}{\partial (\muR)^\alpha{}_\beta} &=  -\fu\wedge (\hodge \hat{\form{\spp}}_H)^\beta{}_\alpha\,,
\\
\nonumber
\frac{\partial \WCS}{\partial \mu} &= -\fu\wedge \hodge  \hat{\fJ}_{BZ}\,, & \frac{\partial \WCS}{\partial (\muR)^\alpha{}_\beta} &=  -\fu\wedge (\hodge \hat{\form{\spp}}_{BZ})^\beta{}_\alpha \,,
\end{align}
which can be easily proven using the definition of $\VP$ and $\WCS$ in~\eqref{E:reminderVPWCS},  along with the expressions~\eqref{E:HallBZ} for the Hall and BZ currents. For instance, the first identity follows from
\beq
\frac{\partial \VP}{\partial \mu} =-\frac{\fu}{2\fomega}\wedge \frac{\partial\hat{\fP}}{\partial\mu} =-\frac{\fu}{2\fomega}\wedge  \frac{\partial \hat{\fP}}{\partial\hat{\fF}}\wedge (2\fomega) = - \fu \wedge \frac{\partial \hat{\fP}}{\partial\hat{\fF}} = - \fu \wedge \hodge \hat{\form{J}}_H\,,
\eeq
where in the last equality we have used~\eqref{E:HallBZ}. We also require the identities
\begin{subequations}
\label{eq:HallBZDecompose}
\begin{align}
	 \frac{\partial \WCS}{\partial \fA} + D\left(\frac{\partial \WCS}{\partial \fB} \right)
 	&=\hodge \fJ_{BZ} - \hodge \hat{\fJ}_{BZ}-\hodge \form{J}_{\fP}\,,   
 	\\
 	\frac{\partial \WCS}{\partial \fGamma^\alpha{}_\beta} +D\left(\frac{\partial \WCS}{\partial (\fBR)^\alpha{}_\beta} \right)
	&= (\hodge \form{\spp}_{BZ})^\beta{}_\alpha - (\hodge \hat{\form{\spp}}_{BZ})^\beta{}_\alpha-(\hodge \form{\spp}_{\fP})^\beta{}_\alpha\,,  
	\\
	D\left(\frac{\partial \VP}{\partial \fB}\right) 
	&= D\hodge \fJ_{\fP} = \hodge \form{J}_H -\hodge \hat{\form{J}}_H\,,  
	\\
	D\left(\frac{\partial \VP}{\partial (\fBR)^\alpha{}_\beta} \right)
	& =  D(\hodge \form{\spp}_{\fP})^\beta{}_\alpha = (\hodge \form{\spp}_H)^\beta{}_\alpha - (\hodge \hat{\form{\spp}}_H)^\beta{}_\alpha \,,
\end{align}
\end{subequations}
which we now turn to prove.

We begin with the first identity of~\eqref{eq:HallBZDecompose}. Using~\eqref{E:HallBZ}, the difference between the hatted and unhatted BZ currents is given by
\begin{align}
\begin{split}
\label{E:DeltaJBZ1}
	\hodge\fJ_{BZ} - \hodge \hat{\fJ}_{BZ} 
	&= \frac{\partial \ICS  }{ \partial \fF}-\frac{\partial \hat{\form{I}}_{CS}  }{ \partial \hat{\fF} } 
	  = d\brk{\frac{\fu}{2\fomega}} \wedge \prn{ \frac{\partial \ICS  }{ \partial \fF}-\frac{\partial \hat{\form{I}}_{CS}  }{ \partial \hat{\fF} } }
	  \\
	&= D\brk{\frac{\fu}{2\fomega} \wedge \prn{ \frac{\partial \ICS  }{ \partial \fF}-\frac{\partial \hat{\form{I}}_{CS}  }{ \partial \hat{\fF} } } }
	+ \frac{\fu}{2\fomega} \wedge \brk{ D\prn{ \frac{\partial \ICS  }{ \partial \fF} } - D\prn{ \frac{\partial \hat{\form{I}}_{CS}  }{ \partial \hat{\fF} } } } \,.
\end{split}
\end{align}
The first term in the second line of~\eqref{E:DeltaJBZ1} is just $D(\partial \WCS/\partial \fB)$, and using~\eqref{E:reminderVPWCS},~\eqref{E:PandICS} together with $\fu\wedge D(\hdots) = \fu \wedge \hat{D}(\hdots)$ we may rewrite the rest to give
\beq
\label{E:DeltaJBZ2}
	\hodge\fJ_{BZ} - \hodge \hat{\fJ}_{BZ} =D\left(\frac{\partial \WCS}{\partial \fB} \right) +  \frac{\partial \WCS}{\partial \fA} +   \frac{\partial \VP }{\partial \fB}\,,
\eeq
which upon using $\hodge \fJ_{\fP} = \partial\VP/\partial\fB$ proves the first equality of~\eqref{eq:HallBZDecompose} as desired. A similar manipulation with Christoffel connection gives the second identity via
\beq
\begin{split}
\label{E:DeltaLBZ}
(\hodge \form{\spp}_{BZ})^\beta{}_\alpha - (\hodge\hat{\form{\spp}}_{BZ})^\beta{}_\alpha &= 
D\left(\frac{\partial \WCS}{\partial (\fBR)^\alpha{}_\beta} \right) +  \frac{\partial \WCS}{\partial \fGamma^\alpha{}_\beta} 
+   \frac{\partial \VP }{\partial (\fBR)^\alpha{}_\beta}\,.
\end{split}
\eeq
The difference between Hall currents in the third identity can also be tied to partial derivatives of $\WCS$ and $\VP$,
\begin{equation}
\begin{split}
\label{E:DeltaJH}
\hodge \form{J}_H -\hodge \hat{\form{J}}_H &= \frac{\partial \fP  }{ \partial \fF}-\frac{\partial \hat{\fP}  }{ \partial \hat{\fF} } 
= d\brk{\frac{\fu}{2\fomega}} \wedge \prn{ \frac{\partial \fP  }{ \partial \fF}-\frac{\partial \hat{\fP}  }{ \partial \hat{\fF} } }\\
&= D\brk{\frac{\fu}{2\fomega} \wedge \prn{ \frac{\partial \fP  }{ \partial \fF}-\frac{\partial \hat{\fP}  }{ \partial \hat{\fF}} } }
+ \frac{\fu}{2\fomega} \wedge \brk{ D\prn{ \frac{\partial \fP  }{ \partial \fF} } - D\prn{ \frac{\partial \hat{\fP}  }{ \partial \hat{\fF} } } } \\
&=D\left(\frac{\partial \VP }{\partial \fB}  \right) \,,
\end{split}
\end{equation}
where in going from the second line to the third we have used~\eqref{E:reminderVPWCS},~\eqref{E:conservedHall}, as well as $\fu\wedge D(\hdots) = \fu \wedge \hat{D}(\hdots)$. In a similar manner, we compute the last identity of~\eqref{eq:HallBZDecompose}
\begin{equation}
\label{E:DeltaLH}
(\hodge \form{\spp}_H)^\beta{}_\alpha - (\hodge \hat{\form{\spp}}_H)^\beta{}_\alpha = D\left(\frac{\partial \VP }{\partial (\fBR)^\alpha{}_\beta}  \right)\,.
\end{equation}
Note that all these partial derivatives treat $\{\fu,2\fomega, \mu, \fA, \fB, \muR, \fGamma, \fBR\}$ as independent variables.

Using \eqref{E:GeneralVariation},~\eqref{eq:muDeriv}, and~\eqref{eq:HallBZDecompose}, along with
\beq
\nonumber
\delta \hat{\fA} = \delta\fA + \delta \fu \mu + \delta \mu \fu\,, \qquad \delta\hat{\fGamma}\ab{\alpha}{\beta} = \delta\fGamma\ab{\alpha}{\beta} +\delta \fu (\muR)\ab{\alpha}{\beta} +  \delta (\muR)\ab{\alpha}{\beta}\fu\,,
\eeq
we can write the variation of $\VP$ as 
\begin{align}
\nonumber
	\delta \VP = & d\left[ \delta\fu \wedge \hodge \form{q}_{\fP} + \delta \fA \wedge\cdot  \hodge \fJ_{\fP} + \delta\fGamma\ab{\alpha}{\beta}\wedge (\hodge \form{\spp}_{\fP})\ab{\beta}{\alpha}\right]
	\\
	\nonumber
	& \,\,\, + \delta \fu\wedge\left[ \frac{\partial\VP}{\partial\fu}+(D+\fa)\hodge\form{q}_{\fP} - \fE\wedge \cdot \hodge \fJ_{\fP} - (\fER)\ab{\alpha}{\beta}\wedge(\hodge \form{\spp}_{\fP})\ab{\beta}{\alpha} \right]
	\\
	\nonumber
	& \,\,\, +\delta \fu \wedge \left[  \mu \cdot \hodge \hat{\fJ}_H + (\muR)\ab{\alpha}{\beta}(\hodge \hat{\form{\spp}}_{H})\ab{\beta}{\alpha}\right]
	\\
	\label{E:deltaVP}
	& \,\,\, + \delta \fA \wedge \cdot \hodge \fJ_H -\delta\hat{\fA} \wedge \cdot \hat{\fJ}_H + \delta \fGamma\ab{\alpha}{\beta} \wedge( \hodge \form{\spp}_H)\ab{\beta}{\alpha} - \delta \hat{\fGamma}\ab{\alpha}{\beta}\wedge (\hodge \hat{\form{\spp}}_H)\ab{\beta}{\alpha}\,.
\end{align}
Similarly, the variation of $\WCS$ takes the form
\begin{align}
	\label{E:deltaWCS}
	\delta \WCS 
	=& d\brk{\delta \fu \wedge \frac{\partial \WCS}{\partial (2\fomega)} +\delta \fA\wedge \cdot  \frac{\partial \WCS}{\partial \fB}
	 + \delta \fGamma^\alpha{}_\beta\wedge  \frac{\partial \WCS}{\partial (\fBR)^\alpha{}_\beta} } 
	 \\ 
	 \nonumber
	& +  \delta \fu\wedge  \left[ \frac{\partial \WCS}{\partial \fu}  + (D+\fa) \frac{\partial \WCS}{\partial (2\fomega)} 	-\fE\wedge \cdot \frac{\partial \WCS}{\partial \fB} - (\fER)^\alpha{}_\beta \wedge  \frac{\partial \WCS}{\partial (\fBR)^\alpha{}_\beta}  \right]
	\\
	\nonumber
	&  +\delta\fu\wedge \left[ \mu \cdot \hodge  \hat{\fJ}_{BZ} +  (\muR)^\alpha{}_\beta   (\hodge \hat{\form{\spp}}_{BZ})^\beta{}_\alpha  \right] + \delta \fA\wedge \cdot \brk{\hodge \fJ_{BZ} -\hodge \fJ_{_\fP} } - \delta \hat{\fA} \wedge \cdot  \hodge  \hat{\fJ}_{BZ} 
	\\
	\nonumber
	& + \delta \fGamma^\alpha{}_\beta \wedge   \brk{  (\hodge \form{\spp}_{BZ})^\beta{}_\alpha -(\hodge \form{\spp}_{_\fP})^\beta{}_\alpha  }
	- \delta \hat{\fGamma}^\alpha{}_\beta\wedge    (\hodge \hat{\form{\spp}}_{BZ})^\beta{}_\alpha  \,.
\end{align}

The expressions~\eqref{E:deltaVP} and~\eqref{E:deltaWCS} match~\eqref{E:variationsmain} except for the terms proportional to $\delta\fu$. These terms may be shown to vanish in the following way. Consider the variation of $\VP + d\WCS = \ICS - \hat{\form{I}}_{CS}$, which using~\eqref{E:deltaICS2} and~\eqref{E:HallBZ} is given by
\begin{align}
	\nonumber
	\delta\ICS - \delta\hat{\form{I}}_{CS}= & d\left[ \delta \fA \wedge\cdot \hodge \fJ_{BZ} - \delta \hat{\fA} \wedge \cdot\hat{\fJ}_{BZ} + \delta \fGamma\ab{\alpha}{\beta} \wedge (\hodge \form{\spp}_{BZ})\ab{\beta}{\alpha} - \delta \hat{\fGamma}\ab{\alpha}{\beta}\wedge (\hodge \hat{\form{\spp}}_{BZ})\ab{\beta}{\alpha}\right]
	\\
	\label{E:lastIdentity}
	& + \delta \fA \wedge \cdot \hodge \fJ_H - \delta \hat{\fA} \wedge \cdot \hodge \hat{\fJ}_H + \delta \fGamma\ab{\alpha}{\beta} \wedge (\hodge \form{\spp}_H)\ab{\beta}{\alpha}-\delta\hat{\fGamma}\ab{\alpha}{\beta}\wedge(\hodge \hat{\form{\spp}}_H)\ab{\beta}{\alpha}\,.
\end{align}
However, this variation is also computed by~\eqref{E:deltaVP} and~\eqref{E:deltaWCS} to be
\begin{align}
	\nonumber
	\delta \VP + d\delta \WCS = & d\left[ \delta \fA \wedge\cdot \hodge \fJ_{BZ} - \delta \hat{\fA} \wedge \cdot\hat{\fJ}_{BZ} + \delta \fGamma\ab{\alpha}{\beta} \wedge (\hodge \form{\spp}_{BZ})\ab{\beta}{\alpha} - \delta \hat{\fGamma}\ab{\alpha}{\beta}\wedge (\hodge \hat{\form{\spp}}_{BZ})\ab{\beta}{\alpha}\right]
	\\
	\nonumber
	& + \delta \fA \wedge \cdot \hodge \fJ_H - \delta \hat{\fA} \wedge \cdot \hodge \hat{\fJ}_H + \delta \fGamma\ab{\alpha}{\beta} \wedge (\hodge \form{\spp}_H)\ab{\beta}{\alpha}-\delta\hat{\fGamma}\ab{\alpha}{\beta}\wedge(\hodge \hat{\form{\spp}}_H)\ab{\beta}{\alpha}
	\\
	\nonumber
	& + \delta \fu\wedge\left[ \frac{\partial\VP}{\partial\fu}+(D+\fa)\hodge\form{q}_{\fP} - \fE\wedge \cdot \hodge \fJ_{\fP} - (\fER)\ab{\alpha}{\beta}\wedge(\hodge \form{\spp}_{\fP})\ab{\beta}{\alpha} \right.
	\\
	\label{E:lastIdentity2}
	&\left.+ \mu \cdot \hodge \hat{\fJ}_H + (\muR)\ab{\alpha}{\beta}(\hodge \hat{\form{\spp}}_{H})\ab{\beta}{\alpha}\right]
	\\
	\nonumber
	& + d\left\{ \delta\fu\wedge \left[  \frac{\partial \WCS}{\partial \fu}  + (D+\fa) \frac{\partial \WCS}{\partial (2\fomega)} 	-\fE\wedge \cdot \frac{\partial \WCS}{\partial \fB} - (\fER)^\alpha{}_\beta \wedge  \frac{\partial \WCS}{\partial (\fBR)^\alpha{}_\beta} \right.\right.
	\\
	\nonumber
	& + \left.\left.  \mu \cdot \hodge  \hat{\fJ}_{BZ} +  (\muR)^\alpha{}_\beta   (\hodge \hat{\form{\spp}}_{BZ})^\beta{}_\alpha  +\hodge \form{q}_{\fP}\right]\right\}\,.
\end{align}
Comparing~\eqref{E:lastIdentity} against~\eqref{E:lastIdentity2} we conclude that the coefficients of $\delta\fu$ and $d\delta \fu$ vanish, giving
\begin{equation}\label{eq:OmegaderivWCS}
\begin{split}
\hodge \form{q}_{\fP} + \frac{\partial \WCS}{\partial \fu}  &+ (D+\fa) \frac{\partial \WCS}{\partial (2\fomega)} 
+\mu \cdot \hodge  \hat{\fJ}_{BZ} +  (\muR)^\alpha{}_\beta   (\hodge \hat{\form{\spp}}_{BZ})^\beta{}_\alpha\\
&=\fE\cdot \frac{\partial \WCS}{\partial \fB} + (\fER)^\alpha{}_\beta   \frac{\partial \WCS}{\partial (\fBR)^\alpha{}_\beta} \,,
\end{split}
\end{equation}
and
\beq
\label{eq:OmegaderivVP}
\frac{\partial \VP}{\partial \fu}  +( D+\fa)\hodge \form{q}_{\fP} 
+ \mu \cdot \hodge  \hat{\fJ}_H +  (\muR)^\alpha{}_\beta   (\hodge \hat{\form{\spp}}_H)^\beta{}_\alpha -\fE\cdot \hodge\fJ_{\fP} - (\fER)^\alpha{}_\beta\  (\hodge \form{\spp}_{\fP})^\beta{}_\alpha = 0\,.
\eeq
Using~\eqref{eq:OmegaderivWCS} and \eqref{eq:OmegaderivVP} we can eliminate the terms proportional to $\delta\fu$ in \eqref{E:deltaVP} and \eqref{E:deltaWCS}. This gives~\eqref{E:variationsmain} which is the main result of this section.  

\section{A consistency check involving the anomalous Ward identities in equilibrium}

The main goal of this paper was to obtain a representative for the anomalous contribution to the covariant current and stress tensor, $J_\mathcal{P}^{\mu}$ and $T_{\mathcal{P}}^{\mu\nu}$  which takes the form given by \eqref{E:VPIntro}, \eqref{E:1ptFns} and  \eqref{E:1ptTP}. The anomaly-induced part of the hydrostatic currents which we have computed in this paper must satisfy the covariant anomalous Ward identities derived in Appendix \ref{A:conCovAnom},
\begin{subequations}
\label{E:WardJpTp}
\begin{align}
\label{E:WardJp}
D_{\mu} J^{\mu}_{\mathcal{P}} &= J_H^{\perp}\,,
\\
\label{E:WardTp}
D_{\nu} T^{\mu\nu}_{\mathcal{P}} & = F\ab{\mu}{\nu}\cdot J_{\mathcal{P}}^{\nu} + D_{\nu}\spp_H^{\perp\mu\nu}\,,
\end{align}
\end{subequations}
where we emphasize that the equalities hold in equilibrium. In this section we will show explicitly that this is indeed the case serving as a consistency check of our computations.

We begin with the covariant current, \eqref{E:WardJp}. Using \eqref{eq:HallBZDecompose} we can write the divergence of the anomalous contribution to the covariant current as
\beq
\label{E:DJP}
D_{\mu}J_{\mathcal{P}}^{\mu} = J_H^{\perp} - \hat{J}_H^{\perp}.
\eeq
In equilibrium, the hatted Hall flavor current $\hodge \hat{\form{J}}_H = \partial\hat{\fP}/\partial\hat{\form{F}}$ is completely transverse to $\form{u}$ which implies
$ \hat{J}_H^{\perp} = 0$. The divergence~\eqref{E:DJP} then reduces to \eqref{E:WardJp}.

To show that the stress tensor satisfies \eqref{E:WardTp} requires some more work. We separate $T_{\mathcal{P}}^{\mu\nu}$ into heat and spin current parts as
\beq
T_{\mathcal{P}}^{\mu\nu} = T_{\mathcal{P},q}^{\mu\nu} + T_{\mathcal{P},\spp}^{\mu\nu}\,,
\eeq
with
\begin{equation}
	 T_{\mathcal{P},q}^{\mu\nu} =u^{\mu}q_{\mathcal{P}}^{\nu} + u^{\nu}q_{\mathcal{P}}^{\mu} \,,
	 \qquad
	   T_{\mathcal{P},\spp}^{\mu\nu} = D_{\rho}\left(\spp_{\mathcal{P}}^{\mu[\nu\rho]}+\spp_{\mathcal{P}}^{\nu[\mu\rho]} - \spp_{\mathcal{P}}^{\rho(\mu\nu)}\right)\,.
\eeq
The divergence of $T_{\mathcal{P},q}^{\mu\nu}$ is rather unilluminating when out of equilibrium, but in equilibrium it takes the simple form
\beq
\label{E:divTqP}
	D_{\nu}T_{\mathcal{P},q}^{\mu\nu} = u^{\mu}(D_{\nu}+a_{\nu})q_{\mathcal{P}}^{\nu}- 2 \omega\ab{\mu}{\nu}q_{\mathcal{P}}^{\nu}\,.
\eeq
The divergence of the spin part of the stress tensor can be written in the form
\beq
\label{E:divTLP}
D_{\nu}T_{\mathcal{P},\spp}^{\mu\nu}  = -D_{\rho}D_{\nu}\spp_{\mathcal{P}}^{\nu\rho\mu} - R\ab{\mu}{\nu\rho\sigma}\spp_{\mathcal{P}}^{\nu\sigma\rho}\,.
\eeq
Using \eqref{eq:HallBZDecompose} once again we rewrite the divergence of the spin current as
\beq
\label{E:divLP}
D_{\nu}\spp_{\mathcal{P}}^{\nu\rho\mu} = \spp_H^{\perp\rho\mu}-\hat{\spp}_H^{\perp\rho\mu}\,.
\eeq
In equilibrium the hatted Hall spin current $\hodge \hat{\form{\spp}}_H = \partial\hat{\fP}/\partial\hat{\form{R}}$ is completely transverse to $\form{u}$ so that $\hat{\spp}_H^{\perp\rho\mu}= 0$. Combining~\eqref{E:divTqP},~\eqref{E:divTLP}, and~\eqref{E:divLP}, we then have the equilibrium relation
\beq
D_{\nu}T^{\mu\nu}_{\mathcal{P}}  = u^{\mu}\left[ (D_{\nu}+a_{\nu})q_{\mathcal{P}}^{\nu} - (E_R)_{\rho\sigma\nu}\spp_{\mathcal{P}}^{\nu\sigma\rho}\right] - 2\omega\ab{\mu}{\nu}q_{\mathcal{P}}^{\nu} - (B_R)_{\rho\sigma\phantom{\mu}\nu}^{\phantom{\rho\sigma}\mu}\spp_{\mathcal{P}}^{\nu\sigma\rho} + D_{\nu}\spp_H^{\perp\mu\nu}\,,
\eeq
where we have decomposed the Riemann curvature into electric and magnetic parts
\beq
R\ab{\mu}{\nu\rho\sigma}=u_{\rho}(E_R)\ab{\mu}{\nu\sigma}-u_{\sigma}(E_R)\ab{\mu}{\nu\rho}+(B_R)\ab{\mu}{\nu\rho\sigma}\,,
\eeq
as described in Section \ref{S:hydrostatics}. If we decompose the the field strength in a similar manner, we find that in equilibrium
\begin{align}
\begin{split}
\label{E:DivTeq}
D_{\nu}T_{\mathcal{P}}^{\mu\nu} -F\ab{\mu}{\nu}\cdot J_{\mathcal{P}}^{\nu} - D_{\nu}\spp_H^{\perp\mu\nu} = \,&u^{\mu}\left[ (D_{\nu}+a_{\nu})q_{\mathcal{P}}^{\nu}-E_{\nu}\cdot J_{\mathcal{P}}^{\nu}-(E_R)_{\rho\sigma\nu}\spp_{\mathcal{P}}^{\nu\sigma\sigma}\right] 
\\
&- 2\omega\ab{\mu}{\nu}q_{\mathcal{P}}^{\nu} - B\ab{\mu}{\nu}\cdot J_{\mathcal{P}}^{\nu}-(B_R)_{\rho\sigma\phantom{\mu}\nu}^{\phantom{\rho\sigma}\mu}\spp_{\mathcal{P}}^{\nu\sigma\rho}\,.
\end{split}
\end{align}

To prove \eqref{E:WardTp} it remains to show that the longitudinal and transverse expressions on the right hand side of \eqref{E:DivTeq} vanish. Using \eqref{eq:OmegaderivVP} we find that 
\beq
\label{E:DivqEqCoord}
(D_{\mu}+a_{\mu})q_{\mathcal{P}}^{\mu} = E_{\mu}\cdot J_{\mathcal{P}}^{\mu} + (E_R)_{\rho\sigma\mu}\spp_{\mathcal{P}}^{\mu\sigma\rho}\,,
\eeq
where we have used that there are transverse volume forms in $2n$ dimensions. Thus, the longitudinal part on the left hand side of \eqref{E:DivTeq} indeed vanishes.

We proceed to study the transverse contribution to the right hand side of~\eqref{E:DivTeq}. Consider the interior product of $\VP$ with an arbitrary vector field $\xi^m$ transverse to $u_m$ which satisfies the property that $\xi^m$ is tangent to the boundary $\partial\mathcal{M}$. Since the interior product is a derivation we have
\begin{align}
\begin{split}
\iota_{\xi}\VP &=  \iota_{\xi}(2\form{\omega})\frac{\partial\VP}{\partial(2\form{\omega})}+ \iota_{\xi}\form{B} \cdot \frac{\partial\VP}{\partial \form{B}} + \iota_{\xi}(\form{B}_R)\ab{a}{b}\frac{\partial\VP}{\partial(\form{B}_R)\ab{a}{b}}\,,
\\
& = \iota_{\xi}(2\form{\omega})\hodge \form{q}_{\mathcal{P}} + \iota_{\xi} \form{B} \cdot \hodge \form{J}_{\mathcal{P}} + \iota_{\xi}(B_R)\ab{a}{b} \hodge \form{\spp}\ab{b}{a}\,.
\end{split}
\end{align}
We now recall that $\VP$ is a top-form in $2n+1$ dimensions and so it has a leg along the $\perp$ direction. Since $\xi^{\perp}$ vanishes on the boundary $\partial\mathcal{M}$, the pullback of $\iota_{\xi}\VP$ to $\partial\mathcal{M}$ vanishes,
\beq
\text{P}[\iota_{\xi}\VP] = 0\,.
\eeq
In coordinates this means that
\beq
\label{E:transverseWardAlmost}
\xi_{\mu}\left(2\omega\ab{\mu}{\nu}q_{\mathcal{P}}^{\nu} + B\ab{\mu}{\nu}\cdot J_{\mathcal{P}}^{\nu} + (B_R)_{\rho\sigma\phantom{\mu}\nu}^{\phantom{\rho\sigma}\mu} \spp_{\mathcal{P}}^{\nu\sigma\rho}\right) = 0\,.
\eeq
The bracketed expression is precisely the transverse part of~\eqref{E:DivTeq}. Now since the bracketed expression in \eqref{E:transverseWardAlmost} is transverse, we find that the bracketed expression and therefore the transverse part of~\eqref{E:DivTeq} vanishes as claimed.

\section{Spin current and torque: Mathisson-Papapetrou-Dixon equations}
\label{A:spin}

As we have seen, the conservation equations for flavor, energy-momentum and angular momentum get modified in a specific way for anomalous field theories. For example, a useful way of thinking about gravitational anomalies is to think of them as Lorentz anomalies leading to a non-conservation of angular momentum via anomalous torques. 

In order to see how this works in practice, it is useful to first understand the dynamics of angular momentum when the anomalies are absent. We will consider the case where the system is coupled to an external medium which applies an external torque on the QFT so that we can understand how torque appears in the conservation laws. Thus in this Appendix, we will remark on some basic results regarding the dynamics of spin and torque that will clarify the physical content of various equations that appear in the paper. 

Given a system with a conserved angular momentum, the split of that angular momentum into an orbital and a spin part is somewhat arbitrary. This is exactly analogous to the statement in electrodynamics of media that the division of a charge current into a  `free' charge current and a magnetization-induced `bound' charge current is arbitrary per se. In both these cases, however, a particular split might be more natural and convenient in a given physical situation. Our aim in the rest of the Appendix is to describe how such splits in angular momentum can be achieved and how one handles the ambiguity inherent to such a split.

The discussion in the rest of this appendix is straightforward if somewhat long. So, for the convenience of the reader, we will summarize the main conclusions: first of all, we will show that different definitions of spin dynamics prevalent in literature can be reconciled into a single set of conservation equations that can directly be derived from the generating functional. We refer to these equations as Mathisson-Papapetrou-Dixon equations. They naturally incorporate the ambiguity mentioned above. We will also show that $\spp^{\mu[\nu\lambda]}$ should be interpreted as \emph{half} the spin current - a corollary that follows is that $\spp_H^{\perp[\mu\nu]}$ is half the spin Hall current that flows into the boundary. Via anomaly inflow of angular momentum, $\spp_H^{\perp[\mu\nu]}$ thus gives the \emph{half}-torque due to the covariant gravitational anomaly.

To prove these statements, we begin by slightly modifying the procedure in the Appendix~\ref{A:conCovAnom}. There, we studied the variation of the path-integral with respect to  sources $\{A_\mu,g_{\mu\nu} \}$. Instead, it is often convenient to treat $\{A_\mu, g_{\mu\nu}, \Gamma^\mu{}_{\nu\lambda} \}$  as independent sources and write \eqref{eq:deltaWdef} in the form 
\beq
\label{eq:deltaWdeftSp}
\delta W  \equiv  \int  d^dx\sqrt{-g} \Bigl\{ \delta A_\mu \cdot J^\mu 
  +\half  \delta g_{\mu\nu}t^{\mu\nu} +\delta \Gamma^\mu{}_{\nu\lambda}  \spp^{\lambda\nu}{}_\mu \Bigr\}  + (\text{boundary terms})\,.
\eeq
with $t^{\mu\nu}=t^{\nu\mu}$. We will impose no further symmetry conditions on the other tensors. We will call $\spp^{\lambda\mu\nu}$ the \emph{canonical spin current} or just \emph{spin current} in short. More precisely, it is \emph{half} of what is usually termed the spin current -- where the additional factor of half is included for convenience. The reason for this terminology will become clear as we proceed.

While the division above is somewhat arbitrary, one can treat this as the penultimate step in computing $\delta W$, prior to the last integration by parts which brings $\delta W$ to the form in \eqref{eq:deltaWdef}. The energy-momentum tensor, obtained after integration by parts, can be written in the form
\beq
\label{eq:TfromSigma}
T^{\mu\nu} = t^{\mu\nu}- D_\lambda  \spp^{\lambda\nu\mu}  
 +  D_\lambda \prn{ \spp^{\mu[\nu\lambda]} +\spp^{\nu[\mu\lambda]} + \spp^{\lambda[\nu\mu]} }\,.
\eeq
We will call the contribution in brackets the \emph{spin energy momentum tensor}
\beq
\label{eq:TSpin}
T_{spin}^{\mu\nu}
\equiv   D_\lambda \prn{ \spp^{\mu[\nu\lambda]} +\spp^{\nu[\mu\lambda]} + \spp^{\lambda[\nu\mu]} }\,.
\eeq

Let us place our quantum field theory in an external medium which can apply an external force or torque. A convenient example is an `flavor' electromagnetic medium with a magnetization-dielectric polarization tensor $\Mag^{\mu\nu}$ . In practice, this means there exists an external sector of the quantum field theory such that
\begin{equation}\label{eq:deltaWext}
\begin{split}
\delta W_{ext}  &= \int  d^dx\sqrt{-g}\ \Bigl\{  \half  \Mag^{\mu\nu} \cdot \delta F_{\mu\nu} \Bigr\}
   + (\text{boundary terms})\,,
\end{split}
\end{equation}
with $\Mag^{\mu\nu} = -\Mag^{\nu\mu}$. We will call $\Mag^{\mu\nu}$ the \emph{magnetization} tensor of the medium. This implies that under an infinitesimal gauge and coordinate transformation $\chi=\{\xi^{\mu},\Lambda\}$, $W_{ext}$ varies as
\begin{align}
	\nonumber
 \diffF W_{ext}  &= \int  d^dx\sqrt{-g}\ \Bigl\{  \half  \Mag^{\mu\nu} \cdot \diffF F_{\mu\nu} \Bigr\}
   + (\text{boundary terms})
      \\
   \nonumber
   &= \int  d^dx\sqrt{-g}\ \Bigl\{  \xi^\alpha\  \half  \Mag^{\mu\nu} \cdot D_\alpha F_{\mu\nu} 
   +\half D_\alpha \xi^\beta \prn{ \Mag^{\alpha\nu} \cdot F_{\beta\nu} + \Mag^{\mu \alpha} \cdot F_{\mu \beta} } \Bigr\}
      \\
   \label{eq:mediumvar}
   &\qquad + \Bigl. (\Lambda+\xi^\alpha A_\alpha ) \cdot \half  [\Mag^{\mu\nu} , F_{\mu\nu} ]
    \Bigr\}
   + (\text{boundary terms})
   \\
   \nonumber
    &\equiv  \int  d^dx\sqrt{-g}\ \Bigl\{  \xi_\alpha f^\alpha_{ext} 
   +\half (D_\alpha \xi_\beta) \tau^{\alpha\beta}_{ext} +  (\Lambda+\xi^\alpha A_\alpha ) \cdot Q_{ext}
    \Bigr\}
   + (\text{boundary terms})\,,
\end{align}
where we have parametrized the contributions from the medium by an external force $f^\alpha_{ext}$, an external point torque $\tau^{\alpha\beta}_{ext}$ and an external charge injection rate $Q_{ext}$. The reason for these names would become clear shortly, once we derive the conservation equations. For a magnetized medium, we have  
\beq
\label{eq:DSeqn}
 f^\alpha_{ext}   = \half  \Mag_{\mu\nu} \cdot D^\alpha F^{\mu\nu}\,,
 \qquad 
 \tau^{\alpha\beta}_{ext} = 2\Mag^{\alpha\sigma} \cdot F^\beta{}_{\sigma} \,,
 \qquad 
  Q_{ext} = \half  [\Mag^{\mu\nu} , F_{\mu\nu} ]\,,
\eeq
where we recognize the Stern-Gerlach force $(\nabla B)\cdot \Mag$ and the familiar $\Mag\times B$ point torque in the antisymmetric part of $\tau^{\alpha\beta}_{ext}$  :
\begin{equation}\label{eq:DSeqn2}
\begin{split}
\tau^{[\alpha\beta]}_{ext} = \Mag^{\alpha\sigma} \cdot F^\beta{}_{\sigma} -\Mag^{\beta\sigma} \cdot F^{\alpha}{}_{\sigma} \,.
\end{split}
\end{equation}
For non-abelian flavor symmetries the magnestization injects flavor charge into the system at the rate given by 
$\half  [\Mag^{\mu\nu} , F_{\mu\nu} ]$.

We want to now rederive the conservation laws by demanding that the joint system $W+W_{ext}$ be diffeomorphism and flavor invariant. In what follows we will not specify a particular form for $W_{ext}$, but instead parameterize it through external forces, torques, and charge injection rates as in~\eqref{eq:mediumvar}. The conservation laws and Noether theorem for the field theory are derived as in Appendix~\ref{A:ward} via an integration by parts. In what follows we split the energy-momentum tensor into $t^{\mu\nu}$ and the spin current $\spp^{\mu\nu\rho}$ as in~\eqref{E:deltaWwithSpin}. Using the transformation of the sources under diffeomorphism and flavor transformations, we get
\begin{align}
\begin{split}
\diffF A_\mu \cdot J^\mu &  + \half \diffF g_{\mu\nu}  t^{\mu\nu}
+ \diffF \Gamma^\mu{}_{\nu\lambda}   \spp^{\lambda\nu}{}_\mu+\xi_\alpha f^\alpha_{ext} 
   +\half (D_\alpha \xi_\beta) \tau^{\alpha\beta}_{ext} +  (\Lambda+\xi^\alpha A_\alpha ) \cdot Q_{ext}\\
&= D_\mu N_{\chi,Canonical}^\mu -   (\Lambda+\xi^\alpha A_\alpha ) \cdot \prn{ D_{\mu} J^{\mu} - Q_{ext} } \\
 &\quad - \xi_{\mu}\Bigl\{ D_{\nu} T_{orbital}^{\mu\nu} - R^\mu{}_{\nu\beta\alpha}\spp^{\nu\alpha\beta}
 - F^\mu{}_\nu \cdot  J^\nu  - f^\mu_{ext} \Bigr\}\,,
\end{split}
\end{align}
where we have defined 
\begin{align}
\label{eq:NoetherXiMod}
\begin{split}
N_{\chi,Canonical}^\mu &\equiv  \prn{\Lambda +\xi^\alpha A_\alpha} \cdot J^\mu
+ \xi_\alpha T_{orbital}^{\alpha\mu}+ (D_\alpha \xi_\beta) \spp^{\mu\alpha \beta}\,,
   \\
T_{orbital}^{\mu\nu} &\equiv   t^{\mu\nu}
  -  \prn{ D_\lambda \spp^{\lambda\nu\mu} -\half \tau_{ext}^{\nu\mu} }\,,
\end{split}
\end{align}
Here, $N_{\chi,Canonical}^\mu$ is the \emph{canonical Noether current} and $T_{orbital}^{\mu\nu}$ is the \emph{orbital} energy-momentum tensor which enters into this Noether current. These differ from the Noether current and the energy momentum tensors we defined previously in~\eqref{eq:NoetherXi} and through the variation of $W_{QFT}$ with respect to the metric. Thus, $\diffF (W + W_{ext})$ evaluates to 
\begin{align}
\begin{split}
\diffF (W +W_{ext}) &= \int  d^dx\sqrt{-g}\ \Bigl\{ \diffF A_\mu \cdot J^\mu  
+ \half \diffF g_{\mu\nu}  t^{\mu\nu}
+  \diffF \Gamma^\mu{}_{\nu\lambda} \spp^{\lambda\nu}{}_\mu\Bigr\}
\\
&\qquad +\int  d^dx\sqrt{-g}\ \Bigl\{ \xi_\alpha f^\alpha_{ext} 
   +\half (D_\alpha \xi_\beta) \tau^{\alpha\beta}_{ext} +  (\Lambda+\xi^\alpha A_\alpha ) \cdot Q_{ext}\Bigr\}
   \\
&\qquad + (\text{boundary terms})
 \\
 &= -\int d^dx \sqrt{-g}\,   \prn{\Lambda +\xi^\alpha A_\alpha}  \cdot \Bigl\{ D_{\mu} J^{\mu}-Q_{ext}  \Bigr\} 
 \\ 
 &\quad  -\int d^dx \sqrt{-g}\, \xi_{\mu}\Bigl\{D_{\nu} T_{orbital}^{\mu\nu} -  R^\mu{}_{\nu\beta\alpha}\spp^{\nu\alpha\beta}
  - F^\mu{}_\nu \cdot  J^\nu  - f^\mu_{ext} \Bigr\}\\
 &\qquad +  (\text{boundary terms})\,,
\end{split}
\end{align}
which gives upon demanding $\diffF (W+W_{ext})=0$ 
\begin{align}\label{eq:MPD1}
\begin{split}
D_{\mu} J^{\mu}&= Q_{ext} \,,
\\
D_{\nu} T_{orbital}^{\mu\nu} &=   R^\mu{}_{\nu\beta\alpha}\spp^{\nu\alpha\beta}
 + F^\mu{}_\nu \cdot  J^\nu
  +  f^\mu_{ext}\,,
  \\
D_\lambda \spp^{\lambda[\mu\nu]} &=T_{orbital}^{[\mu\nu]} 
+ \half \tau_{ext}^{[\mu\nu]}\,,
\end{split}
\end{align}
where the third relation follows from  $t^{[\mu\nu]}=0$. The divergence of the canonical Noether current is 
\begin{align}
\label{eq:MPDNoether1}
\begin{split}
D_\mu N_{\chi,Canonical}^\mu &=  \diffF A_\mu \cdot J^\mu   + \half \diffF g_{\mu\nu} t^{\mu\nu}
+ \diffF \Gamma^\mu{}_{\nu\lambda} \spp^{\lambda\nu}{}_\mu
 \\
&\qquad +\xi_\alpha f^\alpha_{ext} 
   +\half (D_\alpha \xi_\beta) \tau^{\alpha\beta}_{ext} +  (\Lambda+\xi^\alpha A_\alpha ) \cdot Q_{ext}\,.
\end{split}
\end{align}

The equations \eqref{eq:MPD1} are the basic equations that we will need to describe the dynamics of spin currents. Henceforth, we will refer to the conservation equations in \eqref{eq:MPD1} as \emph{Mathisson-Papapetrou-Dixon equations}.\footnote{Mathisson~\cite{Mathisson:1937zz}, Papapetrou~\cite{Papapetrou:1951pa} and Dixon~\cite{Dixon:1970zza} arrived at the particle analogue of these equations while studying the motion of spinning particles in curved spacetime. The Mathisson-Papapetrou-Dixon equations with  $\{f^\mu_{ext} ,\tau^{[\mu\nu]}_{ext}\}$ as given in \eqref{eq:DSeqn} and  \eqref{eq:DSeqn2} are sometimes also called Dixon-Soriau equations.} We can show that the Mathisson-Papapetrou-Dixon equations are equivalent to the usual conservation equations for the standard energy-momentum tensor $T^{\mu\nu}$ in \eqref{E:stressIntoTandSpin}. A direct integration by parts gives
\begin{align}
\delta W  &\equiv  \int  d^dx\sqrt{-g} \Bigl\{ \delta A_\mu \cdot J^\mu 
  +\half \delta g_{\mu\nu} T^{\mu\nu} \Bigr\} + (\text{boundary terms})\,, 
  \\
  \nonumber
\diffF W_{ext}     &\equiv  \int  d^dx\sqrt{-g}\ \Bigl\{  \xi_\alpha \brk{f^\alpha_{ext} -\half D_\beta \tau^{\beta\alpha}_{ext} }
   +  (\Lambda+\xi^\alpha A_\alpha ) \cdot Q_{ext}
    \Bigr\}
   + (\text{boundary terms})\,,
\end{align}
where $T^{\mu\nu}$ is as in \eqref{E:stressIntoTandSpin}. Using~\eqref{E:deltaAg} and integrating the variations in $\delta_{\chi}W$ by parts, this in turn implies that $J^\mu$ and  $T^{\mu\nu}$ satisfy the Ward identities
\beq
\label{eq:MPD2}
D_{\mu} J^{\mu}= Q_{ext} \,,
\qquad
D_{\nu} T^{\mu\nu} =  F^\mu{}_\nu \cdot  J^\nu
  +  f^\mu_{ext} -\half D_\nu \tau_{ext}^{\nu\mu} \,,
\qquad 
T^{\mu\nu} -T^{\nu\mu} =0\,.
\eeq
Note that in this way of describing the dynamics, the point torques just appear as a part of external forces acting on the system.

An alternate way to derive the Mathisson-Papapetrou-Dixon equations is to think of the metric $g_{\mu\nu}$ and the connection $\Gamma^\mu{}_{\nu\lambda}$  as derived from frame fields and spin connection $\{E^{\bar{a}}{}_\mu,\mathring{\Gamma}^{\bar{a}}{}_{\bar{b} \mu}\}$ (in what follows barred English indices are tangent space indices, which are raised and lowered with the Minkowski metric $\eta_{\bar{a}\bar{b}}$ and its inverse $\eta^{\bar{a}\bar{b}}$). The metric is determined in terms of the frame fields
\beq
g_{\mu\nu} = \eta_{\bar{a}\bar{b}} E\ab{\bar{a}}{\mu}E\ab{\bar{b}}{\nu}\,,
\eeq
and the coframe fields $e\ab{\mu}{\bar{a}}$ are defined such that
\beq
E\ab{\bar{a}}{\mu}e\ab{\mu}{\bar{b}} = \delta\ab{\bar{a}}{\bar{b}}\,, \qquad e\ab{\mu}{\bar{a}}E\ab{\bar{a}}{\nu} = \delta\ab{\mu}{\nu}\,.
\eeq
Since we couple our field theory to the Christoffel connection, the spin connection $\mathring{\Gamma}\ab{\bar{a}}{\bar{b}\mu}$ is determined in terms of the metric through the frame fields, but here as before we find it useful to treat $\{E^{\bar{a}}{}_\mu,\mathring{\Gamma}^{\bar{a}}{}_{\bar{b} \mu}\}$ as independent sources.\footnote{To be precise, we take $\mathring{\Gamma}\ab{\bar{a}}{\bar{b}\mu}$ to be arbitrary so as it satisfies metric compatibility $D_{\mu}g_{\nu\rho}=0$. In terms of the frame fields and spin connection, this just means that we impose antisymmetry of $\mathring{\Gamma}^{\bar{a}}{}_{\bar{b} \mu}$ in its frame indices.} At the end of the day, we convert variations of the spin connection into variations of the frame fields.

One may verify that the holonomic (coordinate frame) connection $\Gamma\ab{\mu}{\nu\rho}$ is determined in terms of the spin connection as
\beq
\label{E:holoToSpin}
\Gamma\ab{\mu}{\nu\rho} = e\ab{\mu}{\bar{a}}\mathring{\Gamma}\ab{\bar{a}}{\bar{b}\rho} E\ab{\bar{b}}{\nu} + e\ab{\mu}{\bar{a}}\partial_{\rho}E\ab{\bar{a}}{\nu}\,.
\eeq
This is equivalent to the condition that the frame fields are covariantly constant under the spin covariant derivative $\mathring{D} = D + \mathring{\Gamma}$,
\begin{equation*}
\mathring{D}_{\mu} E\ab{\bar{a}}{\nu} = \partial_{\mu} E\ab{\bar{a}}{\nu} - \Gamma\ab{\rho}{\nu\mu}E\ab{\bar{a}}{\rho} + \mathring{\Gamma}\ab{\bar{a}}{\bar{b}\mu}E\ab{\bar{b}}{\nu} = 0\,.
\end{equation*}
Solving~\eqref{E:holoToSpin} for $\mathring{\Gamma}\ab{\bar{a}}{\bar{b}\mu}$ and using that $\Gamma\ab{\mu}{\nu\rho}$ is the Christoffel connection and so is determined in terms of the metric~\eqref{E:christoffel}, the spin connection is given by
\beq
\label{E:spinCon}
\mathring{\Gamma}\ab{\bar{a}}{\bar{b}\mu} = E\ab{\bar{a}}{\alpha} g^{\alpha\lambda} e\ab{\nu}{\bar{b}}e\ab{\sigma}{\bar{c}}\left(  g_{\sigma\mu} \partial_{[\nu}E\ab{\bar{c}}{\lambda]} + g_{\sigma\nu}\partial_{[\mu}E\ab{\bar{c}}{\lambda]} + g_{\sigma\lambda}\partial_{[\nu}E\ab{\bar{c}}{\mu]}\right)\,.
\eeq

Treating the variations of the frame fields and spin connection as independent for now, $W$ varies as
\begin{align}
\label{eq:deltaWdefTCanSp}
\begin{split}
\delta W  &\equiv  \int  d^dx\sqrt{-g} \left(\delta A_\mu\cdot J^\mu   
  +\delta E^{\bar{a}}{}_\nu E^{\bar{b}}{}_{\mu} \eta_{\bar{a}\bar{b}}  T_{orbital}^{\mu\nu}  
  +  \delta \mathring{\Gamma}^{\bar{a}}{}_{\bar{b} \lambda}E^{\bar{b}}{}_\mu E^{\bar{c}}{}_{\nu} \eta_{\bar{a}\bar{c}}  \spp^{\lambda[\mu\nu]}  
\right)
\\
&\qquad + (\text{boundary terms})\,,
\end{split}
\end{align}
which can be taken as an alternate way to define the tensors $\{T_{orbital}^{\mu\nu}, \spp^{\lambda[\mu\nu]} \}$. Note that the antisymmetry of $\mathring{\Gamma}^{\bar{a}}{}_{\bar{b} \mu}$ means that the variation of $W$ with respect to $\mathring{\Gamma}^{\bar{a}}{}_{\bar{b} \mu}$ yields $\spp^{\lambda[\mu\nu]}$: an antisymmetric tensor in its last two indices. Of course, the variation of the spin connection is given by derivatives of the frame fields as in~\eqref{E:spinCon}, and so its variation is given by
\begin{equation}
\delta \mathring{\Gamma}^{\bar{a}}{}_{\bar{b} \mu} = 
E^{\bar{a}}{}_\alpha g^{\alpha\lambda} e^{\nu}{}_{\bar{b}} e^{\sigma}{}_{\bar{c}}
\left( g_{\sigma\mu} \mathring{D}_{[\nu} \delta E^{\bar{c}}{}_{\lambda]}
+ g_{\sigma\nu} \mathring{D}_{[\mu} \delta E^{\bar{c}}{}_{\lambda]}
+  g_{\sigma\lambda} \mathring{D}_{[\nu} \delta E^{\bar{c}}{}_{\mu]} \right)\,.
\end{equation}
This, after an integration of parts, gives 
\beq
\label{eq:deltaWTNonSym}
\delta W  =  \int  d^dx\sqrt{-g} \left(  \delta A_\mu\cdot J^\mu  
  +\delta E^{\bar{a}}{}_\nu E^{\bar{b}}{}_{\mu} \eta_{\bar{a}\bar{b}}  T_{Non-Sym}^{\mu\nu}  
\right) + (\text{boundary terms})\,,
\eeq
with $T_{Non-Sym}^{\mu\nu}$ given by 
\beq
T_{Non-Sym}^{\mu\nu} \equiv T_{orbital}^{\mu\nu} 
+  D_\lambda \left( \spp^{\mu[\nu\lambda]} +\spp^{\nu[\mu\lambda]} + \spp^{\lambda[\nu\mu]} \right)\,.
\eeq
This is the equivalent of \eqref{E:stressIntoTandSpin} when we work with the frame fields and spin connection.

We now wish to compute the conservation equations and Ward identities in this context. To do so we follow the same algorithm as above. We parameterize the variation of the external medium under a gauge and coordinate transformation as
\begin{align}
\nonumber
 \diffF W_{ext}  
    &\equiv  \int  d^dx\sqrt{-g}\ \left(  \xi_\alpha f^\alpha_{ext} 
   + \half E^{\bar{a}}{}_\mu \eta_{\bar{b}\bar{c}} E^{\bar{c}}{}_{\nu}
 \prn{\theta^{\bar{b}}{}_{\bar{a}}+\xi^\alpha \mathring{\Gamma}^{\bar{b}}{}_{\bar{a}\alpha} } \tau_{ext}^{[\mu\nu]}
 +  (\Lambda+\xi^\alpha A_\alpha ) \cdot Q_{ext}
    \right)
    \\
    \label{eq:mediumvarLor}
   &\qquad + (\text{boundary terms})\,.
\end{align}
The derivation of conservation equations then proceeds as before. We have 
\begin{align}
\nonumber
\diffF A_\mu& \cdot J^\mu   
+ \diffF E^{\bar{a}}{}_{\nu} E^{\bar{b}}{}_{\mu} \eta_{\bar{a}\bar{b}} T_{orbital}^{\mu\nu} 
  + \diffF \mathring{\Gamma}^{\bar{b}}{}_{\bar{a} \lambda} E^{\bar{a}}{}_\mu \eta_{\bar{b}\bar{c}} E^{\bar{c}}{}_{\nu}\spp^{\lambda[\mu\nu]}  
\\
\nonumber
&\qquad +\xi_\alpha f^\alpha_{ext} 
    + \half E^{\bar{a}}{}_\mu \eta_{\bar{b}\bar{c}} E^{\bar{c}}{}_{\nu}
 \prn{\theta^{\bar{b}}{}_{\bar{a}}+\xi^\alpha \mathring{\Gamma}^{\bar{b}}{}_{\bar{a}\alpha} } \tau_{ext}^{[\mu\nu]}
   + (\Lambda+\xi^\alpha A_\alpha ) \cdot Q_{ext}
   \\
&= D_\mu \tilde{N}_{\chi,Canonical}^\mu -   (\Lambda+\xi^\alpha A_\alpha ) \cdot \prn{ D_{\mu} J^{\mu} - Q_{ext} } 
\\
\nonumber
 &\quad - \xi_{\mu}\Bigl\{ D_{\nu} T_{orbital}^{\mu\nu} - R^\mu{}_{\nu\beta\alpha}\spp^{\nu\alpha\beta}
 - F^\mu{}_\nu \cdot  J^\nu  - f^\mu_{ext} \Bigr\} 
 \\
 \nonumber
 &\quad  - \half E^{\bar{a}}{}_\mu \eta_{\bar{b}\bar{c}} E^{\bar{c}}{}_{\nu}
 \prn{\theta^{\bar{b}}{}_{\bar{a}}+\xi^\alpha \mathring{\Gamma}^{\bar{b}}{}_{\bar{a}\alpha} }   
 \prn{ 2D_\lambda \spp^{\lambda[\mu\nu]} -\prn{T_{orbital}^{\mu\nu} -T_{orbital}^{\nu\mu} } 
-  \tau_{ext}^{[\mu\nu]} }\,,
\end{align}
where the new Canonical Noether current is defined by
\beq
\label{eq:NoetherXiMod2}
\tilde{N}_{\chi,Canonical}^\mu \equiv  \prn{\Lambda +\xi^\alpha A_\alpha} \cdot J^\mu
+ \xi_\alpha T_{orbital}^{\alpha\mu}+ E^{\bar{a}}{}_\alpha \eta_{\bar{b}\bar{c}} E^{\bar{c}}{}_{\beta}
 \prn{\theta^{\bar{b}}{}_{\bar{a}}+\xi^\lambda \mathring{\Gamma}^{\bar{b}}{}_{\bar{a}\lambda} } \spp^{\mu[\alpha\beta]}\,.
\eeq
It follows that the Mathisson-Papapetrou-Dixon equations in \eqref{eq:MPD1} are reproduced by demanding that bulk terms in $\diffF (W +W_{ext})$ vanish. The corresponding Noether identity is 
\begin{align}
\label{eq:WNoetherMod2}
D_\mu \tilde{N}_{\chi,Canonical}^\mu &= \diffF A_\mu \cdot J^\mu 
+ \diffF E^{\bar{a}}{}_\nu E^{\bar{b}}{}_{\mu} \eta_{\bar{a}\bar{b}}  T_{orbital}^{\mu\nu} 
  + \diffF \mathring{\Gamma}^{\bar{b}}{}_{\bar{a} \lambda} E^{\bar{a}}{}_\mu \eta_{\bar{b}\bar{c}} E^{\bar{c}}{}_{\nu}
  \spp^{\lambda[\mu\nu]}  
\\
\nonumber
&\qquad +\xi_\alpha f^\alpha_{ext} 
    + \half E^{\bar{a}}{}_\mu \eta_{\bar{b}\bar{c}} E^{\bar{c}}{}_{\nu}
 \prn{\theta^{\bar{b}}{}_{\bar{a}}+\xi^\alpha \mathring{\Gamma}^{\bar{b}}{}_{\bar{a}\alpha} } \tau_{ext}^{[\mu\nu]}
   + (\Lambda+\xi^\alpha A_\alpha ) \cdot Q_{ext}\,.
\end{align}
As is characteristic of the formalism in terms of frame-fields and spin connection, Lorentz transformations are treated almost in par with the flavor transformations: $\spp^{\lambda[\mu\nu]}$ plays the role of the current associated with Lorentz symmetry and $\tau_{ext}^{[\mu\nu]}$ is the external injection rate of the `Lorentz charge.' If we had started from \eqref{eq:deltaWTNonSym}, then the corresponding conservation laws would have been
\begin{equation}\label{eq:MPD3}
D_{\mu} J^{\mu}= Q_{ext} \,, \qquad
D_\nu T_{Non-Sym}^{\mu\nu} = F^\mu{}_\nu \cdot  J^\nu + f^\mu_{ext}\,, \qquad
-T_{Non-Sym}^{\mu\nu}+T_{Non-Sym}^{\nu\mu} = \tau_{ext}^{[\mu\nu]} \,.
\end{equation}

To summarize, we have four different descriptions, depending on our choice of sources.
\begin{enumerate}
 \item With $\{A_\mu, g_{\mu\nu} ,\Gamma^\mu{}_{\nu\lambda}\}$ as independent sources, 
 we have a description in terms of the currents  $\{J^\mu, t^{\mu\nu},\spp^{\lambda\mu\nu}\}$. With $\{ A_\mu , E^{\bar{a}}{}_\mu, \Gamma^{\bar{a}}{}_{\bar{b}\lambda} \}$  as independent sources, we have a description in terms of the currents $\{J^\mu, T_{orbital}^{\mu\nu},\spp^{\lambda[\mu\nu]}\}$ where 
 \begin{equation}
  T_{orbital}^{\mu\nu} = t^{\mu\nu} -  \prn{ D_\lambda \spp^{\lambda\nu\mu}- \half\tau_{ext}^{\nu\mu} }  \,.
 \end{equation}
 In both these cases, the conservation equations are the Mathisson-Papapetrou-Dixon equations
\begin{align}
\label{eq:MPD1alt}
\begin{split}
D_{\mu} J^{\mu}&= Q_{ext}\,, \\
D_{\nu} T_{orbital}^{\mu\nu} &=  R^\mu{}_{\nu\beta\alpha}\spp^{\nu\alpha\beta}
 + F^\mu{}_\nu \cdot  J^\nu
  +  f^\mu_{ext} \,,\\
D_\lambda \spp^{\lambda[\mu\nu]} &=T_{orbital}^{[\mu\nu]} 
+ \half \tau_{ext}^{[\mu\nu]}\,.
\end{split}
\end{align}
\item  With $\{A_\mu, E^{\bar{a}}{}_\mu \}$ as independent sources, we have a description in terms of the currents  $\{J^\mu, T_{Non-Sym}^{\mu\nu}\}$ where 
\begin{equation}
\begin{split}
T_{Non-Sym}^{\mu\nu} &\equiv T_{orbital}^{\mu\nu} 
+ D_\lambda \prn{ \spp^{\mu[\nu\lambda]} +\spp^{\nu[\mu\lambda]} + \spp^{\lambda[\nu\mu]} }\\
 &= t^{\mu\nu}-  \prn{ D_\lambda  \spp^{\lambda\nu\mu}  -\half\tau_{ext}^{\nu\mu} }
 +  D_\lambda \prn{ \spp^{\mu[\nu\lambda]} +\spp^{\nu[\mu\lambda]} + \spp^{\lambda[\nu\mu]} }\\
 &= T^{\mu\nu}+ \half \tau_{ext}^{\nu\mu} \,.\\
\end{split}
\end{equation}
The  conservation equations are 
\begin{align}\label{eq:MPD3Alt}
\begin{split}
D_{\mu} J^{\mu} &= Q_{ext} \,,
\\
D_\nu T_{Non-Sym}^{\mu\nu} &= F^\mu{}_\nu \cdot  J^\nu + f^\mu_{ext}\,,
\\
0 & = \left( T_{Non-Sym}^{\mu\nu}-T_{Non-Sym}^{\nu\mu} \right)+ \tau_{ext}^{[\mu\nu]} \,,
\end{split}
\end{align}
which are the Mathisson-Papapetrou-Dixon equations with zero spin currents.
 \item  With $\{A_\mu, g_{\mu\nu} \}$ as independent sources, we have a description in terms of
 the currents  $\{J^\mu, T^{\mu\nu}\}$ where 
 \begin{align}\label{eq:TfromSigmaAlt}
\begin{split}
T^{\mu\nu}
 &= t^{\mu\nu}-D_\lambda  \spp^{\lambda\nu\mu}  
 +  D_\lambda \prn{ \spp^{\mu[\nu\lambda]} +\spp^{\nu[\mu\lambda]} + \spp^{\lambda[\nu\mu]} }
 \\
 &= T_{orbital}^{\mu\nu} +  \brk{ D_\lambda \prn{ \spp^{\mu[\nu\lambda]} +\spp^{\nu[\mu\lambda]} + \spp^{\lambda[\nu\mu]} } -
\half \tau_{ext}^{\nu\mu} } \,.
\end{split}
\end{align}
The  conservation equations are 
\begin{align}\label{eq:MPD2Alt}
\begin{split}
D_{\mu} J^{\mu}&= Q_{ext} \,,\qquad
D_{\nu} T^{\mu\nu} =  F^\mu{}_\nu \cdot  J^\nu
  +  f^\mu_{ext} -\half D_\nu \tau_{ext}^{\nu\mu} \,,\qquad 
T^{\mu\nu} -T^{\nu\mu} =0 \,,
\end{split}
\end{align}
which are the Mathisson-Papapetrou-Dixon equations with zero spin currents and zero point torque.
\end{enumerate}

Note that all these different descriptions -- be it in terms of  $\{T^{\mu\nu}_{orbital}, \spp^{\sigma\mu\nu}\}$ obeying \eqref{eq:MPD1}, or in terms of $T_{Non-Sym}^{\mu\nu}$ obeying \eqref{eq:MPD3}, or in terms of the symmetric energy momentum tensor $T^{\mu\nu}$ -- all of them are equivalent descriptions of the same system related by various redefinitions. In fact, the Mathisson-Papapetrou-Dixon equations are invariant under a broader set of Belinfante-Rosenfeld transformations which shift the spin current $\spp^{\sigma\mu\nu}$ and torques $\tau_{ext}^{\mu\nu}$. These transformations may be regarded as an ambiguity involved in the definition of angular momentum 
and point torques, and under them we have
\begin{align}\label{eq:Belinfante}
\begin{split}
\tau_{ext}^{\mu\nu} &\mapsto \tau_{ext}^{\mu\nu} - 2\mathcal{A}^{\mu\nu}\,,\quad
f_{ext}^\mu \mapsto f_{ext}^\mu - D_\nu  \mathcal{A}^{\nu\mu} \,,\quad
 \spp^{\sigma\mu\nu} \mapsto \spp^{\sigma\mu\nu} - \mathcal{B}^{\sigma\mu\nu}\,,
 \\
 T^{\mu\nu}_{orbital} &\mapsto T^{\mu\nu}_{orbital}
 +  \brk{ D_\lambda \prn{ \mathcal{B}^{\mu\nu\lambda} +\mathcal{B}^{\nu\mu\lambda} + \mathcal{B}^{\lambda\nu\mu} }
 - \mathcal{A}^{\nu\mu}}\,, \\
\end{split}
\end{align}
where $\mathcal{B}^{\lambda\mu\nu}$ must satisfy $\mathcal{B}^{\lambda\mu\nu}= -\mathcal{B}^{\lambda\nu\mu}$. This result follows by using the identity 
\begin{align}
\begin{split}
  D_\nu D_\lambda &\prn{ \mathcal{B}^{\mu\nu\lambda} +\mathcal{B}^{\nu\mu\lambda} + \mathcal{B}^{\lambda\nu\mu} }
  +R^\mu{}_{\nu\beta\alpha} \mathcal{B}^{\nu\alpha\beta}
  \\
  &=  [D_\nu ,D_\lambda]\mathcal{B}^{\mu\nu\lambda} +R^\mu{}_{\nu\beta\alpha} \mathcal{B}^{\nu\alpha\beta}
  -\half [D_\nu ,D_\lambda] \prn{\mathcal{B}^{\mu\nu\lambda} +\mathcal{B}^{\lambda\mu\nu}+\mathcal{B}^{\nu\lambda\mu}  } 
  \\
   &=  \half \prn{R^\mu{}_{\nu\beta\alpha}+R^\mu{}_{\beta\alpha\nu}+R^\mu{}_{\alpha\nu\beta} } \mathcal{B}^{\nu\alpha\beta}
   \\
 &= 0\,,
\end{split}
\end{align}
which holds for any tensor $\mathcal{B}^{\lambda\mu\nu}$ provided $\mathcal{B}^{\lambda\mu\nu} = -\mathcal{B}^{\lambda\nu\mu}$. This shows that the Mathisson-Papapetrou-Dixon equations  automatically incorporate the ambiguity involved in the definition of spin angular momentum and point torques. Further, it is easily checked that all the cases considered above are related to each other via a Belinfante-Rosenfeld shift. We will use these descriptions interchangeably in the rest of this article.

In the rest of this Appendix, we will remark on some important features of the Mathisson-Papapetrou-Dixon equations. As mentioned before, the external forces and torques $\{ f^\mu_{ext}, \tau^{\mu\nu}_{ext} \}$ appearing in \eqref{eq:MPD1} denote the external forces and torques per unit volume acting on the system, injecting canonical energy-momentum and spin into the system. Note that $\tau^{[\mu\nu]}_{ext}$ includes only the point torques (or the `spin' part of torques) and excludes `orbital' $r\times f$ torques that arise from $f^\mu_{ext}$ itself. For example, in an electrodynamic medium with magnetization-polarization tensor $\Mag^{\mu\nu}$ and a background field strength  $F_{\alpha\beta}$, our derivation gives \eqref{eq:DSeqn}. 

Further, we note that the force corresponding to the canonical energy momentum tensor includes, in addition to $f^\mu_{ext}$ and the Lorentz force on the flavor current $J^\nu$, an additional term $R^\mu{}_{\nu\beta\alpha} \spp^{\nu\alpha\beta}$ which is sometimes called the Mathisson force. Mathisson force is crucial, for example, in explaining why free spinning particles do not follow geodesics in a curved spacetime (or more colloquially, why free fall of gyroscopes are affected by how fast they are spinning). In the solid state context, where disclinations in the solid can be modeled as a background curvature, Mathisson force is the force on a spin as it crosses a disclination. 

The Mathisson-Papapetrou-Dixon equations exhibit various other properties which make them an appropriate description of 
spin dynamics.
\begin{itemize}
\item They reduce to the correct equations for describing spinning particles moving in curved spacetimes. If $Q,P^\mu$ and $S^{\mu\nu}$ denote the flavor charge, canonical momentum and the spin of a particle, then the Mathisson-Papapetrou-Dixon equations become
\begin{align}
\label{eq:MPDParticle}
\begin{split}
 \frac{DQ}{D\tau}  &= ( \text{charge injection rate})\,,
 \\
 \frac{DP^\mu}{D\tau} &=  Q\cdot F^\mu{}_\nu\frac{dx^\nu}{d\tau} + R^\mu{}_{\nu\beta\alpha} \frac{dx^\nu}{d\tau} S^{\alpha\beta} +
 (\text{force})^\mu_{ext}\,,
  \\
 \frac{DS^{[\mu\nu]}}{D\tau}  &=   P^{\mu} \frac{dx^\nu}{d\tau}-P^\nu \frac{dx^\mu}{d\tau} +(\text{torque})^{[\mu\nu]}_{ext} \,,
\end{split}
\end{align}
where $x^\mu$ is the position of the particle,  $\tau$ denotes its proper time and $\frac{D}{D\tau}$ denotes the covariant derivative take along the worldline. We have used $\{(\text{force})^\mu_{ext}, (\text{torque})^{\mu\nu}_{ext}\}$ respectively to denote the external forces and point torques acting on the particle and we have used the fact that $\spp^{\mu\nu\lambda}$ denotes \emph{half} of the spin current. In the absence of flavor charges, we obtain the original form in which these equations were derived by Mathisson, Papapetrou and Dixon.
\item In flat spacetime and in cartesian co-ordinates, they give rise to familiar energy-momentum and angular momentum conservation equations
\begin{align}
\begin{split}
 \partial_\mu J^\mu &= Q_{ext}\,,
  \\
 \partial_\nu T^{\mu\nu}_{orbital}&= F^\mu{}_\nu \cdot  J^\nu +f^\mu_{ext}\,,
 \\
 \partial_\sigma \Bigl[ x^\mu T^{\nu\sigma}_{orbital} \Bigr. &\Bigl.-x^\nu T^{\mu\sigma}_{orbital}  +2 \spp^{\sigma[\mu\nu]} \Bigr] \\
  &= x^\mu \prn{F^\nu{}_\alpha \cdot  J^\alpha + f^\nu_{ext}}-x^\nu \prn{F^\mu{}_\alpha \cdot  J^\alpha + f^\mu_{ext}} + \tau^{[\mu\nu]}_{ext}\,,
\end{split}
\end{align}
where we note that $2\spp^{\sigma[\mu\nu]}$ appears as the `spin' part of angular momentum current as expected whereas the `orbital' part of the angular momentum current is constructed from the `orbital' part of the energy-momentum
$T^{\mu\nu}_{orbital}$. 
\item In QFTs without spin-orbit interaction, the conservation of the total angular momentum breaks up into separate conservation equations for orbital and spin angular momentum, i.e., $D_\lambda \spp^{\lambda[\mu\nu]} = T_{orbital}^{[\mu\nu]} + \frac{1}{2} \tau_{ext}^{[\mu\nu]} $ breaks up into $D_\lambda \spp^{\lambda[\mu\nu]} \approx  \frac{1}{2}\tau_{ext}^{[\mu\nu]} $ and $T_{orbital}^{[\mu\nu]}\approx 0$. In other words, $T_{orbital}^{[\mu\nu]} $ appearing in the third Mathisson-Papapetrou-Dixon equation is \emph{half} the `spin-orbit torque' arising because of orbital angular momentum seeping into spin angular momentum..
\end{itemize}

Thus, we have concluded that irrespective of the starting point the conservation laws for angular momentum always take the Mathisson-Papapetrou-Dixon form \eqref{eq:MPD1} as claimed. This discussion of a non-anomalous QFT in contact with a medium which injects charge and angular momentum has many similarities with an anomalous QFT where charge or angular momentum is injected by the anomaly instead. These similarities are especially clear in the anomaly inflow picture where charge or angular momentum can be thought of as injected from  a system with one dimension higher. To conclude this Appendix, we note that both the consistent~\eqref{E:conWard} and covariant Ward identities~\eqref{E:covWard} take the Mathisson-Papapetrou-Dixon form
\begin{align}\label{eq:MPDFinal}
\begin{split}
D_{\mu} J^{\mu}&= Q_{ext} \,,
\\
D_{\nu} T_{orbital}^{\mu\nu} &=   R^\mu{}_{\nu\beta\alpha}\spp^{\nu\alpha\beta}
 + F^\mu{}_\nu \cdot  J^\nu
  +  f^\mu_{ext}\,,
  \\
D_\lambda \spp^{\lambda[\mu\nu]} &=T_{orbital}^{[\mu\nu]} 
+ \half \tau_{ext}^{[\mu\nu]}\,.
\end{split}
\end{align}
Written this way, the effects of the anomaly on the consistent Ward identities~\eqref{E:conWard} can be accounted for by an external force, external torque, and charge injection rate
\beq
\label{E:conAnomTorque}
f_{ext}^{\alpha} = - A^{\alpha} \cdot \mathcal{J} - g^{\alpha\beta}\Gamma\ab{\mu}{\beta\nu}\mathcal{T}\ab{\nu}{\mu}\,, \qquad \tau_{ext}^{\alpha\beta} = 2 \mathcal{T}\ab{\alpha\beta}\,, \qquad Q_{ext} = \mathcal{J}\,,
\eeq
where $\mathcal{J}$ and $\mathcal{T}\ab{\mu}{\nu}$ were defined in~\eqref{E:defConAnom}. Similarly, the effects of the anomalies in the covariant Ward identities~\eqref{E:covWard} can be accounted for by
\beq
\label{E:covAnomTorque}
 f^\alpha_{ext}   = 0\,,
 \qquad 
 \tau^{\alpha\beta}_{ext} = 2\spp_H^{\perp\alpha\beta} \,,
 \qquad 
  Q_{ext} = J_H^{\perp}\,,
\eeq
where the Hall current $J_H^M$ and spin current $\spp_H^{MNP}$ were defined in~\eqref{E:HallBZ}.

\section{The relativistic Boltzmann weight}
\label{A:Boltzmann}

The description of equilibrium physics in the transverse gauge is intimately related to the traditional presentation of Euclidean thermal field theory. However there are a couple of subtle differences which we elicit in this Appendix. The essential observable in thermal field theory, in the absence of anomalies, is the thermal partition function
\beq
\label{E:ZE}
Z_E = \text{tr} \, \exp(-\beta \mathcal{H})\,,
\eeq
where $\beta$ is the parametric length of the thermal circle and $\mathcal{H}$ is the generator of time translations in the background~\eqref{E:transversegaugeexplicit}.\footnote{In writing $Z_E$ as in~\eqref{E:ZE}, it is clear that the usual presentation of thermal field theory is non-covariant. We have separated time and space from the outset, in exactly the same way as in the transverse gauge.}
As a result $Z_E$ is a functional of the background metric and gauge field. For theories with a functional integral description, this partition function is equal to the functional integral of the Euclidean weight $\exp(-S_E)$ on the Euclidean background~\eqref{E:eucBack} with appropriate boundary conditions around the thermal circle. The temperature and flavor chemical potential are usually defined through observables which is local in the spatial $x$-directions, but non-local in Euclidean time. The temperature is the inverse length of the thermal circle, as a function of space, and the flavor chemical potential is defined through the Wilson line of the gauge field around the thermal circle. The inverse length is
\beq
L^{-1} = \left( \int_0^{\beta}d\lambda \sqrt{-K^2(\lambda)}\right)^{-1} = \frac{1}{\beta \sqrt{-K^2}}\,,
\eeq
which agrees with the local temperature defined in the covariant context in~\eqref{E:covHydroVar}.  In the same way, we could define the spin chemical potential through the Wilson line of the Christoffel connection. Denoting the thermal circle at a position $x$ on the spatial slice as $\mathcal{C}_x$, the Wilson lines around $\mathcal{C}_x$ in the transverse gauge are\footnote{We remind the reader that, in anti-hermitian flavor basis, the usual Wilson lines are defined without an $i$ in the exponential.} 
\begin{align}
\begin{split}
\label{E:wilsonLines}
P\exp\left( - \int_{\mathcal{C}_x}\mathbf{A}\right)&= \exp(-\beta (A_0)_E) = \exp(i \beta A_0) = \exp \left(i \frac{\mu}{T}\right)\,,
\\
P\exp\prn{- \int_{\mathcal{C}_x} \fGamma }\ab{\mu}{\nu}& =  \exp\prn{- \beta(\Gamma_0)_E }\ab{\mu}{\nu}
=   \exp\prn{i\beta \Gamma_0}\ab{\mu}{\nu} = \exp\prn{i\frac{\muR}{T} }\ab{\mu}{\nu}\,,
\end{split}
\end{align}
so that the logarithms of the Wilson lines are  $i\mu/T$ and $i(\muR)\ab{\mu}{\nu}/T$ respectively. As an aside, we note that the Wilson lines of the hatted connections are trivial in transverse gauge are given by
\begin{equation*}
P\exp\left( - \int_{\mathcal{C}_x} \hat{\fA}\right) = 1\,, \qquad P\exp\left( - \int_{\mathcal{C}_x} \hat{\form{\Gamma}}\right)\ab{\mu}{\nu} = \delta\ab{\mu}{\nu}\,.
\end{equation*}

Actually, we should be careful. In writing down the Wilson lines above in the way they usually appear in certain textbook discussions of field theory, we have written down unphysical objects. The proper observables are the holonomies perceived by $G$ and matrix-valued tensors when parallel transported around the thermal circle. That is, the physical quantities are the phases perceived by charged states when they go around the circle. These phases receive two contributions, one from the integrals above~\eqref{E:wilsonLines} and another from twisted boundary conditions that may be imposed around the circle. The latter are implemented by $\Lambda_K$ and $\partial_{\nu}K^{\mu}$. Only the combination is physical, which at this stage should be unsurprising to the reader: in the main text, we defined the flavor and spin chemical potentials to be the covariant objects $(K^{\alpha}A_{\alpha} + \Lambda_K)/\sqrt{-K^2}$ and $D_{\nu}K^{\mu}/\sqrt{-K^2} = (K^{\alpha}\Gamma\ab{\mu}{\nu\alpha} + \partial_{\nu}K^{\mu})/\sqrt{-K^2}$ respectively. The terms involving the connection essentially give the integrals~\eqref{E:wilsonLines} of $\fA$ and $\fGamma\ab{\mu}{\nu}$ around the thermal circle, while the terms $\{\Lambda_K,\partial_{\nu}K^{\mu}\}$ provide the twisted boundary conditions.

It is tempting to think of the quantities $\{T,\mu,(\muR)\ab{\mu}{\nu}\}$ more generally in terms of these non-local quantities around the thermal circle. In fact, just this identification was made for $T$ and $\mu$ in~\cite{Jensen:2012jh}. However this is a little misleading. We stress that the quantities $\{T,\mu,(\muR)\ab{\mu}{\nu}\}$ defined in our covariant analysis are \emph{local} on spacetime, whereas the integrals around the thermal circle are only valued on the spatial slice. However, in a transverse gauge, all fields and background fields are independent of time, and so may be regarded as local fields on the spatial slice. This is why the traditional thermal field theory presentation is related to the transverse gauge. 

For a non-anomalous theory, the thermal partition function $Z_E$ should be gauge and coordinate-reparametrization invariant. In order for this to be true, we must be a little careful by what we mean by the Hamiltonian $\mathcal{H}$. There are several Noether-like currents that we may define and so several potential Hamiltonians. In the rest of this Appendix we revisit the canonical Noether current defined in~\eqref{eq:NoetherXiMod} and its relation to $Z_E$, relying heavily on results derived in the previous Appendix. In that context, we were interested in the Ward identities and Noether currents when coupling a field theory to a general external sector which dumped energy-momentum, angular momentum, and charge into the field theory. We related that formalism to anomalous field theories at the end of that Appendix, in particular showing that it describes the dynamics of the consistent currents obeying the Ward identities~\eqref{E:conWard}. The corresponding external force, torque, and charge injection rate were given in~\eqref{E:conAnomTorque} to be
\begin{equation*}
f_{ext}^{\alpha} = - A^{\alpha} \cdot \mathcal{J} - g^{\alpha\beta}\Gamma\ab{\mu}{\beta\nu}\mathcal{T}\ab{\nu}{\mu}\,, \qquad \tau_{ext}^{\alpha\beta} = 2 \mathcal{T}\ab{\alpha\beta}\,, \qquad Q_{ext} = \mathcal{J}\,,
\end{equation*}
where $\mathcal{J}$ and $\mathcal{T}\ab{\mu}{\nu}$ were defined in~\eqref{E:defConAnom}. Here and in what follows, we build Noether currents out of the consistent currents and specialize to the symmetry transformation $\{K^{\mu},\Lambda_K\}$. Redefining the Noether current in~\eqref{eq:NoetherXiMod} by a minus sign (in which case, for the energy-momentum tensor and flavor current of ideal hydrodynamics, it becomes the energy current), we define
\begin{align}
\begin{split}
\label{E:JKdef}
J_{K,canonical}^{\mu} &\equiv  -K_{\nu}T_{orbital}^{\nu\mu} - (K^{\alpha}A_{\alpha}+\Lambda)\cdot J^{\mu} - D_{\beta}K^{\alpha}\spp\ab{\mu\beta}{\alpha} 
\\
&= -K_{\nu}T_{orbital}^{\nu\mu} - \sqrt{-K^2}\left( \mu \cdot J^{\mu} + (\mu_R)\ab{\alpha}{\beta}\spp\ab{\mu\beta}{\alpha}\right)\,,
\end{split}
\end{align}
where $T_{orbital}^{\mu\nu}$ was also defined in~\eqref{eq:NoetherXiMod}. Its divergence when $\delta_K$ generates a symmetry is given by~\eqref{eq:MPDNoether1} to be
\beq
D_{\mu}J_{K,canonical}^{\mu} = - K_{\mu}f_{ext}^{\mu} - \frac{1}{2}(D_{\alpha}K_{\beta})\tau_{ext}^{\alpha\beta} - (K^{\alpha}A_{\alpha}+\Lambda_K)\cdot Q_{ext}\,.
\eeq
Substituting the external force, torque, and charge injection rates~\eqref{E:conAnomTorque} relevant for describing the consistent anomaly, we find
\beq
\label{E:divJK}
D_{\mu}J_{K,canonical}^{\mu} = - \Lambda_K \cdot \mathcal{J} - \partial_{\nu}K^{\mu} \mathcal{T}\ab{\nu}{\mu}\,,
\eeq
We see that in the absence of anomalies, $J_{K,canonical}^{\mu}$ is always conserved. Further, it remains conserved even in the presence of anomalies, provided that we are in transverse gauge.

As an aside, consider the related current
\beq
\label{E:tildeJK}
\tilde{J}_K^{\mu} \equiv  -K_{\nu}T_{orbital}^{\nu\mu} - (K^{\alpha}A_{\alpha} \cdot J^{\mu} + D_{\beta}K^{\alpha}\spp\ab{\mu\beta}{\alpha}) = J_{K,canonical}^{\mu} + \Lambda_K \cdot J^{\mu} \,.
\eeq
From~\eqref{E:divJK}, we see that its divergence is
\beq
\label{E:divTildeJK}
D_{\mu}\tilde{J}_K^{\mu} = D_{\mu} \Lambda_K \cdot J^{\mu} -\partial_{\nu}K^{\mu}\mathcal{T}\ab{\nu}{\mu}\,,
\eeq
which vanishes when $\Lambda_K$ is covariantly constant and $\partial_{\rho}K^{\nu}$ vanishes. In the absence of anomalies, the common convention   seems to be to choose the generator $\mathcal{H}$ appearing in the thermal partition function~\eqref{E:ZE} to be conjugate to the current $\tilde{J}_K^{\mu}$ defined in~\eqref{E:tildeJK}. However, $\tilde{J}_K^{\mu}$ is not generally conserved; for a non-anomalous theory, its divergence~\eqref{E:divTildeJK} only vanishes when $\Lambda_K$ is covariantly constant. In that instance, one can go to a gauge and coordinate choice where $\Lambda_K$ is independent of time and space, in which case we may relabel it as the position-independent constant $\Lambda_K \equiv \mu_0$. One can then define the Boltzmann weight
\beq
\label{E:tildeBoltzmann}
\exp\left[ - \beta\left( \tilde{\mathcal{H}}_K - \mu_0 \cdot \mathcal{Q} \right)\right]\,,
\eeq
where $\tilde{\mathcal{H}}_K$ is the operator conjugate to the current $\tilde{J}_K^{\mu}$, and$\mathcal{Q}$ is the flavor charge operator conjugate to the flavor current. Let us call the corresponding partition function $\tilde{\mathcal{Z}}_E$. From $\tilde{\mathcal{Z}}_E$ one can define thermodynamic energy, and flavor charge through derivatives with respect to $\beta$ and $\mu_0$. The Boltzmann weight~\eqref{E:tildeBoltzmann} is reminiscent of textbook statistical mechanics. It is also written in a way to remind us what the chemical potentials do: they twist the weighting of charged states in the sum over states. We note that one can go through a similar transformation of the Noether current involving the spin chemical potential and spin current. To do so, it is most convenient to work with the frame fields and spin connection, and reformulate the gravitational anomalies in terms of Lorentz anomalies. We do not do so here in a momentary attempt at brevity.

The Boltzmann weight~\eqref{E:tildeBoltzmann} is written in an extremely non-covariant way. In order to write $Z_E$ in a manifestly gauge and coordinate-invariant way, we can choose the Boltzmann weight to instead be
\beq
\exp\left( - \beta \mathcal{H}_K\right)\,,
\eeq
where $\mathcal{H}_K$ is the operator conjugate to the current $J_{K,canonical}^{\mu}$ defined in~\eqref{E:JKdef}. When $\Lambda_K$ is constant, this weight is equal to that defined before in~\eqref{E:tildeBoltzmann}. However, unlike $\tilde{Z}_E$, the partition function $Z_E = \text{tr}\,\exp\left( - \beta \mathcal{H}_K\right)$ is gauge-invariant as it ought to be.

So far we have discussed the Boltzmann weight for non-anomalous theories. Almost all of that discussion carries over when the underlying theory is anomalous. The only real change comes with the Boltzmann weights and the thermal partition function, which is no longer invariant under gauge and coordinate transformations. However, by~\eqref{E:divJK}, the current $J_K^{\mu}$ remains conserved even when there are anomalies provided that we are in transverse gauge. As a result the thermal partition function may still be understood as a sum over states with Boltzmann weight $\exp(-\beta \mathcal{H}_K)$ as before, but $\mathcal{H}_K$ is akin to a Hamiltonian (in the sense that it generates a symmetry) only in transverse gauge.

\section{Notation}
\label{ss:forms}

It is often useful to shift to the language of differential forms (which we will denote by bold 
letters ) which is a more efficient way of dealing with fully antisymmetric tensors. In  
this Appendix, we will summarize our conventions for differential forms.

\begin{itemize}
 \item We begin with our convention for the wedge product which is fixed by demanding that
\begin{equation}
\label{E:defWedge}
dx^{\mu_1} \wedge dx^{\mu_2} \wedge \ldots \wedge dx^{\mu_p} \equiv p!\ dx^{[\mu_1} \otimes dx^{\mu_2} \otimes \ldots \otimes dx^{\mu_p]}\,,
\end{equation}
where $\{dx^\mu\}$ are the basis 1-forms associate with co-ordinates $x^\mu$. Further, $[\mu_1\ldots \mu_p]$ indicates a projection to the antisymmetric part and $\otimes$ is the ordinary tensor product. For example,
\begin{equation}
dx^{\mu} \wedge dx^{\nu}  \equiv 2!\  dx^{[\mu} \otimes dx^{\nu]} = dx^{\mu} \otimes dx^{\nu}-  dx^{\nu} \otimes dx^{\mu}\,.
\end{equation}
Since we know how the wedge product acts on basis forms, we can linearly extend the definition to arbitrary forms.
\item A $p$-form $\form{V}$ is a fully antisymmetric $p$-tensor whose components are given by $V_{\mu_1..\mu_p}$. As a tensor it is 
\begin{equation}
\begin{split}
 \form{V} &\equiv V_{\mu_1\ldots \mu_p}\ dx^{\mu_1} \otimes  \ldots \otimes dx^{\mu_p} = V_{\mu_1\ldots \mu_p}\ dx^{[\mu_1} \otimes  \ldots \otimes dx^{\mu_p]} \\
 &= \frac{1}{p!} V_{\mu_1\ldots \mu_p}\ dx^{\mu_1} \wedge dx^{\mu_2} \wedge \ldots \wedge dx^{\mu_p}\,.
\end{split}
\end{equation}
We will also encounter tensors with a number of fully antisymmetric covariant indices. These may be regarded as just another tensor, or as tensor-valued $p$-forms like $\form{V}\ab{\alpha}{\beta}$. In components, these has $p$ fully antisymmetric covariant indices along with additional `free' tensor indices (for $\form{V}\ab{\alpha}{\beta}$ these would be $\alpha$ and $\beta$). For example, 
\beq
 \form{V}^\alpha{}_\beta  \equiv  \frac{1}{p!} V^\alpha{}_{\beta \mu_1\ldots \mu_p} dx^{\mu_1} \wedge dx^{\mu_2} \wedge \ldots \wedge dx^{\mu_p}\,.
\eeq
\item Given a 1-form $\form{A}$ and a $p$-form $\form{V}$, their wedge product is then defined by the wedge product of the basis 1-forms~\eqref{E:defWedge} and linearity, which gives
\begin{equation}
\begin{split}
 \form{A}\wedge \form{V} &\equiv \frac{1}{p!} A_{\lambda}  V_{\mu_1\ldots \mu_p}\ 
 dx^\lambda \wedge dx^{\mu_1} \wedge dx^{\mu_2} \wedge \ldots \wedge dx^{\mu_p} \\
  &= \frac{1}{(p+1)!} \Bigl\{ A_{\mu_1}  V_{\mu_2\ldots \mu_{p+1}}+(-1)^p A_{\mu_2}  V_{\mu_3\ldots \mu_{p+1}\mu_1} \Bigr.\\
  &\Bigl.\qquad 
  + (-1)^{2p}A_{\mu_3}  V_{\mu_4\ldots \mu_{p+1}\mu_1\mu_2} 
  +\ldots +(-1)^{p^2} A_{\mu_{p+1}}  V_{\mu_1\ldots \mu_p} \Bigr\} \\
  &\qquad    
  dx^{\mu_1} \wedge dx^{\mu_2} \wedge \ldots \wedge dx^{\mu_p} \wedge dx^{\mu_{p+1}}\,. \\
\end{split}
\end{equation}
Hence the components of the $(p+1)$-form $\form{A}\wedge \form{V}$ are given by 
\begin{equation}
\begin{split}
(\form{A}\wedge \form{V})_{\mu_1\mu_2\ldots \mu_{p+1}} &\equiv  A_{\mu_1}  V_{\mu_2\ldots \mu_{p+1}}+(-1)^p A_{\mu_2}  V_{\mu_3\ldots \mu_{p+1}\mu_1} \Bigr.\\
  &\Bigl.\qquad 
  + (-1)^{2p}A_{\mu_3}  V_{\mu_4\ldots \mu_{p+1}\mu_1\mu_2} 
  +\ldots +(-1)^{p^2} A_{\mu_{p+1}}  V_{\mu_1\ldots \mu_p}\\
  &= \sum_{k=1}^{p+1}  (-1)^{p(k-1)} A_{\mu_k}  V_{\mu_{k+1}\mu_{k+2}\ldots\mu_{p+1}\mu_1\mu_2\ldots \mu_{k-1}}\,.
\end{split}
\end{equation}
\item The exterior derivative $d$ is a derivation that maps $p$-forms to $p+1$-forms. Our convention for $d$ in components are given by 
\begin{equation}
\begin{split}
 d\form{V} &\equiv \frac{1}{p!} \partial_{\lambda}  V_{\mu_1\ldots \mu_p}\ 
 dx^\lambda \wedge dx^{\mu_1} \wedge dx^{\mu_2} \wedge \ldots \wedge dx^{\mu_p} \\
  &= \frac{1}{(p+1)!} \Bigl\{ \partial_{\mu_1}  V_{\mu_2\ldots \mu_{p+1}}+(-1)^p \partial_{\mu_2}  V_{\mu_3\ldots \mu_{p+1}\mu_1} \Bigr.\\
  &\Bigl.\qquad 
  + (-1)^{2p}\partial_{\mu_3}  V_{\mu_4\ldots \mu_{p+1}\mu_1\mu_2} 
  +\ldots +(-1)^{p^2} \partial_{\mu_{p+1}}  V_{\mu_1\ldots \mu_p} \Bigr\} \\
  &\qquad    
  dx^{\mu_1} \wedge dx^{\mu_2} \wedge \ldots \wedge dx^{\mu_p} \wedge dx^{\mu_{p+1}}\,. \\
\end{split}
\end{equation}
Hence the components of the $(p+1)$-form $d\form{V}$ are given by 
\begin{equation}
\begin{split}
(d\form{V})_{\mu_1\mu_2\ldots \mu_{p+1}} &\equiv  \partial_{\mu_1}  V_{\mu_2\ldots \mu_{p+1}}+(-1)^p \partial_{\mu_2}  V_{\mu_3\ldots \mu_{p+1}\mu_1} \\
  &\qquad 
  + (-1)^{2p}\partial_{\mu_3}  V_{\mu_4\ldots \mu_{p+1}\mu_1\mu_2} 
  +\ldots +(-1)^{p^2} \partial_{\mu_{p+1}}  V_{\mu_1\ldots \mu_p} \\
 &= \sum_{k=1}^{p+1}  (-1)^{p(k-1)} \partial_{\mu_k}  V_{\mu_{k+1}\mu_{k+2}\ldots\mu_{p+1}\mu_1\mu_2\ldots \mu_{k-1}}\,.
\end{split}
\end{equation}
The covariant exterior derivative $D$ is defined similarly using the covariant derivative $D_\lambda$  instead of the ordinary partial derivative $\partial_\lambda$, giving for instance
\begin{equation}
\begin{split}
(D\form{V})_{\mu_1\mu_2\ldots \mu_{p+1}} &\equiv  D_{\mu_1}  V_{\mu_2\ldots \mu_{p+1}}+(-1)^p D_{\mu_2}  V_{\mu_3\ldots \mu_{p+1}\mu_1} \\
  &\qquad 
  + (-1)^{2p}D_{\mu_3}  V_{\mu_4\ldots \mu_{p+1}\mu_1\mu_2} 
  +\ldots +(-1)^{p^2} D_{\mu_{p+1}}  V_{\mu_1\ldots \mu_p}  \\ 
   &= \sum_{k=1}^{p+1}  (-1)^{p(k-1)} D_{\mu_k}  V_{\mu_{k+1}\mu_{k+2}\ldots\mu_{p+1}\mu_1\mu_2\ldots \mu_{k-1}}\,.
\end{split}
\end{equation}
When $\form{V}$ is a flavor singlet $p$-form, we have $d\form{V}=D\form{V}$ by the torsionlessness of the Christoffel connection.
\item The interior product $\iota_{\xi}$ is a derivation that takes $p$-forms to $p-1$ forms given a vector $\xi\equiv \xi^{\mu}\partial_{\mu}$. In components it acts via 
\beq
 (\iota_{\xi}\form{V})_{\mu_1\ldots \mu_{p-1}} \equiv  \xi^\lambda V_{\lambda \mu_1\ldots \mu_{p-1}}\,,
\eeq
so that 
\beq
 \iota_{\xi} \form{V} \equiv \frac{1}{(p-1)!} \xi^\lambda V_{\lambda \mu_1\ldots \mu_{p-1}}\ 
  dx^{\mu_1} \wedge dx^{\mu_2} \wedge \ldots \wedge dx^{\mu_{p-1}} \,.\\
\eeq

\item The Lie derivative of a $p$-form $\form{V}$ along a vector $\xi=\xi^{\mu}\partial_{\mu}$ in components is given by
\beq
(\lieD_\xi \form{V})_{\mu_1\ldots\mu_p}
\equiv \xi^\lambda \partial_\lambda V_{\mu_1\mu_2\ldots \mu_p} + \sum_{k=1}^p (\partial_{\mu_k} \xi^\lambda)V_{\mu_1\ldots\mu_{k-1}\lambda \mu_{k+1}\ldots\mu_p}\,.
\eeq
We can rewrite this using
\begin{equation}
\begin{split}
(\partial_{\mu_k} \xi^\lambda)V_{\mu_1\ldots\mu_{k-1}\lambda \mu_{k+1}\ldots\mu_p}
&= (-1)^{pk} \xi^\lambda \partial_{\mu_k} V_{\mu_{k+1} \ldots \mu_p \lambda \mu_1\ldots \mu_{k-1}} \\
&\qquad + (-1)^{(p-1)(k-1)} \partial_{\mu_k} \brk{ \xi^\lambda V_{\lambda \mu_{k+1}\ldots \mu_p\mu_1\ldots \mu_{k-1}} }\,,
\end{split}
\end{equation}
so that 
\begin{equation}
\begin{split}
(\lieD_\xi \form{V})_{\mu_1\ldots\mu_p}
&=  \xi^\lambda \partial_\lambda V_{\mu_1\mu_2\ldots \mu_p} 
+ \sum_{k=1}^p (-1)^{pk} \xi^\lambda \partial_{\mu_k} V_{\mu_{k+1} \ldots \mu_p \lambda \mu_1\ldots \mu_{k-1}} \\
&\qquad + \sum_{k=1}^p  (-1)^{(p-1)(k-1)} \partial_{\mu_k} \brk{ \xi^\lambda V_{\lambda \mu_{k+1}\ldots \mu_p\mu_1\ldots \mu_{k-1}} } \\
&= (\iota_\xi d\form{V}+d\iota_\xi \form{V})_{\mu_1\ldots\mu_p}
\end{split}
\end{equation}
where the final expression is Cartan's identity. The Lie derivative is important when computing the variation of tensors under infinitesimal coordinate and gauge transformations $\{\xi^{\mu},\Lambda\}$. For example, the variation of a tensor-valued form $\form{\fTheta}\ab{\alpha}{\beta}$ is given by
\begin{equation}\label{eq:deltafmTheta}
\begin{split}
\diffF \fTheta^\alpha{}_\beta&=  \lieD_\xi \fTheta^\alpha{}_\beta + [\fTheta^\alpha{}_\beta,\Lambda]\\
& = \prn{d \iota_\xi + \iota_\xi d}\fTheta^\alpha{}_\beta 
-\prn{\partial_\sigma\xi^\alpha} \fTheta^\sigma{}_\beta+\fTheta^\alpha{}_\sigma \partial_\beta\xi^\sigma
+ [\fTheta^\alpha{}_\beta,\Lambda]\\
& =  \prn{\tilde{D} \iota_\xi + \iota_\xi \tilde{D}}\fTheta^\alpha{}_\beta 
-\prn{\tilde{\nabla}_\sigma\xi^\alpha-\tilde{\mathrm{T}}^\alpha{}_{\sigma\nu}\xi^\nu} \Theta^\sigma{}_\beta\\
&\quad +\Theta^\alpha{}_\sigma \prn{\tilde{D}_\beta\xi^\sigma-\tilde{\mathrm{T}}^\sigma{}_{\beta\nu}\xi^\nu}
+ [\Theta^\alpha{}_\beta,\Lambda+ \xi^\sigma A_\sigma]\,,
\end{split}
\end{equation}
where in going from the second line to the third we have exchanged ordinary partial derivatives for covariant ones $\tilde{D}_{\mu}$ in terms of arbitrary connections $\{\tilde{\fA},\tilde{\form{\Gamma}}\ab{\alpha}{\beta}\}$. We have also defined the torsion $\tilde{T}\ab{\mu}{\nu\rho} \equiv -\tilde{\Gamma}\ab{\mu}{\nu\rho} + \tilde{\Gamma}\ab{\mu}{\rho\nu}$.

\item Given a metric $g$ we may define a volume form on spacetime, whose explicit expression is given by
\begin{equation}
 d^d x\sqrt{g\ \text{Sign}[g]} = \frac{\text{Sign}[g]}{d!} \varepsilon_{\mu_1\mu_2\ldots \mu_d} dx^{\mu_1}\wedge dx^{\mu_2}\wedge\ldots \wedge dx^{\mu_p}
\end{equation}
where $g$ denotes the determinant of the metric and $\text{Sign}[g]$ is its signature. Thus, 
the components of the volume form are given by $\varepsilon_{\mu_1\mu_2\ldots \mu_d}= \pm \sqrt{g\ \text{Sign}[g]} $ .

For pseudo-Riemannian metrics describing spacetime, we have $\text{Sign}[g]=-1$. We think of all other metrics as being obtained from such a pseudo-Riemannian metric via analytic continuation a la Wick rotation. Unfortunately, the standard signature function $\text{Sign}[g]$ is not analytic. We will fix this by taking $\text{Sign}[g]=-1$ even for complex metrics obtained by Wick rotation. Note that this  means, for example, that when $g$ is real and positive (as in the case of static Euclidean metrics), $\varepsilon_{\mu_1\mu_2\ldots \mu_d}$ is purely imaginary. The square root for complex Euclidean metrics is determined via analytic continuation under Wick rotation which fixes $\sqrt{-g}=-i\sqrt{g}$.
In particular, in Lorentzian signature we take $\varepsilon_{012\ldots(d-1)}\equiv  - \sqrt{-g} $
and  in the Euclidean signature, we take $\prn{\varepsilon_{012\ldots(d-1)} }_E\equiv  i\sqrt{g} $ .

\item We can use these components to write down a formula for the projector which projects covariant $p$-tensors to $p$-forms. In components it is
\begin{equation}
\begin{split}
\delta^{[\mu_1}_{[\nu_1} \delta^{\text{}\mu_2}_{\text{}\nu_2} \ldots \delta^{\mu_p]}_{\nu_p]} 
&= \frac{\text{Sign}[g]}{p!(d-p)!}\, 
\varepsilon^{\mu_1\mu_2\ldots\mu_p}{}_{\alpha_1 \ldots\alpha_{d-p}}\,
\varepsilon_{\nu_1\mu_2\ldots \nu_p}{}^{\alpha_1 \ldots\alpha_{d-p}} \\
&= \frac{\text{Sign}[g](-1)^{p(d-p)}}{p!(d-p)!}\,
\varepsilon^{\mu_1\mu_2\ldots\mu_p}{}_{\alpha_1 \ldots\alpha_{d-p}}\,\varepsilon^{\alpha_1 \ldots\alpha_{d-p}}{}_{\nu_1\mu_2\ldots \nu_p}\,.
\end{split}
\end{equation}
\item Given a metric $g$ on a $d$-dimensional manifold, we may define the Hodge star, which is a linear map that takes $p$-forms to $d-p$-forms. We define the action of the Hodge star on a $p$-form $\form{V}$ in components via
\begin{equation}
(\hodge \form{V})_{\mu_1 \mu_2\ldots \mu_{d-p}} 
\equiv \frac{\text{Sign}[g]}{p!} V_{\nu_1\nu_2\ldots \nu_p} \varepsilon^{\nu_1\nu_2\ldots\nu_p}{}_{\mu_1 \mu_2\ldots \mu_{d-p}}\,,
\end{equation}
or
\begin{equation}
\hodge \form{V} 
\equiv \frac{\text{Sign}[g]}{p!(d-p)!}\ V_{\nu_1\nu_2\ldots \nu_p}\ \varepsilon^{\nu_1\nu_2\ldots\nu_p}{}_{\mu_1 \mu_2\ldots \mu_{d-p}}\ 
dx^{\mu_1} \wedge dx^{\mu_2} \ldots \wedge dx^{\mu_{d-p}}\,.
\end{equation}
Note that acting on a $p$-form 
\[ {\hodge}^2 = \text{Sign}[g] (-1)^{p(d-p)} \,,\]
or alternately 
\[ {\hodge}^{-1} = \text{Sign}[g] (-1)^{p(d-p)}\ \hodge \,.\]

\item The definition above is equivalent to 
\beq
 \hodge \prn{ dx_{\nu_1}\wedge dx_{\nu_2}\ldots dx_{\nu_p} }
\equiv \frac{\text{Sign}[g]}{(d-p)!}\  \varepsilon_{\nu_1\nu_2\ldots\nu_p\mu_1 \mu_2\ldots \mu_{d-p}}\ 
dx^{\mu_1} \wedge dx^{\mu_2} \ldots \wedge dx^{\mu_{d-p}}\,,
\eeq
or 
\begin{equation}
\begin{split}
 dx^{\nu_1}\wedge dx^{\nu_2}\ldots dx^{\nu_p} 
 &\equiv \frac{ 1 }{(d-p)!}\  \varepsilon^{\mu_1 \mu_2\ldots \mu_{d-p}\nu_1\nu_2\ldots\nu_p}\
 \hodge\prn{dx_{\mu_1} \wedge dx_{\mu_2} \ldots \wedge dx_{\mu_{d-p}} } \\
&\equiv \frac{  (-1)^{p(d-p)} }{(d-p)!}\  \varepsilon^{\nu_1\nu_2\ldots\nu_p\mu_1 \mu_2\ldots \mu_{d-p}}\ 
\hodge\prn{dx_{\mu_1} \wedge dx_{\mu_2} \ldots \wedge dx_{\mu_{d-p}} }\,.
\end{split}
\end{equation}

\item One of the main uses of the last formula is in translating expressions of the following form 
into components
\begin{equation}
\hodge \form{V} = \form{A}_1 \wedge \form{A}_2 \wedge\ldots \wedge \form{A}_k \,.
\end{equation}
Here $\form{V}$ is a $d-p$-form, $\form{A}_1 $ is a $q_1$-form, $\form{A}_2 $ is a $q_2$-form etc. such that $q_1+q_2 + \ldots +q_k = p$. We have 
\begin{equation}
\begin{split}
\hodge \form{V} &= \form{A}_1 \wedge \form{A}_2 \wedge\ldots \wedge \form{A}_k \\
&= \frac{1}{q_1 ! q_2 ! \ldots q_k !} 
(A_1)_{\alpha_1\ldots \alpha_{q_1}} (A_2)_{\beta_1\ldots \beta_{q_2}} \ldots (A_k)_{\lambda_1\ldots \lambda_{q_k}} \\
&\qquad  \qquad dx^{\alpha_1}\wedge \ldots dx^{\alpha_{q_1}}\wedge dx^{\beta_1}\ldots dx^{\beta_{q_2}}\wedge \ldots  dx^{\lambda_1}\wedge\ldots 
dx^{\lambda_{q_k}}  \\
&= \frac{ 1 }{q_1 ! q_2 ! \ldots q_k !(d-p)!} 
 \varepsilon^{\mu_1\mu_2\ldots\mu_{d-p} \alpha_1\ldots \alpha_{q_1}\beta_1\ldots 
\beta_{q_2}\ldots \lambda_1\ldots \lambda_{q_k} } \\
&\qquad  \qquad \ 
(A_1)_{\alpha_1\ldots \alpha_{q_1}} (A_2)_{\beta_1\ldots \beta_{q_2}} \ldots (A_k)_{\lambda_1\ldots \lambda_{q_k}}
\hodge\prn{dx_{\mu_1} \wedge dx_{\mu_2} \ldots \wedge dx_{\mu_{d-p}} } \,,
\end{split}
\end{equation}
so that 
\begin{equation}\label{eq:starToComponents}
\begin{split}
\form{V}_{\mu_1\mu_2\ldots\mu_{d-p}} &= \frac{ 1 }{q_1 ! q_2 ! \ldots q_k !} 
 \varepsilon_{\mu_1\mu_2\ldots\mu_{d-p}}{}^{\alpha_1\ldots \alpha_{q_1}\beta_1\ldots 
\beta_{q_2}\ldots \lambda_1\ldots \lambda_{q_k} } \\
&\qquad  \qquad \ 
(A_1)_{\alpha_1\ldots \alpha_{q_1}} (A_2)_{\beta_1\ldots \beta_{q_2}} \ldots (A_k)_{\lambda_1\ldots \lambda_{q_k}}\,.
\end{split}
\end{equation}

\item Given two $p$-forms $\form{V}_1$ and $\form{V}_2$, $\form{V}_1\wedge \hodge \form{V}_2$ is a top form given by 
\begin{equation}
\form{V}_1 \wedge \hodge \form{V}_2 = d^dx \sqrt{-g}\ \frac{1}{p!}\ (V_1)_{\mu_1\mu_2\ldots\mu_p} (V_2)^{\mu_1\mu_2\ldots \mu_p} \,.
\end{equation}
We may then regard $\int \form{V}_1\wedge \hodge \form{V}_2$ as an inner product on the space of $p$-forms.

Given a $p$-form $\form{V}_1$ and a $q$-form $\form{V}_2$ with $q \geq p$, we have
\begin{equation}
 \form{V}_1 \wedge \hodge \form{V}_2 = 
\frac{1}{q!(q-p)!}\ \prn{V_1}_{\nu_1\nu_2\ldots\nu_p} \prn{V_2}^{\mu_1\mu_2\ldots \mu_{q-p}\nu_1\nu_2\ldots\nu_p } 
\hodge\prn{dx_{\mu_1} \wedge dx_{\mu_2} \ldots dx_{\mu_{q-p}} }\,.
\end{equation}
\item Under this inner product, we can define the co-exterior derivative, which takes flavor singlet $p$-forms to singlet $p-1$-forms. In components it acts as 
\beq
 (d^\dag \form{V})_{\mu_1\ldots \mu_{p-1}} \equiv  D^\lambda V_{\mu_1\ldots \mu_{p-1}\lambda} 
 = \frac{1}{\sqrt{-g}} \partial_\sigma \brk{\sqrt{-g}\  g^{\lambda\sigma} 
V_{\mu_1\ldots \mu_{p-1}\lambda}   }\,,
\eeq
so that 
\beq
 d^\dag \form{V} \equiv \frac{1}{(p-1)!} D^\lambda V_{ \mu_1\ldots \mu_{p-1}\lambda}\ 
  dx^{\mu_1} \wedge dx^{\mu_2} \wedge \ldots \wedge dx^{\mu_{p-1}}\,.
\eeq
Note that the co-exterior derivative obeys  $\hodge d^\dag =  d\hodge  $. This, in particular means that for a $p$-form $\form{V}$ we have
\begin{equation}
\begin{split}
\frac{1}{(p-1)!}D_{\mu_p} V^{\mu_1 \mu_2 \ldots \mu_{p-1} \mu_p}\
 &\hodge \prn{dx_{\mu_1}dx_{\mu_2}\ldots dx_{\mu_{p-1}}}\\
 &= \hodge d^\dag \form{V} =  d\hodge \form{V}  \\
 &= d\brk{ \frac{1}{p!} V^{\mu_1 \mu_2 \ldots \mu_{p-1} \mu_p}\
 \hodge \prn{dx_{\mu_1}dx_{\mu_2}\ldots dx_{\mu_p}} }\,.
\end{split}
\end{equation}

\item It is useful to write various currents in terms of forms using the Hodge star. To do it, let us begun by defining the hypersurface volume forms $d^{d-1}\mathrm{S}_\mu$ : the $d-1$ forms in $d$ dimensional spacetime which when pulled back and integrated over a hypersurface give the volume of that hypersurface. More precisely they are the Hodge-duals of the basis 1-forms $dx^\mu$. If $g_{\mu\nu}$ represents the metric on spacetime, we can define $d^{d-1}\mathrm{S}_\mu$ via the relations 
\beq
d^{d-1}\mathrm{S}_{\nu}\equiv \hodge dx_{\nu}\,,
\eeq
or
\beq
dx^\mu\wedge  d^{d-1}\mathrm{S}_\nu=\delta^\mu_{\nu}\ d^dx\sqrt{-g} \,.
\eeq
Using these forms, we can define the Hodge-duals of the currents which are tensor valued $(d-1)$-forms
and the Hodge-dual of energy-momentum tensor which is a tensor valued $d$-form. We have
\begin{equation}
\begin{split}
\hodge \fJ  &\equiv   (d^{d-1}\mathrm{S}_\lambda)\ J^\lambda\ \ ,\hspace{.42in}
\hodge \form{\spp}^\mu{}_\nu  \equiv   (d^{d-1}\mathrm{S}_\lambda)\ \spp^{\lambda\mu}{}_\nu\,,\\
\hodge \form{T}^{\mu\nu}  &\equiv   (d^dx\sqrt{-g})\ T^{\mu\nu}\ \ ,\qquad
\hodge \form{t}^{\mu\nu}  \equiv   (d^dx\sqrt{-g})\ t^{\mu\nu}\,,
\end{split}
\end{equation}
so that we can write 
\begin{equation}\label{eq:deltaWdefForms}
\begin{split}
\delta W  &\equiv  \int \Bigl\{  \delta \fA  \cdot \hodge \fJ 
+ \half \delta \fGamma^\mu{}_{\nu} \ \hodge \form{\spp}^\nu{}_\mu +\half \delta g_{\mu\nu} \hodge \form{t}^{\mu\nu}\Bigr\} + (\text{boundary terms}) \\
&=  \int \Bigl\{  \delta \fA  \cdot \hodge \fJ 
+ \half \delta g_{\mu\nu} \hodge \form{T}^{\mu\nu}\Bigr\} + (\text{boundary terms})\,.
\end{split}
\end{equation}
\end{itemize}

\end{appendix}

\bibliographystyle{JHEP}
\bibliography{refs}
\end{document}